\documentclass[preprint]{aastex62}
\usepackage{graphicx}
\usepackage{color}
\usepackage{amsmath,amssymb}
\usepackage{ulem}
\usepackage{bm}

\def\vec#1{\mbox{\boldmath $#1$}}

\received{September	 2, 2019}
\revised{November 22, 2019}
\accepted{November 26, 2019}
\submitjournal{ApJ}
\shorttitle{Matter mixing}
\shortauthors{Ono et al.}
\begin{document}

\title{Matter Mixing in Aspherical Core-collapse Supernovae: Three-dimensional Simulations with Single Star and Binary Merger Progenitor Models for SN 1987A}

\correspondingauthor{Masaomi Ono}
\email{masaomi.ono@riken.jp}

\author[0000-0002-0603-918X]{Masaomi Ono}
\affil{Astrophysical Big Bang Laboratory, RIKEN Cluster for Pioneering Research, 2-1 Hirosawa, Wako, Saitama 351-0198, Japan}
\affil{RIKEN Interdisciplinary Theoretical and Mathematical Science Program (iTHEMS), 2-1 Hirosawa, Wako, Saitama 351-0198, Japan}
\author{Shigehiro Nagataki}
\affil{Astrophysical Big Bang Laboratory, RIKEN Cluster for Pioneering Research, 2-1 Hirosawa, Wako, Saitama 351-0198, Japan}
\affil{RIKEN Interdisciplinary Theoretical and Mathematical Science Program (iTHEMS), 2-1 Hirosawa, Wako, Saitama 351-0198, Japan}
\author{Gilles Ferrand}
\affil{Astrophysical Big Bang Laboratory, RIKEN Cluster for Pioneering Research, 2-1 Hirosawa, Wako, Saitama 351-0198, Japan}
\affil{RIKEN Interdisciplinary Theoretical and Mathematical Science Program (iTHEMS), 2-1 Hirosawa, Wako, Saitama 351-0198, Japan}
\author{Koh Takahashi}
\affil{Max-Planck-Institut f\"{u}r Gravitationsphysik (Albert-Einstein-Institute) Am M\"{u}hlenberg 1, D-14476 Potsdam-Golm, Germany}
\author{Hideyuki Umeda}
\affil{Department of Astronomy, Graduate School of Science, The University of Tokyo, 7-3-1 Hongo, bunkyo-ku, Tokyo 113-0033, Japan}
\author{Takashi Yoshida}
\affil{Department of Astronomy, Graduate School of Science, The University of Tokyo, 7-3-1 Hongo, bunkyo-ku, Tokyo 113-0033, Japan}
\author{Salvatore Orlando}
\affil{INAF -- Osservatorio Astronomico di Palermo, Piazza del Parlamento 1, 90134 Palermo, Italy}
\author{Marco Miceli}
\affil{Dipartimento di Fisica e Chimica, Universit$\grave{a}$ degli Studi di Palermo, Piazza del Parlamento 1, 90134 Palermo, Italy}
\affil{INAF -- Osservatorio Astronomico di Palermo, Piazza del Parlamento 1, 90134 Palermo, Italy}

\begin{abstract}
We perform three-dimensional hydrodynamic simulations of aspherical core-collapse supernovae focusing on the matter mixing in SN 1987A. 
The impacts of four progenitor (pre-supernova) models and parameterized aspherical explosions are investigated. 
The four pre-supernova models include a blue supergiant (BSG) model based on a slow merger scenario developed recently for the progenitor 
of SN 1987A \citep{2018MNRAS.473L.101U}. The others are a BSG model based on a single star evolution and two red supergiant (RSG) 
models. Among the investigated explosion (simulation) models, a model with the binary merger progenitor model and with an asymmetric bipolar-like 
explosion, which invokes a jetlike explosion, 
best reproduces constraints on the mass of high velocity $^{56}$Ni, as inferred from the observed [Fe II] line profiles. 
%
%
The advantage of the binary merger progenitor model for the matter mixing is the flat and less extended $\rho \,r^3$ profile of the C+O core and the helium layer, 
which may be characterized by the small helium core mass. 
%
%
From the best explosion model, the direction of the bipolar explosion axis (the strongest explosion direction), the neutron star (NS) kick velocity, and 
its direction are predicted. 
Other related implications and future prospects are also given.

\end{abstract}

\keywords{Supernova --- SN 1987A}

\section{Introduction} \label{sec:intro}

From the discovery of supernova 1987A (SN 1987A) in the Large Magellanic Cloud on February 23, 1987, more than 30 years have passed and 
it has been in a phase of young supernova remnant. So far, there have been many observations of SN 1987A in a wide range of wavelengths 
\citep[for a review of observational features of SN 1987A, see e.g.,][]{1989ARA&A..27..629A,1993ARA&A..31..175M,2016ARA&A..54...19M}. 
Then, SN 1987A provides a unique opportunity to investigate the evolution from the supernova to the supernova remnant thanks to its age 
($\sim$ 30 yr) and proximity ($\sim$ 50 kpc). In order to extract information from the observations of the supernova remnant. 
theoretical modeling of the evolution from the explosion to its supernova remnant is indispensable. In this paper, we theoretically investigate an 
early evolution of SN 1987A (up to a few days) focusing on the matter mixing and related observables; we have a plan to 
link the evolution to phases of the supernova remnant (Orlando et al. 2019, in preparation). 

Large-scale matter mixing has been indicated from observations of SN 1987A as follows. Early detections of hard X-ray emission 
\citep{1987Natur.330..230D, 1987Natur.330..227S} and direct $\gamma$-ray lines from the decay of $^{56}$Co 
\citep{1988Natur.331..416M,1990MNRAS.245..570V} 
have revealed the existence of radioactive $^{56}$Ni in high velocity outer layers in the expanding ejecta consisting of helium and hydrogen. 
It is noted that the $^{56}$Co was the decay product of $^{56}$Ni (in the sequence of $^{56}$Ni$\,\rightarrow\,$$^{56}$Co$\,\rightarrow\,$$^{56}$Fe) 
that had been synthesized by the explosive nucleosynthesis during the explosion \citep[the half-lives of the sequence, 
$^{56}$Ni$\,\rightarrow\,$$^{56}$Co$\,\rightarrow\,$$^{56}$Fe, are 6.1 days and 77 days, respectively:][]{1994ApJS...92..527N}. The fine structure 
developed in the ${\rm H}_{\alpha}$ line \citep[the so-called Bochum event:][]{1988MNRAS.234P..41H} has also implied the existence of clumps of 
high velocity ($\sim$ 4700 km s$^{-1}$) $^{56}$Ni with a few 10$^{-3} M_{\odot}$ \citep{1995A&A...295..129U}. Observed emission lines of [Fe II] 
(18 $\mu$m and 26 $\mu$m) from SN 1987A at $\sim$ 400 days after the explosion \citep{1990ApJ...360..257H} have shown that the tails of the 
distribution of Doppler velocities reach $\sim$ 4000 km s$^{-1}$ and the centroids of the lines are redshifted (for 18 $\mu$m and 26 $\mu$m, the 
centroids are at 450 $\pm$ 200 km s$^{-1}$and 680 $\pm$ 200 km s$^{-1}$, respectively). It has been interpreted that between 4\% to 17\% of the 
iron had a high velocity of $\gtrsim$ 3000 km s$^{-1}$ \citep{1990ApJ...360..257H}. Later observations of [Ni II] lines from SN 1987A at $\sim$ 640 
days have also indicated similar high velocity of iron ($\sim$ 3000 km s$^{-1}$) \citep{1994ApJ...427..874C}. 
The spectral modeling of the late phase (200 -- 2000 days) of SN 1987A has revealed inward mixing of hydrogen 
down to velocities of $\lesssim$ 700 km s$^{-1}$ \citep{1998ApJ...497..431K}. 
%
Theoretical studies based on one-dimensional hydrodynamical simulations with radiative transfer have shown that the early appearance of hard X-ray 
emission and $\gamma$-ray lines, optical light curves cannot be explained without some degree of mixing of $^{56}$Ni into fast moving outer layers 
\citep{1988ApJ...329..820P, 1988ApJ...330..218W, 1990ApJ...360..242S}. To reproduce the observed optical light curves, artificial mixing of $^{56}$Ni 
up to velocities of 3000--4000 km s$^{-1}$, is necessary \citep{1990ApJ...360..242S, 2000ApJ...532.1132B}. In addition to $^{56}$Ni, 
\citet{1990ApJ...360..242S} and \citet{2000ApJ...532.1132B} have 
also 
insisted on inward mixing of hydrogen down to velocities of 800 and 1300 km s$^{-1}$, 
respectively, 
although the values are higher than that deduced from the spectral modeling \citep{1998ApJ...497..431K}.

What is the mechanism of the mixing? Rayleigh-Taylor (RT) instability has been considered to be one of possible mechanisms of the matter mixing 
in core-collapse supernovae. The RT unstable condition is described as 
$\nabla P \cdot \nabla \rho < 0$ \citep{1976ApJ...207..872C}, where $P$ is the pressure and $\rho$ is the density. 
One-dimensional hydrodynamical simulations with a progenitor model for SN 1987A have shown that  the unstable condition could be realized during 
the shock propagation around the interface between C+O core and helium layer (C+O/He) and one between helium and hydrogen layers (He/H) 
\citep{1989ApJ...344L..65E, 1990ApJ...348L..17B}. 

Motivated by the observational evidence of the matter mixing in SN 1987A, multi-dimensional hydrodynamical simulations of the propagation 
of the supernova shock wave have been performed focusing on the development of RT instabilities 
\citep{1989ApJ...341L..63A, 1990ApJ...358L..57H, 1992ApJ...390..230H, 1991ApJ...367..619F, 1991A&A...251..505M, 1991ApJ...370L..81H, 
1992ApJ...387..294H}. All the studies have assumed spherical symmetry in the explosions and among the investigated models the obtained 
maximum velocity of $^{56}$Ni is only $\sim$ 2000 km s$^{-1}$ \citep{1991ApJ...370L..81H, 1992ApJ...387..294H}. 
Hence, aspherical explosions which had not been considered in those studies above might be necessary to explain the observations. 

Recent observations of emission lines of [Si I] $+$ [Fe II] and He I (1.644 $\mu$m and 2.058 $\mu$m, respectively) by HST and VLT 
\citep{2010A&A...517A..51K,2013ApJ...768...89L,2016ApJ...833..147L} have revealed that the three-dimensional (3D) morphology of the inner ejecta of 
SN 1987A is globally elliptical/elongated (the ratio of the major to minor axes of the inner ejecta is 1.8 $\pm$ 
0.17 
\citep{2010A&A...517A..51K}. 
It has been known that in the nebula around SN 1987A there is a triple ring structure consisting of an inner equatorial ring (ER) and two outer rings (ORs) and 
the configurations with respect to the Earth have been deduced \citep[e.g.,][]{2011A&A...527A..35T,2005ApJ...627..888S}. 
The optical spectroscopy of the light echos of SN 1987A has also indicated asymmetries in the line profiles of H$_{\alpha}$ and Fe II, which is consistent 
with the elongated ejecta, and two-sided distribution of $^{56}$Ni \citep{2013ApJ...767...45S}. 
New spots and diffuse emission outside the ER found by more recent HST observations may provide additional 
insights into the evolution of the ER and ejecta \citep{2019arXiv191009582L}. 
Recent observations of spatially resolved 3D distributions of the rotational transition lines of CO and SiO molecules 
by AMLA have indicated that the distributions 
are not spherical at all and clumpy \citep{2017ApJ...842L..24A}. 
Further observations by ALMA have revealed that dust emission from the the inner ejecta is also clumpy 
and asymmetric \citep{2019arXiv191002960C}.

Direct emission lines from the decay of long-lived radioisotope $^{44}$Ti \citep[the half-life of the decay sequence, 
$^{44}$Ti$\,\rightarrow\,$$^{44}$Sc$\,\rightarrow\,$$^{44}$Ca, is 58.9 $\pm$ 0.3 yr:][]{2006PhRvC..74f5803A}, which was the product of the 
explosive nucleosynthesis, have been observed in SN 1987A \citep{2012Natur.490..373G,2015Sci...348..670B}. The initial mass of $^{44}$Ti 
was estimated as (3.1 $\pm$ 0.8) $\times$ 10$^{-4}$ $M_{\odot}$ in \citet{2012Natur.490..373G}. 
Recent observations by NuSTAR \citep{2015Sci...348..670B} have also revealed that $^{44}$Ti gamma-ray lines have been redshifted with 
a velocity of 700 $\pm$ 400 km s$^{-1}$, which invokes a large-scale asymmetry in the explosion of SN 1987A, and the initial mass of $^{44}$Ti 
was estimated as (1.5 $\pm$ 0.3) $\times$ 10$^{-4}$ $M_{\odot}$ \citep{2015Sci...348..670B}. 

Despite searching for more than 30 years, the compact object of SN 1987A has not been detected yet. From the fact that 
there has been no detection from millimeter, near-infrared, optical, ultraviolet, and X-ray observations, several constraints on the compact object have 
been argued and it has been inferred that the compact object is a thermally emitting NS obscured by dust \citep{2015ApJ...810..168O,2018ApJ...864..174A}. 
Meanwhile, X-ray observations of nearby core-collapse supernova remnants, 
e.g., Cassiopeia A (Cas A), 
have revealed that the direction of a NS motion relative to the explosion center is opposite to the gaseous intermediate elements in the supernova ejecta 
and it has been inferred that the NS kicks stem from asymmetric explosive mass ejections \citep{2018ApJ...856...18K}. 
Theoretically, NS kicks are expected by neutrino-driven core-collapse supernova explosions thanks to their aspherical nature 
\citep{2010ApJ...725L.106W, 2013A&A...552A.126W}. 
{Recent ALMA observations of dust emission from the ejecta of SN 1987A have insisted that a dust peak found at the 
northeast of the center of the remnant could be an indirect detection of the compact object \citep{2019arXiv191002960C}. 

The mechanisms of core-collapse supernova explosions have not been elucidated yet. Theoretically, it has been considered that a canonical core-collapse 
supernova could be triggered by the delayed neutrino heating aided by convection \citep{1994ApJ...435..339H} and/or standing accretion shock instability 
(SASI) \citep{2003ApJ...584..971B}, where multi-dimensional effects are essential 
\citep[for a review of the mechanism of core-collapse supernovae, see][]{2012PTEP.2012aA301K,2012AdAst2012E..39K,2012ARNPS..62..407J,
2012PTEP.2012aA309J,2013RvMP...85..245B,2016PASA...33...48M}. Hitherto, based on the delayed neutrino heating mechanism many two-dimensional 
(2D) and 3D hydrodynamical simulations with an approximate neutrino transport have been performed for a few decades 
\citep{1995ApJ...450..830B,2003A&A...408..621K, 2006A&A...457..963S, 2006A&A...453..661K, 2008A&A...477..931S, 2009ApJ...694..664M, 2010PASJ...62L..49S, 2010ApJ...720..694N, 
2012ApJ...756...84M, 2012ApJ...761...72M, 2012ApJ...749...98T, 2012ApJ...755...11K, 2012A&A...537A..63M, 2013ApJ...767L...6B, 2013A&A...552A.126W, 
2013ApJ...778L...7C, 2013ApJ...770...66H, 2013ApJ...765..110D, 2013ApJ...768..115O, 2014ApJ...782...91N, 2014ApJ...786...83T, 2014ApJ...785..123C, 
2015ApJ...799....5C, 2016ApJ...818..123B, 2016ApJ...817...72P, 2016ApJ...820...76R, 2017ApJS..229...42N, 2017ApJ...850...43R, 2017MNRAS.472..491M, 
2019MNRAS.482..351V, 2019MNRAS.485.3153B}. 
%
%
%
%
Since the involved physical effects, e.g., neutrino transport, general relativity, and nuclear equation of state, 
are rather complicated and multi-dimensional ab initio hydrodynamical simulations of core-collapse supernovae are rather demanding in the viewpoint of numerical 
costs, the adopted physical effects and their approximations in particular for the neutrino transport have been rather varied among simulations; a consensus has not 
been reached yet. Actually, comparisons of the results between 2D and 3D simulations have been made and the explodabilities in 3D relative 
to those in 2D are controversial \citep{2010ApJ...720..694N, 2013ApJ...770...66H, 2013ApJ...765..110D, 2014ApJ...786...83T}. 
It is noted that a strong sloshing mode ($l = 1$) of SASI seen in 2D, which makes an asymmetric dipolar morphology of the shock, 
tends to be less evident in 3D at later phases of the shock revival \citep{2010ApJ...720..694N, 2013ApJ...770...66H}. 

On the other hand, magnetohydrodynamical (MHD) simulations of core-collapse supernovae have demonstrated jetlike magnetorotationally-driven explosions 
\citep{2004ApJ...608..391K,2005ApJ...631..446S,2007ApJ...664..416B,2009ApJ...691.1360T,2013ApJ...764...10S,2014ApJ...785L..29M,2016ApJ...817..153S}. 
For a successful launch of a jet, generally both strong magnetic field and rapid rotation before the core-collapse are necessary; 
however, it has not been unveiled yet from which evolutionary paths the both conditions are fulfilled \citep[for an example of stellar evolution calculations of a single 
massive star with magnetic field, see e.g.,][]{2005ApJ...626..350H}. 
The magnetorotational instability \citep{1998RvMP...70....1B} could play a significant role for the amplification of magnetic field during the core-collapse and shock 
revival but high resolution simulations are necessary to capture the fastest growing mode and it is difficult to assess its role by global hydrodynamical simulations. 
See e.g., \citet{2016ApJ...817..153S} for an attempt to investigate the impact of the magnetorotational instability on core-collapse supernovae. 

In the context of matter mixing in core-collapse supernovae, possible effects of aspherical core-collapse supernova explosions have been investigated 
based on multi-dimensional hydrodynamic simulations. 
%
Effects of mildly jetlike explosions on matter mixing have been studied based on 2D hydrodynamical simulations with a progenitor model for SN 1987A 
\citep{1991ApJ...382..594Y,1998ApJ...495..413N, 2000ApJS..127..141N}. 
\citet{1998ApJ...495..413N} and \citet{2000ApJS..127..141N} have obtained high velocity $^{56}$Ni corresponding to the tails (up to $\sim$ 3000 km s$^{-1}$) 
of [Fe II] lines with a large amplitude (30\%) of perturbations in velocities at the phase when the shock wave reaches the He/H interface. In the context of jetlike 
explosions, \citet{1997ApJ...486.1026N,1998ApJ...492L..45N} have suggested that a jetlike explosion enhances the amount of $^{44}$Ti synthesized by the 
explosive nucleosynthesis thanks to a strong alpha-rich freezeout. 
\citet{2003ApJ...594..390H,2005ApJ...635..487H} have investigated the effects of jetlike and single-lobe explosions on the $\gamma$-ray lines using 
a 3D smoothed particle hydrodynamical (SPH) code.
\citet{2000ApJ...531L.123K,2003A&A...408..621K,2006A&A...453..661K} have investigated matter mixing with more realistic explosion models 
based on 2D high resolution hydrodynamical simulations (with adaptive mesh refinement: AMR) of neutrino-driven core-collapse supernovae aided by convection 
and/or SASI.
The authors have found that a globally aspherical explosion dominated by low-order unstable modes ($l = 1, 2$) with the explosion energy of 2 $\times$ 10$^{51}$ erg produces 
high velocity $^{56}$Ni clumps ($\sim$ 3300 km s$^{-1}$) \citep{2006A&A...453..661K}. 
\citet{2009ApJ...693.1780J,2010ApJ...709...11J,2010ApJ...723..353J} have studied the development of RT instabilities in spherical core-collapse 
supernovae of solar-metallicity, metal-poor, and zero-metallicity massive stars based on 2D and 3D hydrodynamical simulations. 
If a star ends its life as a compact BSG, the mixing by RT instability is significantly reduced and fallback is enhanced compared with those of RSGs. 
Thus, the structure of the progenitor star could be essential for matter mixing.
%
%
%
%
\citet{2012ApJ...755..160E} studied RT mixing in a series of aspherical 
core-collapse supernova explosions using a 3D SPH code and sizes of arising clumps were studied 
based on Fourier transformations. 

The effects of the dimensionality of hydrodynamical simulations on the matter mixing in core-collapse supernovae have been controversial. 
The growth of RT instabilities in 3D simulations of a spherical supernova explosion is faster than that in corresponding 2D simulations 
but the widths of the mixed regions at the time of the saturation are similar in 2D and 3D in the end \citep{2010ApJ...723..353J}. 
On the other hand, \cite{2010ApJ...714.1371H} has demonstrated an effective mixing in 3D due to the 
faster growth of RT fingers and the less deceleration of metal-rich clumps compared with that in the corresponding 2D simulation. 
Generally, the resolutions of 2D hydrodynamical simulations can be higher than those of 3D ones; however,  
axisymmetric 2D Eulerian hydrodynamical simulations could introduce numerical artifacts around the polar axis \citep{2010A&A...521A..38G}. 
In keeping with the different behaviors between 2D and 3D and the defect of possible numerical artifacts in axisymmetric 2D simulations, 3D high resolution 
simulations are necessary for a study of matter mixing.

In our previous papers \citep[][hereafter Paper~I and Paper~II, respectively]{2013ApJ...773..161O,2015ApJ...808..164M}, we have systematically investigated the 
matter mixing in SN 1987A based on 2D hydrodynamical simulations with an AMR code with an ad hoc way of the initiation of explosions. 
In Paper~I, we explored parametrically the 
impact of mildly aspherical explosions with a clumpy structure on the distribution of the radial velocities of $^{56}$Ni and the line of sight velocity distribution of $^{56}$Ni, 
which corresponds to the observed velocity profiles of [Fe II] lines. It was found that the maximum velocity 
of $^{56}$Ni is at most $\sim$ 3000 km s$^{-1}$. In Paper~II, possible effects of large perturbations in the density of the progenitor star were explored and at most 
$\sim$ 4000 km s$^{-1}$ of $^{56}$Ni can be obtained by an asymmetric bipolar explosions with radially coherent perturbations (amplitude of 50\%) in the density of 
the progenitor star. The obtained line of sight velocity distribution of $^{56}$Ni, however, seems to be 
different from those of the observed [Fe II] line profiles.

\citet{2015A&A...577A..48W} investigated the dependence of matter mixing on progenitor models based on 3D hydrodynamical simulations of neutrino-driven 
core-collapse supernovae from the shock revival to the shock breakout. It was found that the extent of mixing depends sensitively on the density structure of the 
progenitor model, i.e., the sizes of C+O core and helium layer and the density gradient at the He/H interface. 
In RSG models, high velocity $^{56}$Ni of 4000--5000 km s$^{-1}$ is obtained. 
In a 15 $M_{\odot}$ BSG model, relatively high velocity $^{56}$Ni ($\sim$ 3500 km s$^{-1}$) is obtained. 
On the other hand, in a 20 $M_{\odot}$ BSG model, the maximum $^{56}$Ni velocity is only $\sim$ 2200 km s$^{-1}$ 
because of the strong deceleration of inner ejecta by the reverse shock and insufficient time for the growth of RT instabilities at the He/H interface. 
\citet{2015A&A...581A..40U} modeled optical light curves based on part of 3D hydrodynamical models above. 
Among the investigated models, only one BSG model reproduces 
the dome-like shape of the light curve maximum of SN 1987A. As the authors mentioned, the mass of the helium core of the progenitor model is, however, only 
$\sim$ 4 $M_{\odot}$, which is less than the value for the progenitor star of SN 1987A 
\citep[6 $\pm$ 1 $M_{\odot}$:][see below for the details]{1989ARA&A..27..629A}.

The properties of the progenitor star of SN 1987A have been obtained from observations \citep[for a review, see][]{1989ARA&A..27..629A}. The progenitor was 
identified as a compact B3 Ia BSG, Sanduleak $-$69$^{\circ}$ 202 (hereafter Sk $-$69$^{\circ}$ 202) \citep{1987A&A...177L...1W,1987ApJ...321L..41W}. The 
estimated intrinsic bolometric magnitude is translated into the luminosity of (3--6) $\times$ 10$^{38}$ erg s$^{-1}$. The effective temperature is $\sim$ 16,000 K 
\citep{1984ApJ...284..565H} with a probable range of 15,000--18,000 K \citep{1989ARA&A..27..629A}. 
From models of massive stars, the helium core mass of Sk $-$69$^{\circ}$ 202 could be in the range of 6 $\pm$ 1 $M_{\odot}$ \citep[e.g,][]{1988ApJ...330..218W}. 
%
Another notable feature related to the progenitor of SN 1987A is the triple ring structure discovered around Sk $-$69$^{\circ}$ 202 after the supernova event 
\citep{1990ApJ...362L..13W,1995ApJ...452..680B}, which invokes an axisymmetric but non-spherical 
mass ejection during the stellar evolution. 
The expansion velocities of the rings, the inner ER and the two ORs, have been deduced as $\sim$ 10 km s$^{-1}$ and $\sim$ 26 km s$^{-1}$, respectively 
\citep{1991Natur.350..683C,2000ApJ...528..426C}, which are consistent with wind velocities of RSGs; it has been interpreted that the three 
rings were ejected at least $\sim$ 20,000 yr ago, i.e., Sk $-$69$^{\circ}$ 202 could have been a RSG about 20,000 yr ago 
\citep{1991Natur.350..683C,1995ApJ...452..680B}. 
Additionally, anomalous abundances of helium and CNO-processed elements in the circumstellar material including the rings have been reported from observations of emission lines, 
i.e., He/H (number ratio) = 0.25 $\pm$ 0.05 
\citep{1996ApJ...464..924L}, He/H = 0.17 $\pm$ 0.06 \citep{2010ApJ...717.1140M}, 
He/H = 0.14 $\pm$ 0.06 \citep{2011ApJ...743..186F}, N/C = 7.8 $\pm$ 4 \citep{1989ApJ...336..429F}, N/C = 5.0 $\pm$ 2.0 \citep{1996ApJ...464..924L}, 
N/O = 1.6 $\pm$ 0.8 \citep{1989ApJ...336..429F}, N/O = 1.1 $\pm$ 0.4 \citep{1996ApJ...464..924L}, and N/O = 1.5 $\pm$ 0.7 \citep{2010ApJ...717.1140M}. 
These abundance ratios indicate an enhancement of material that underwent hydrogen burning through CNO cycle in the nebula. The problem, however, is how 
the products of the hydrogen burning had been mixed into the hydrogen envelope and the nebula in the end.

Hitherto, there have been many attempts to construct single star evolution models which satisfy at least a part of the requirements for Sk $-$69$^{\circ}$ 202 
mentioned above. A major issue in single star models, however, is that extreme fine tuning of parameters related to specific assumptions, e.g., reduced metallicity 
\citep{1989ARA&A..27..629A}, enhancements of the mass-loss and helium abundance in the hydrogen envelope \citep{1988ApJ...331..388S}, restricted 
convection \citep{1988ApJ...324..466W}, and rotationally-induced mixing \citep{1988A&A...197L..11W}, is necessary in order for the progenitor to end as a BSG 
and/or to obtain the abundance anomalies. 
Another unignorable issue in single star scenarios is how the triple ring nebula could be formed in this context. 
If a progenitor star is rapidly rotating, the envelope could obtain considerable angular momentum by a spin-up mechanism \citep{1998A&A...334..210H}. 
\citet{2008A&A...488L..37C} performed 2D hydrodynamical simulations of the evolution of the wind nebula of a 12 $M_{\odot}$ star with a blue loop 
(red-blue-red evolution) in the Hertzsprung-Russell (HR) diagram and the formation of a triple ring structure was demonstrated during the blue phase thanks to the 
spin-up mechanism \citep{1998A&A...334..210H}. The star, however, ends its life as a RSG. To date there has been no single star model which 
satisfies all the observational features of Sk $-$69$^{\circ}$ 202 
\citep[for reviews on the progenitor of SN 1987A, see,][]{1989ARA&A..27..629A,1992PASP..104..717P,2009ARA&A..47...63S}. 
On the other hand, evolution models for Sk $-$69$^{\circ}$ 202 based on binary mergers through a common envelope interaction have been proposed 
\citep[][]{1990A&A...227L...9P,1992ApJ...391..246P} \citep[see][for a related common envelop model]{1989A&A...219L...3H} as alternative (and probably 
more natural) explanations of the red-to-blue evolution, the abundance anomalies in the nebula, and the formation of the triple ring nebula 
(for the overall binary merger scenario, see Section~\ref{subsec:pre-sn}). 
Along this scenario, \citet{2002MNRAS.334..819I} demonstrated the penetration of the material from the secondary into the core of 
the primary based on 2D hydrodynamical simulations. 
Later, 
\citet{2007Sci...315.1103M,2009MNRAS.399..515M} successfully reproduced the formation of a triple ring structure very similar to the observed one 
based on 3D SPH simulations. 
Recently, progenitor models for SN 1987A based on the binary merger scenario have been developed by two independent 
groups \citep{2017MNRAS.469.4649M,2018MNRAS.473L.101U}. They has successfully found appropriate models that satisfy all the observational 
features of Sk $-$69$^{\circ}$ 202 mentioned above. 
Compared with \citet{2017MNRAS.469.4649M}, \citet{2018MNRAS.473L.101U} included the effects of the spin-up of the envelope due to the angular momentum 
transfer from the orbit. 
Additionally, recent light curve modeling for SN 1987A \citep{2019MNRAS.482..438M} based on the binary merger models \citep{2017MNRAS.469.4649M} 
have shown that the models better fit to the observed optical light curves than single star models. 
Recently, direct $\gamma$-rays from the decay of $^{56}$Ni and the scattered X-rays have been theoretically 
investigated based on 3D hydrodynamical models of neutrino-driven core-collapse supernovae with some binary merger progenitor models 
\citep{2019ApJ...882...22A}. 
%
%
%
%
%

In the context of matter mixing in SN 1987A, the studies that have obtained high velocity $^{56}$Ni ($\gtrsim$ 3000 km s$^{-1}$) have investigated only 
single progenitor star models. \citet{2006A&A...453..661K}, \citet{2010ApJ...714.1371H}, and \citet{2015A&A...577A..48W} have used a 15 $M_{\odot}$ BSG model B15 
\citep{1988ApJ...324..466W}\citep[denoted as W15 in][]{2016ApJ...821...38S} to obtain the high velocity $^{56}$Ni, but the 
luminosity of the pre-supernova model is outside the observational constraints. Whereas, with a BSG model \citep{1988PhR...163...13N,1990ApJ...360..242S} 
corresponding to the main sequence mass of 20 $M_{\odot}$ \citep[denoted as N20 in][]{2015A&A...577A..48W}, only lower velocity of $^{56}$Ni 
($\lesssim$ 3000 km s$^{-1}$) has been achieved \citep[e.g., Paper~I;][]{2015A&A...577A..48W}, although the model satisfies the final position in the 
HR diagram.\footnote{For the positions of the two BSG models in the HR diagram, see the points denoted as W15 and 
N20 in the Figure~2 in \citet{2016ApJ...821...38S}.} 
Hitherto, there has been no consistent hydrodynamical model that explains the observed high velocity $^{56}$Ni with a single progenitor star model that fulfills all the 
observational requirements for Sk $-$69$^{\circ}$ 202. Recently, \citet{2019A&A...624A.116U} revisited the modeling of light curves for larger variety of BSG 
models than that in \citet{2015A&A...581A..40U}; it was confirmed that there is no single star model that matches all observational features. 
Therefore, it is worth revisiting the matter mixing in SN 1987A with a binary merger model. 

Motivated by recent observations of the supernova remnant of SN 1987A, 3D hydrodynamical/MHD simulations of the interaction of the expanding ejecta with 
the ER have been performed, focusing on the X-ray and/or radio emission \citep{2014ApJ...794..174P, 2015ApJ...810..168O, 2019A&A...622A..73O}. 
Recently, \citet{2019NatAs...3..236M} compared the 3D hydrodynamical model \citep{2015ApJ...810..168O} with observed X-ray spectra of the 
remnant of SN 1987A. 
Although the morphology of the inner ejecta of SN 1987A is obviously non-spherical \citep[e.g.,][]{2016ApJ...833..147L}, 
in those studies, spherical symmetry has been assumed in the explosions and no realistic stellar evolution model has been used. 
In order to maximize the information which can be extracted by comparing theories with observations of the remnant, 3D 
hydrodynamical models of aspherical explosions with a realistic stellar evolution model are imperative. 

The purpose of this paper is to investigate the impact of progenitor models and parameterized aspherical explosions on the matter mixing in SN 1987A 
and related observational outcomes, in particular the line of sight velocity distribution of $^{56}$Ni corresponding to the [Fe II] line profiles, which 
may provide non-trivial information on the morphology of the inner ejecta and the configuration relative to the triple ring nebula. 
In order to accomplish this, we perform 3D hydrodynamical simulations of core-collapse supernova explosions with four pre-supernova models, 
two are BSG models and the other two are RSG models. 
It is noted that a recent binary merger BSG model \citep{2018MNRAS.473L.101U} is adopted 
for the study of the matter mixing for the first time. 
First, we perform many lower resolution simulations to explore a wide range of parameters related to the asphericities of the explosion and the progenitor dependence. 
In Paper~II, the impact of large density perturbations in the progenitor star was investigated; however, in order to focus on the purpose above, 
such effects are not considered in this paper. 
As a result, we find the best parameter set related to aspherical explosions and with the parameter set, 
high velocity $^{56}$Ni of $\sim$ 4000 km s$^{-1}$ is obtained with the binary merger model. 
Then, regarding the best parameter set as a fiducial one, we discuss the parameter and progenitor model dependences. 
Next, fixing the parameter set as the fiducial one, we perform two high resolution simulations for the two BSG progenitor models 
and the differences between the two models are presented. 
We plan to use the results of part of the models in this paper as the initial conditions for 3D MHD simulations of the later evolution of SN 1987A 
(Orlando et al. 2019, in preparation), which will be a natural extension of our previous studies on spherical explosions for SN 1987A 
\citep{2015ApJ...810..168O,2019A&A...622A..73O,2019NatAs...3..236M}. 
 
This paper is organized as follows. Section~\ref{sec:method} is dedicated to the description of the method of computations and the initial conditions. 
In Section~\ref{sec:models}, the models and related parameters are delineated. In Section~\ref{sec:results}, the results of one-dimensional and 3D simulations 
are presented. Section~\ref{sec:discussion} is devoted for the discussion on related topics. 
Finally, the study in this paper is summarized in Section~\ref{sec:summary}. 

\section{Method and Initial Conditions} \label{sec:method}

In this section, the numerical method for hydrodynamical simulations and initial conditions including the pre-supernova models are described in detail.

\subsection{Numerical Method} \label{subsec:method}

In this paper, three-dimensional hydrodynamic simulations of core-collapse supernova explosions are performed. The method is based on our previous papers, 
Paper~I and Paper~II on the matter mixing with two-dimensional hydrodynamic simulations. Here, we briefly summarize the method and stress points which are 
different from the previous ones. The numerical code is FLASH \citep{2000ApJS..131..273F} as in our previous papers (Paper~I,~II). In this paper 3D Cartesian 
coordinates, ($x$, $y$, $z$), are adopted, whereas in the Paper~I,~II, 2D spherical coordinates, ($r$, $\theta$), were adopted. 

In the simulation, we do not follow the process from the core-collapse to a successful shock revival but 
the shock wave propagation from around the interface between the Fe core and the Si layer ($\sim$ 1000 km) 
to a radius ($\gtrsim$ 10$^{14}$ cm) larger than the stellar one ($\sim$ 10$^{12}$--10$^{13}$ cm) is followed. 
In order to follow such a large difference of the spatial scales, the computational domain is gradually expanded as the shock wave propagates outward. 
The computational domain is initially set to be $-5000$ km $\leq$ $x$, $y$, $z$ $\leq$ $5000$ km, i.e., $x_{\rm min}$, $y_{\rm min}$, $z_{\rm min} = - 5000$ km and 
$x_{\rm max}$, $y_{\rm max}$, $z_{\rm max} = 5000$ km. First, physical quantities of a pre-supernova model (see, Section~\ref{subsec:pre-sn}), are mapped 
to the computational domain so that the center of the star is at the origin of the coordinates. 
When the shock wave approaches the computational boundaries, the simulation is stopped once and the 
domain is expanded by a factor of 1.2 for each dimension ($x_{\rm min}$,  $y_{\rm min}$, $z_{\rm min}$, $x_{\rm max}$, $y_{\rm max}$, and $z_{\rm max}$ 
are all multiplied by a factor of 1.2) as in Paper~I,~II (in Paper~I,~II, computational domains are expanded only in radial direction, i.e., $r_{\rm max}$ is
multiplied by a factor of 1.2). Then the physical quantities are remapped to the new (expanded) computational domain. 
During the remapping process, in the cells not covered in the previous simulation, the quantities of either the pre-supernova model (see Section~\ref{subsec:pre-sn}) or 
the profile of an ambient matter are mapped depending on the radius. 
If the cells correspond to the ambient matter, the profile of a spherical steady stellar wind is mapped, 
where the density profile follows $\rho \,(r)$ $\propto$ $r^{-2}$ and the mass loss rate and the wind velocity adopted are 
$\dot{M}_{\rm wind}$ = $10^{-7}$ $M_{\odot}$ yr$^{-1}$, $v_{\rm wind}$ = 500 km s$^{-1}$, respectively, as in \citet{2007Sci...315.1103M}. 
After the remapping process, the simulation is restarted again; to cover the large spatial scales, about 75 remappings are necessary. 

Explosions are initiated by injecting thermal and kinetic energies artificially around the interface between the Fe core and the Si layer of the mapped pre-supernova profile. 
The total injected energy, $E_{\rm in}$, is an initial parameter of the models. The ratio of the injected thermal energy to the kinetic energy is set to be unity. 
The range of the values of $E_{\rm in}$ is (1.5--3.0) $\times$ 10$^{51}$ erg (see Section~\ref{subsec:model-1d} for the range). It is noted that $E_{\rm in}$ is 
not the explosion energy, $E_{\rm exp}$, which should be obtained as a result of the simulation (see Eq.~(\ref{eq:exp}) for the definition of $E_{\rm exp}$ and 
see Table~\ref{table:results} for the obtained values of the explosion energy). In this paper, we consider aspherical explosions, which are obtained by distributing 
initial radial velocities in non-spherical ways. As such non-spherical explosions, bipolar-like explosions along the $z$-axis (polar axis) with asymmetries across the 
$x$-$y$ plane (equatorial plane) are considered, where fluctuations in the initial radial velocities for making clumpy structures are also taken into account. 
For the details of the description on the distributions of initial radial velocities, see Appendix~\ref{sec:app1} and~\ref{sec:app2}. 

The inner regions centered at the origin that correspond to a compact object (could be a proto-neutron star) are excluded from the cells to be solved. 
The size of the inner regions corresponding to the compact object is kept as larger than either 0.005 times $x_{\rm max}$ or three times $\it\Delta x$ 
($\it\Delta x$: the size of the inner cells), whichever is larger, along each dimension. 
The excluded cells are treated as a boundary condition (BC), i.e., the physical quantities on the cells are replaced to meet the BC to adjacent cells 
at every time-step. During an early phase of the simulation ``reflection" BC is adopted for the excluded cells and later it is switched to ``diode" BC as in Paper~I,~II. 
The timing of the switching is arbitrary but it should not affect the major results (see, Paper~I for details). The mass initially in the excluded cells 
is regarded as a point mass at the origin and masses flowing into the excluded cells are added to the point mass at every time-step. 
In the simulation, the point mass gravity and the spherically symmetric self-gravity are taken into account. 
The former is the gravity due to the time-dependent point mass and the latter is obtained from the spherically averaged density profiles. 

In order to reduce the computational costs, the resolution of the computational grids are adaptively refined (such method is called Adaptive Mesh 
Refinement: AMR) with the PARAMESH \citep{2000CoPhC.126..330M} package implemented in the FLASH code. For lower resolution simulations 
in this paper, the maximum and minimum refinement levels are initially set to be 7 and 5, respectively. For high resolution simulations, the 
initial maximum and minimum refinement levels are 8 and 5, respectively. 
If the maximum refinement level is $n$, at most $2^{\,(n-1)}$ blocks can be created for each dimension. 
Here, the number of grid points in one block for each dimension is 8. Then, the effective resolution of the lower (higher) 
resolution simulations is $\left[ 2^{\,6} \times 8 \right]^3 = 512^{\, 3}$ ($\left[ 2^{\,7} \times 8 \right]^{\,3} = 1024^{\,3}$). 
Since a simulation with the maximum refinement level for all computational regions is rather demanding from the point of view of the numerical cost, 
in order to reduce the cost, we manually control the regions where the maximum refinement is allowed. 
The regions around the forward shock (FS) should be solved at the highest resolution because the regions are numerically severe to be solved 
by a shock-capturing scheme and dominate the overall dynamics due to their highest fluid velocities. 
Additionally, of interest are regions where instabilities develop, and actually in the regions around the FS, RT instabilities first start to grow. 
Then, for both lower and high resolution simulations, regions only around the FS are allowed to be at the maximum refinement 
(as mentioned later, starting immediately before the shock breakout, the maximum refinement level is increased and the regions allowed to be at the maximum refinement are changed), 
i.e., the effective resolutions of other regions for lower and high resolution simulations are 256$^{\,3}$ and 512$^{\,3}$, respectively. 
The FS surface (FS radius, $r_{\rm FS}$) is approximately traced at every time step by searching for the cell which has the maximum radial velocity along each radial direction. 
The regions of $r_{\rm FS} \,- \,0.05 \,x_{\rm max} \leq r \leq r_{\rm FS} \,+ \,0.075 \,x_{\rm max}$ are allowed to be at the maximum refinement. 
After the shock breakout, the FS is accelerating rapidly due to the steep pressure gradients, whereas the inner ejecta 
(originally inside the He core) is left far behind the FS. Then, the complex structures of the inner ejecta introduced at earlier phases are numerically 
lost after the shock breakout without a special treatment for the refinement. Therefore, starting just before the shock breakout, the maximum 
refinement levels in inner regions are increased. In the inner regions of approximately 
$r \leq x_{\rm max}/8$, 
the maximum refinement levels are set to be 8 (effective res.: 1024$^{\,3}$) and 9 (effective res.: 2048$^{\,3}$) for lower and high resolution simulations, 
respectively. In the regions of approximately 
$x_{\rm max}/8 \leq r \leq x_{\rm max}/4$ 
or regions around the FS, the maximum refinement levels are set to be 7 (effective res.: 512$^{\,3}$) and 8 
(effective res.: 1024$^{\,3}$) for lower and high resolution simulations, respectively. The resolutions of other regions are the same as before the 
shock breakout. 

Since the density and pressure of the ambient matter are rather small compared with those in the expanding ejecta, the shape of the FS is affected 
by the grid structure of the Cartesian coordinates after the shock breakout, i.e., the shape of the FS tends to be like a square. 
In order to reduce such numerical artifacts on the shape of the FS, starting just before the shock breakout,  
the system is rotated by an arbitrary angle about each axis during the remmaping process; after that, all the physical quantities are remapped. 
The angles are randomly determined within the range from 
$-\pi/2$ to $\pi/2$ for each axis. Due to the randomness of the selection of the arbitrary rotation angles, the effects of the grid structure of the 
Cartesian coordinate are washed out after several remappings. 
Actually, we confirmed that the shape of the FS becomes more roundish (natural) than that without such rotations. 
Since the rotations affect only the outer most ejecta (mostly composed of hydrogen) after the shock breakout, 
the main results of this paper (the spatial distribution of metals and their velocities) except for the shape of the FS should not change with or without the rotations. 

As in Paper~I, perturbations of pre-supernova origins are taken into account in the simulation. When the shock wave reaches around the 
composition interfaces of C+O/He and He/H, perturbations of the amplitude of 5 \% are introduced in the radial velocities. The perturbations 
are functions of the angular position ($\theta$, $\phi$). 
We take $l+1$ sampling points for random numbers along the $\theta$ direction at 
$\theta = 0, \pi/l, 2/(l-1), \ldots, \pi$ and $m+1$ sampling points along the $\phi$ direction at $\theta = 0, \pi/m, 2/(m-1), \ldots, 
2\pi$, where $l$ and $m$ are integers, and $l = 128$ and $m = 256$ are 
adopted. 
%
Then, at each sampling point, one random number is assigned. A factor for the perturbations to the radial velocities at an angular position ($\theta$, $\phi$) is obtained 
by $1 + \epsilon \, {\rm rand}$($\theta$, $\phi$), where $\epsilon$ is the parameter for the amplitude of the perturbations and set to be 5\%; 
${\rm rand}$($\theta$, $\phi$) is a function of the angular position ($\theta$, $\phi$) obtained by the interpolation of the assigned random numbers of the adjacent sampling points around the effective angular position 
($\theta$, $\phi'$) $\equiv$ ($\theta$, $\phi \sin \theta$)\,\footnote{Without the factor of $\sin \theta$ in $\phi' = \phi \sin \theta$, the wave lengths of the perturbations 
around the polar axis become too small compared with those around the equatorial plane.}.
In this paper, we do not discuss the impact of the perturbations of pre-supernova origins (for the impact, see Paper~I). 

As in Paper~I,~II, the explosive nucleosynthesis is taken into account with a small approximate nuclear reaction network \citep{1978ApJ...225.1021W} 
coupled with the FLASH code. Elements, n, p, $^1$H, $^3$He, $^4$He, $^{12}$C, $^{14}$N, $^{16}$O, $^{20}$Ne, $^{24}$Mg, $^{28}$Si, $^{32}$S, 
$^{36}$Ar, $^{40}$Ca, $^{44}$Ti, $^{48}$Cr, $^{52}$Fe, $^{54}$Fe, and $^{56}$Ni, are included. The feedback from the nuclear energy generation is 
also taken into account. The advection of elements is followed by solving an advection equation for the mass fraction of each element (See, Paper~I 
for the details). 

At early phases of the simulation, the Helmholtz EOS \citep{2000ApJS..126..501T}, which includes contributions from the radiation, completely 
ionized ions, and degenerate/relativistic electrons and positrons, is used. The EOS can cover the physical regions of 10$^{-10}$ g cm$^{-3}$ $< \rho <$ 
10$^{11}$ g cm$^{-3}$ and 10$^{4}$ K $< T <$ 10$^{11}$ K. For a later phase when $\rho \lesssim 10^{-8}$ g cm$^{-3}$, another EOS that consists of 
ideal gas of fully ionized ions, electrons, and the radiation is used. For a transition region of 
10$^{-8}$ g cm$^{-3}$ $< \rho <$ 10$^{-7}$ g cm$^{-3}$
, the Helmholtz 
EOS and the EOS mentioned just above are smoothly blended. As for the latter EOS, the contribution to the pressure from the radiation is 
suppressed depending on the density and temperature in an optically thin regime (see, Paper~I). 
As in Paper~I,~II, energy depositions rate from the decays of $^{56}$Ni and $^{56}$Co are also implemented (see, Paper~I for the details). 

\subsection{Initial Conditions: Pre-supernova Models} \label{subsec:pre-sn}

In this subsection, the pre-supernova models used as the initial conditions of the hydrodynamical simulations are described. Here, before the 
description, some properties of Sk $-$69$^{\circ}$ 202 which are closely related to the study in this paper (the matter mixing) are briefly summarized as follows. 
The luminosity and effective temperature of Sk $-$69$^{\circ}$ 202 are (3--6) $\times$ 10$^{38}$ erg s$^{-1}$ and 15,000--18,000 K, respectively \citep{1989ARA&A..27..629A}. 
Since at the time of explosion, energy generation from hydrogen shell burning is generally negligible, the helium core mass is closely related to the luminosity; from the 
evolution models, it is in the range of 6 $\pm$ 1 $M_{\odot}$ for the case of the single star evolution \citep[][]{1988ApJ...330..218W}. 
With the ranges of the luminosity and the effective temperature, the radius is estimated as (2--4) $\times$ 10$^{12}$ cm \citep{1989ARA&A..27..629A}. 

In this paper, four pre-supernova models (denoted as n16.3, b18.3, s18.0, and s19.8) are adopted. 
Important properties of the models are summarized in Table~\ref{table:pre-sn}, where $M$ is the stellar mass, $M_{\rm C+O, c}$ 
is the C+O core mass, $M_{\rm He, c}$ is the helium core mass, $M_{\rm env}$ is the hydrogen envelope mass, $R$ is the stellar radius (listed in 
both the units of cm and $R_{\odot}$ in 6th and 7th columns, respectively), and $q \equiv M_{\rm He, c}/M$ is the ratio of the helium core mass to 
the stellar mass. Here the values are all the ones at the time of the explosion. The 9th and 10th columns,``Type" and ``evolution", denote the types of 
the models, i.e., ``BSG" or ``RSG" and the evolution scenario, i.e., ``single" star evolution or ``binary" merger evolution. 
The models n16.3 and b18.3 are BSGs, whereas the other two, s18.0 and s19.8, are RSGs. As mentioned in Section~\ref{sec:intro}, the progenitor of SN 1987A, 
Sk $-$69$^{\circ}$ 202, was a compact BSG at the time of the explosion and the two RSG models are not appropriate for Sk $-$69$^{\circ}$ 202 from the point of view of 
the effective temperature (stellar radius: see the 6th column in Table~\ref{table:pre-sn}). The two models, however, are included to see the dependence on the progenitor models because the two 
RSG models have distinct properties compared with those of the BSG models. 

\begin{deluxetable*}{lccccccccc}
\tabletypesize{\footnotesize}
\tablewidth{0pt}
\tablenum{1}
\tablecolumns{10}
\tablecaption{Properties of pre-supernova models.}
\label{table:pre-sn}
\tablehead
{
\multicolumn{1}{l}{{Model}} & 
\multicolumn{1}{c}{$M$} &
\multicolumn{1}{c}{$M_{\rm C+O, c}$\tablenotemark{a}} &
\multicolumn{1}{c}{$M_{\rm He, c}$\tablenotemark{b}} &
\multicolumn{1}{c}{$M_{\rm env}$\tablenotemark{c}} &
\multicolumn{1}{c}{$R$} &
\multicolumn{1}{c}{$R$} &
\multicolumn{1}{c}{$q \equiv M_{\rm He, c}/M$} &
\multicolumn{1}{c}{Type\tablenotemark{d}} &
\multicolumn{1}{c}{Evolution\tablenotemark{e}} 
\\
& 
\multicolumn{1}{c}{\scriptsize ($M_{\odot}$)} & 
\multicolumn{1}{c}{\scriptsize ($M_{\odot}$)} &
\multicolumn{1}{c}{\scriptsize ($M_{\odot}$)} &
\multicolumn{1}{c}{\scriptsize ($M_{\odot}$)} &
\multicolumn{1}{c}{\scriptsize (cm)} &
\multicolumn{1}{c}{\scriptsize ($R_{\odot}$)} &
\multicolumn{1}{c}{} &
\multicolumn{1}{c}{} &
\multicolumn{1}{c}{} 
}	
\startdata
b18.3 & 18.3 & 2.87 & 3.98 & 14.3 & \,\,2.12 (12)\tablenotemark{f} & 30.7 & 0.22 & BSG & binary  \\ 
n16.3 & 16.3 & 3.76 & 5.99 & 10.3 & 3.39 (12) & 48.7 & 0.37 & BSG & single  \\ 
\hline
s18.0\tablenotemark{g} & 14.9 & 4.19 & 5.49 & 9.45 & 6.76 (13) & 972 & 0.37 & RSG & single \\ 
s19.8\tablenotemark{g} & 15.9 & 4.89 & 6.24 & 9.61 & 7.36 (13) & 1058 & 0.39 & RSG & single  
\enddata
\scriptsize{
\tablenotetext{a}{Mass of the C+O core.}
\tablenotetext{b}{Mass of the helium core.}
\tablenotetext{c}{Mass of the hydrogen-rich envelope.}
\tablenotetext{d}{Type of the presupernova model, i.e., ``RSG" or ``BSG".}
\tablenotetext{e}{Evolution scenario, i.e., ``binary" (``single") denotes a binary merger (single star) evolution.}
\tablenotetext{f}{Number in parenthesis denotes the power of ten.}
\tablenotetext{g}{The number in the name denotes the zero-age main sequence mass.}
}
\end{deluxetable*}

The model b18.3 is a newly developed \citep{2018MNRAS.473L.101U} compact BSG model based on the binary merger scenario \citep[][]{1990A&A...227L...9P,1992ApJ...391..246P,
2007Sci...315.1103M}. 
The overall binary merger scenario is as follows. A binary system with a large mass ratio consisting of a primary RSG ($\sim$ 15 $M_{\odot}$) and a secondary 
 main sequence star ($\sim$ 5 $M_{\odot}$)
forms a common envelope through dynamical mass transfer from the primary to the secondary  
\citep[here, masses of the two merging stars are taken from][]{2007Sci...315.1103M}\footnote{In Refs.~\citet{1990A&A...227L...9P,1992ApJ...391..246P}, 
the masses of the primary and secondary stars are 16 $M_{\odot}$ and 3 $M_{\odot}$, respectively. In recent binary merger models for the progenitor of SN 1987A 
\citep{2017MNRAS.469.4649M,2018MNRAS.473L.101U}, the masses of two stars are in the range 14--17 $M_{\odot}$ and 4--9 $M_{\odot}$, respectively.}. 
The spiral-in of the core of the 
primary and the secondary due to the friction with the common envelope causes spin-up of the envelope and partial (aspherical) mass ejection from the 
envelope. Then, the secondary starts to transfer its mass to the core of the primary after the Roche lobe radius of the secondary becomes relatively 
smaller than its own stellar radius. During the mass transfer, part of the material from the secondary (composed of hydrogen-rich material) penetrates into 
the helium core of the primary, which triggers additional hydrogen burning and mixing of helium and CNO-processed material into the envelope. Eventually, 
the secondary is completely dissolved into the envelope of the primary to form a single rapidly rotating BSG.
The properties of the model b18.3 are listed in Table~1 in \citet{2018MNRAS.473L.101U} 
(the model is labeled as ``a" with a footnote); it is the outcome of the merger of two massive stars of 14 $M_{\odot}$ and 9.0 $M_{\odot}$. 
This model satisfies all the observational constrains of Sk $-$69$^{\circ}$ 202, i.e., the final position in the HR diagram 
(the observed luminosity and effective temperature), the red-to-blue transition about 20,000 yr ago, the required surface abundances of helium and 
CNO-processed elements, and an ability to form a triple ring structure in the nebula. Hitherto, this model has not been investigated in the study of 
matter mixing. 

The model n16.3 was obtained by combining an evolved 6 $M_{\odot}$ He core corresponding to the zero-age main sequence mass of 20 $M_{\odot}$ 
\citep{1988PhR...163...13N} with  a 10.3 $M_{\odot}$ hydrogen envelope. The hydrogen envelope was taken from an independent stellar evolution 
calculation \citep{1988Natur.334..508S} in which an enhanced mass loss rate and artificial mixing of helium-rich material into the hydrogen envelope were 
implemented to make a compact BSG which satisfies the observed luminosity and the effective temperature 
\citep{1990ApJ...360..242S}. The model n16.3 has also been used in our previous studies on the matter mixing (Paper~I and Paper~II). 
It is noted that the model n16.3 has been denoted as N20 in several studies \citep[e.g.,][]{2015A&A...577A..48W,2019A&A...624A.116U}. 
In previous studies on matter mixing, this progenitor model has had difficulties to reproduce the high velocity $^{56}$Ni of $\gtrsim$ 3000 km s$^{-1}$ 
\citep[e.g., Paper~I;][]{2015A&A...577A..48W}.

A distinct difference of the properties between the two BSG models, b18.3 and n16.3, is the ratios of core mass to envelope mass 
(see Table~\ref{table:pre-sn}). As one can see, both the helium core mass ($\sim$ 4 $M_{\odot}$) and C+O core mass ($\sim$ 3 $M_{\odot}$) of b18.3 are 
smaller than those of n16.3 ($\sim$ 6 $M_{\odot}$ and 4 $M_{\odot}$, respectively). On the contrary, the mass of the hydrogen envelope of b18.3 
($\sim$ 14 $M_{\odot}$) is larger than that of n16.3 ($\sim$ 10 $M_{\odot}$). In other words, the ratio of the helium core mass to the stellar 
mass of b18.3 ($q = 0.22$) is smaller than that of n16.3 ($q = 0.37$). The radius of b18.3 is also smaller than that of n16.3 by a factor of about 0.6.

The pre-supernova models s18.0 and s19.8 are taken from the supplementary 
data\footnote{The supplementary data is taken from https://iopscience.iop.org/article/10.3847/0004-637X/821/1/38/meta.} 
in \citet{2016ApJ...821...38S}. The number in each name does not denote the final stellar mass but the corresponding zero-age main sequence 
mass as in the paper. The two models are favored for SN 1987A in the point of view of the helium core mass ($\sim$ 6 $M_{\odot}$) but the 
radius (in the order of 10$^{13}$ cm) is very different from the observational constraints ($\sim$ 3 $\times$ 10$^{12}$ cm) by a factor of more 
than ten. The two models are essentially the same as the corresponding models calculated in \citet{2002RvMP...74.1015W}. Between the two 
RSG models, there are slight (but non-negligible) differences in the properties. The ratios of the helium core mass to the stellar mass 
($q \sim$ 0.4) have similar values as the model n16.3 but the stellar radii are rather different from those of the BSG models. 

In Figure~\ref{fig:prog1}, $\rho\,r^{3}$ profiles of the four models are shown, where $\rho$ is the density and $r$ is the radius. Top left, top 
right, bottom left, and bottom right panels are the profiles of b18.3, n16.3, s18.0, and s19.8, respectively. Solid vertical lines indicate the 
composition interfaces. Dashed vertical lines denote the transition between convective and radiative regions in the C+O layer. The gradient 
of $\rho\,r^{3}$ provides useful information on where and when the supernova shock wave is accelerated or decelerated. In the self-similar 
solution of a point explosion in a power-law density profile of $\rho (r) \propto r^{-\omega}$ \citep[][]{1959sdmm.book.....S}, the velocity of the 
blast wave can be expressed as $v_{\rm sh}~(t) \propto t^{(\omega -3)/(5-\omega)}$. From the relation, one finds that if the power of the density 
profile $- \omega$ is $- 3$, the velocity of the blast wave is constant. Equivalently, if the gradient of $\rho\,r^{3}$ is positive (corresponding to 
the case of $\omega < 3$), the velocity of the blast wave decreases (the blast wave is decelerated) at the position, and vice versa. In this 
way, the density structure affects how the supernova shock wave propagates in the pre-supernova star. Since the radial velocity of the supernova 
ejecta is very roughly proportional to the radius, basically it is difficult for the inner ejecta to catch up with the higher velocity outer ejecta. But 
depending on the complicated density structure as seen in Figure~\ref{fig:prog1}, the propagation of the blast wave and the expansion of the 
inner ejecta could drastically change among the progenitor models. Additionally, the condition of the RT instability is 
$\nabla P \cdot \nabla \rho < 0$ \citep{1976ApJ...207..872C} and the structure of the density gradient is also important for the growth of the RT 
instability. For the comparison among the models, the $\rho\,r^{3}$ profiles are shown on a single plot (Figure~\ref{fig:prog2}). As one can see, 
among the models there are large differences in the structures of the C+O layer, the helium layer, and the hydrogen envelope. 
%
%
The binary merger model, b18.3, has the flattest $\rho\,r^{3}$ gradient in the C+O layer. 
The structures of the RSG models in the helium layer and the hydrogen envelope are similar between the two RSG models. 
In the RSG models, the blast wave overall accelerates inside the helium layer but the situation is opposite in the hydrogen envelope. 
On the other hand, the $\rho \,r^3$ gradient of the BSG models in the helium layer is overall positive except for a thin region at the outer layer 
and the structures in the hydrogen envelope are rather different from those of the RSG models because of the large differences in the radius. 

\begin{figure*}[htb]
\begin{tabular}{cc}
\begin{minipage}{0.5\hsize}
\begin{center}
\includegraphics[width=7cm,keepaspectratio,clip]{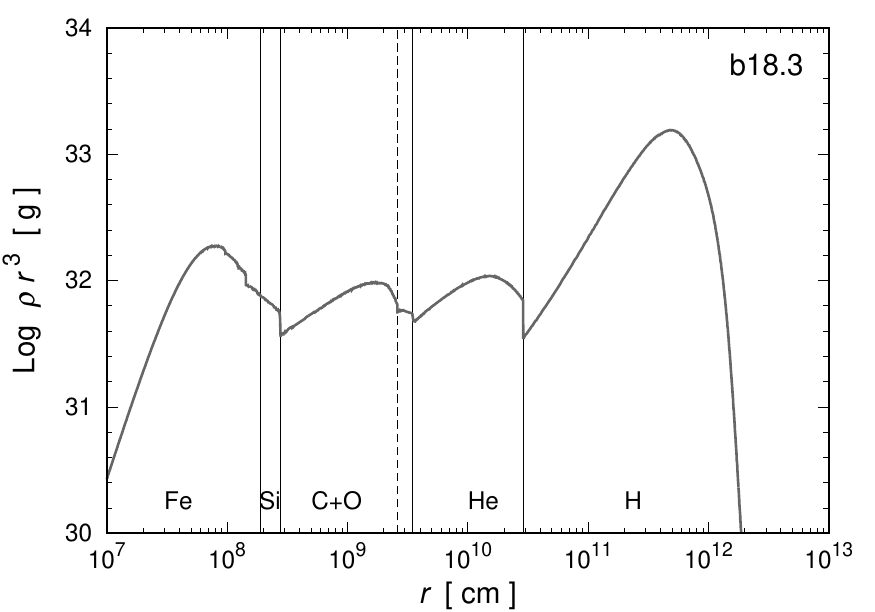}
\end{center}
\end{minipage}
\begin{minipage}{0.5\hsize}
\begin{center}
\includegraphics[width=7cm,keepaspectratio,clip]{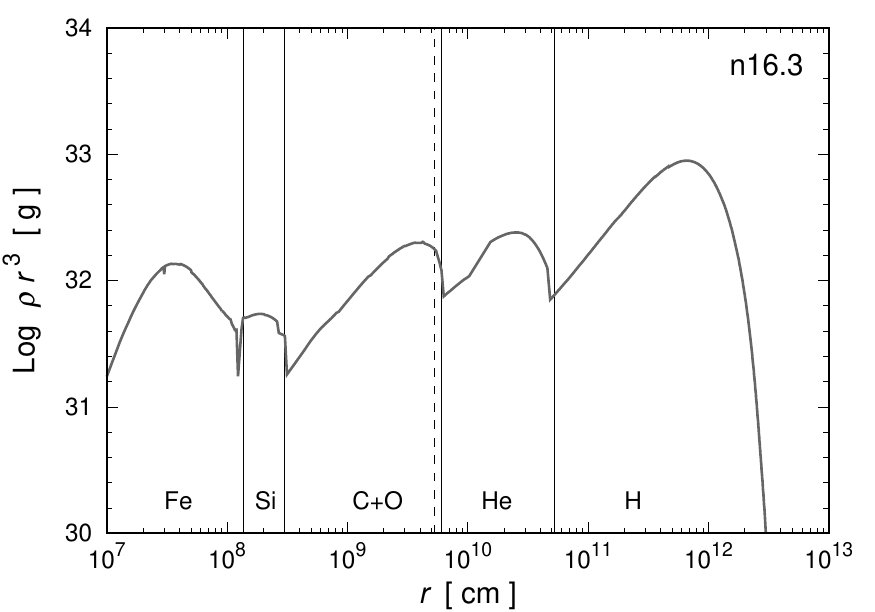}
\end{center}
\end{minipage}
\\
\begin{minipage}{0.5\hsize}
\begin{center}
\includegraphics[width=7cm,keepaspectratio,clip]{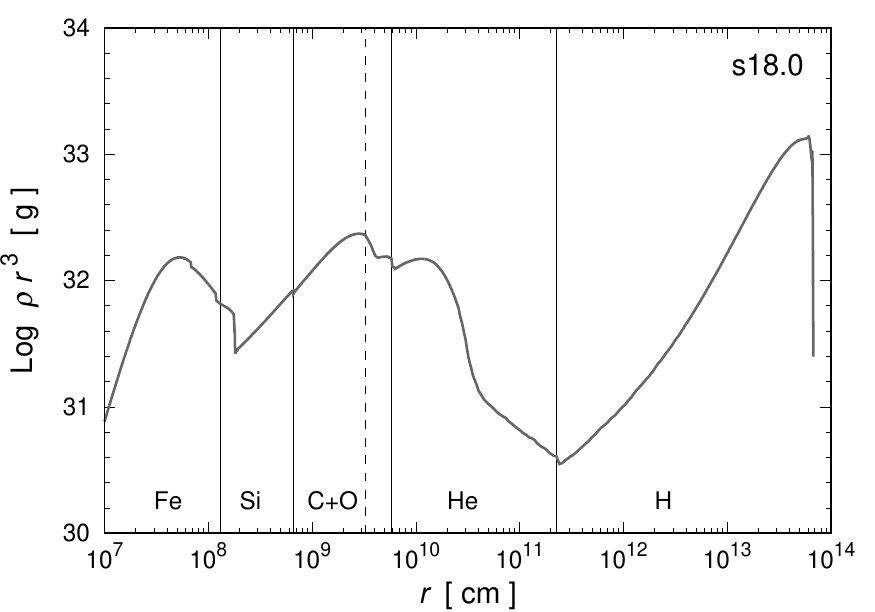}
\end{center}
\end{minipage}
\begin{minipage}{0.5\hsize}
\begin{center}
\includegraphics[width=7cm,keepaspectratio,clip]{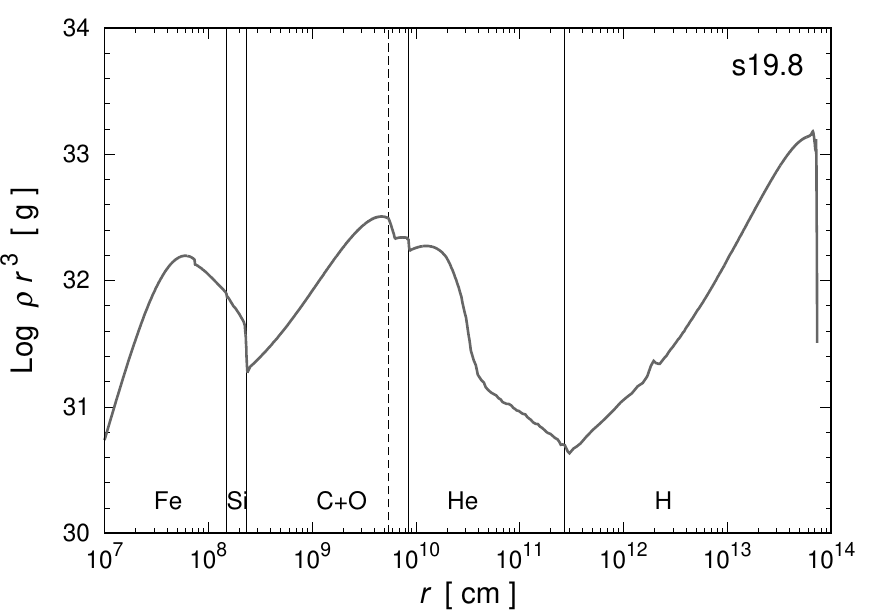}
\end{center}
\end{minipage}
\end{tabular}
\caption{$\rho \,r^3$ profiles of the four progenitor models b18.3 (top left), n16.3 (top right), s18.0 (bottom left), and s19.8 (bottom right), 
where $\rho$ is the density and $r$ is the radius. Solid vertical lines indicate the composition interfaces. Dashed lines denote the transition 
between convective and radiative regions in the C+O layer.}
\label{fig:prog1}
\end{figure*}

\begin{figure}[htb]
\begin{center}
\includegraphics[width=8cm,keepaspectratio,clip]{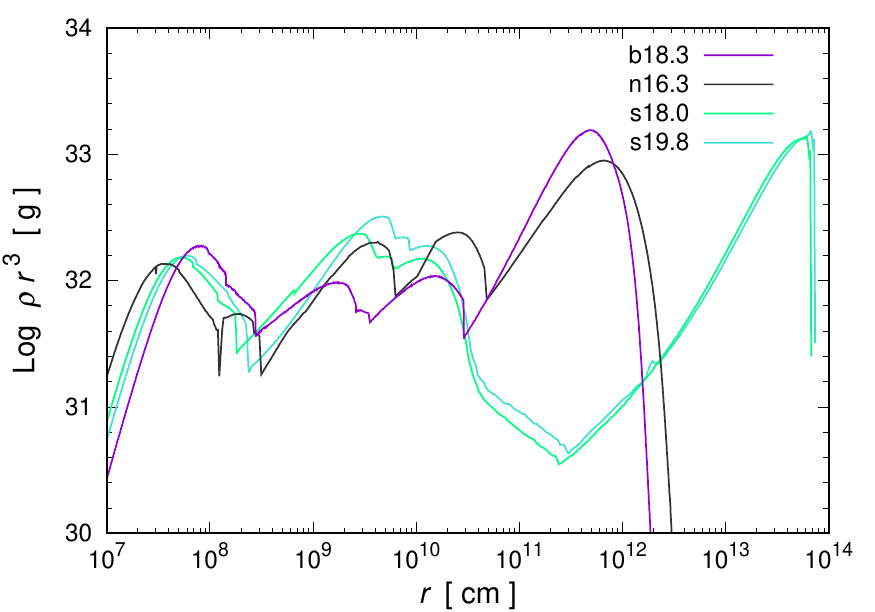}
\end{center}
\caption{Same as Figure~\ref{fig:prog1} but the four $\rho \,r^3$ profiles are plotted on the same figure.}
\label{fig:prog2}
\end{figure}

\section{Simulation Models} \label{sec:models}

In this paper, we perform one-dimensional and 3D hydrodynamical simulations. Here, the hydrodynamical models and related model 
parameters are described in detail.

\subsection{Models of One-dimensional Simulations} \label{subsec:model-1d}

In order to assess the dependence of the matter mixing on the progenitor models, first, we perform one-dimensional hydrodynamical simulations. As in previous 
papers \citep[e.g.,][]{1989ApJ...344L..65E,1990ApJ...348L..17B,1991A&A...251..505M} including Paper~I, from spherical one-dimensional 
hydrodynamical simulations of the blast wave propagation in an expanding star, stability analyses of instabilities for the progenitor models can be 
done. In order to evaluate the time-integrated growths of instabilities (growth factors), several one-dimensional simulations are performed for each 
progenitor model changing the initial injection energy, $E_{\rm in}$. For the simulations, the same numerical code, FLASH, is used and the basic 
method is the same as described in Section~\ref{subsec:method}. But the adopted coordinate system is the spherical coordinate here and the 
resolution of the simulations, the treatments of inner regions corresponding to the compact object, are different because of the differences between 
the two coordinate systems. For the treatments in the spherical coordinate, see Paper~I. 
For one-dimensional simulations, the model parameter on explosions is only $E_{\rm in}$. From the observations of the optical light curves and 
theoretical modeling of the light curves \citep[e.g.,][]{1988ApJ...330..218W,1990ApJ...360..242S}, the explosion energy of SN 1987A has been 
deduced. For example, the range of the explosion energy, $E_{\rm exp}$, was estimated as $E_{\rm exp}/M_{\rm env} = (1.1 \pm 0.3) \times 
10^{50}$ erg $M_{\odot}^{-1}$ in \citet{1990ApJ...360..242S}. Then, the explosion energy depends on the hydrogen envelope mass in this case. 
By substituting the value of the envelope mass of the binary merger model b18.3 ($M_{\rm env}$ = 14.3 $M_{\odot}$) into the equation, 
$E_{\rm exp}/M_{\rm env} = (1.1 \pm 0.3) \times 10^{50}$ erg $M_{\odot}^{-1}$, one can obtain the explosion energy as 
$E_{\rm exp}$ = (1.1--2.0) $\times$ 10$^{51}$ erg. The deduced values of the explosion energy have not converged among the studies 
but overall the values are within the range of $E_{\rm exp}$ = (0.8--2.0) $\times$ 10$^{51}$ erg\,\footnote{The range of the explosion energy was 
summarized in Table 1 in \citet{2014ApJ...783..125H}.}. From the results in Paper~I and Paper~II, it has been empirically found that the final 
explosion energy is roughly approximated as $E_{\rm exp}$ $\simeq$ ($E_{\rm in}$ - 0.5 $\times$ 10$^{51}$) erg. Then, as the values of the 
parameter, $E_{\rm in}$, the four values, (1.5, 2.0, 2.5, and 3.0) $\times$ 10$^{51}$ erg, 
are adopted in this paper. The last value is outside the range above but we include it as an extreme case.

Here, we briefly review the method of the stability analysis. For the analysis, two kind of instabilities are considered. 
One is the RT instability for an incompressible fluid and the other is an instability for a compressible fluid (convection). 
The condition of the RT instability \citep{1976ApJ...207..872C} is expressed as
\begin{equation}
\frac{\mathcal{R}}{\mathcal{P}} < 0 \,,
\label{eq:RT}
\end{equation}
where $\mathcal{R} \equiv \partial \ln \rho / \partial r$ and $\mathcal{P} \equiv \partial \ln P / \partial r$ are the reciprocals of the density and 
pressure scale heights, respectively. Here, $\rho$ is the density, $r$ is the radius, and $P$ is the pressure. The criterion of the convective 
instability for a compressible fluid (Schwarzschild criterion) \citep[e.g.,][]{1984A&A...139..368B} is
\begin{equation}
\frac{\mathcal{R}}{\mathcal{P}} < \frac{1}{\gamma} \,,
\label{eq:SC}
\end{equation}
where $\gamma$ is the adiabatic index.
%
%
The growth rate of the RT instability is written as
\begin{equation}
\sigma_{\rm i} = \sqrt{-\frac{P}{\rho} \mathcal{P} \mathcal{R}} \,.
\label{eq:RT-sigma}
\end{equation}
The growth rate of the convective instability is
\begin{equation}
\sigma_{\rm c} = \frac{c_{\rm s}}{\gamma} \sqrt{{\mathcal P}^2 - \gamma \mathcal{P} \mathcal{R}} \,,
\label{eq:SC-sigma}
\end{equation}
where $c_{\rm s}$ is the sound speed. 
From the growth rate, the time-integrated growth (growth factor) for each instability is calculated as
\begin{equation}
\left. \frac{\zeta}{\zeta_0} \,\right|_{t} = \exp\,\left(\int_0^{t} {\rm Re} \left[ \, \sigma \, \right] \, dt^{\prime} \right) \,,
\end{equation}
where $\zeta_0$ is the initial amplitude of a perturbation, $\zeta$ is the amplitude of the perturbation at time $t$, $\sigma$ = $\sigma_{\rm i}$ 
($\sigma$ = $\sigma_{\rm c}$) for an incompressible (a compressible) fluid. Based on the results of the one-dimensional simulations described in 
Section~\ref{subsec:model-1d}, the growth factors are deduced at each mass coordinate at time $t$. 
The growth factors are based on a local linear analysis of instabilities. Then, once the instabilities enter a non-linear regime, the 
growth rates are no longer followed by Eqs.~(\ref{eq:RT-sigma}) and (\ref{eq:SC-sigma}) in a realistic multi-dimensional situation. 
Actually, the growth rate of the RT instability is proportional to the square root of the wavenumber of the perturbation \citep[e.g.,][]{1989ApJ...344L..65E} and 
after the non-linear regime, merging of fingers (inverse cascading) may occur to form larger scale structures \citep{1992ApJ...390..230H}. 
Therefore, the values of the growth factors should not be taken quantitatively but only qualitatively. Nevertheless, the growth rates can be useful to 
grasp where instabilities are easy to grow and the dependence on the progenitor models. 

\subsection{Models of Three-dimensional Simulations} \label{subsec:model-3d}

As noted in Section~\ref{sec:intro}, first we perform 3D lower resolution simulations to explore a wide range of parameters related to aspherical 
explosions and the progenitor models. Among the models, with the best parameter set of the explosion asphericity, two 3D high resolution 
simulations are performed for the two BSG models, i.e., b18.3 and n16.3. For the parameters related to the aspherical explosions, see 
Appendix~\ref{sec:app1} and Appendix~\ref{sec:app2}. The models of the 3D simulations and the adopted values of the related parameters 
are summarized in Table~\ref{table:models}, where, $\beta \equiv v_{\rm pol}/v_{\rm eq}$ is the ratio of the initial radial velocities along the 
polar to the equatorial ($x$-$y$ plane) directions, $\alpha \equiv v_{\rm up}/v_{\rm down}$ is the ratio of the initial radial velocities along the 
positive to the negative $z$ directions. The 4th column denotes the type of the distribution of the initial radial velocities described in 
Appendix A, i.e., ``cos" or ``exponential" or ``power" or ``elliptical", which correspond to the shapes of the functions, $f(\theta)$, in 
Eqs.,~(\ref{eq:vcos}), (\ref{eq:vexp}), (\ref{eq:vpwr}), and (\ref{eq:vell}), respectively. The 5th column, $\epsilon$, indicates the amplitude of the 
fluctuations in the initial radial velocities. The angular dependence of the fluctuations is described in Appendix~\ref{sec:app2}. 

The nomenclature of the models is as follows. For example, in the case of ``b18.3-mo13", the former part, ``b18.3", before the 
hyphen denotes the adopted pre-supernova model and the latter part ``mo13" indicates the properties of the adopted parameter set related to 
the initial asphericity of the explosion and the injected energy. 
The models with``mo13" 
adopt the parameter set corresponding to the ones in the best model, AM2, in Paper~I \citep{2013ApJ...773..161O}. The models with ``fid" adopt 
the fiducial (the best) parameter set in this paper.  The models with ``beta2", ``beta4", and ``beta8", have the values of the parameter $\beta 
\equiv v_{\rm pol}/v_{\rm eq}$ as 2, 4, and 8, respectively. In those models, only the values of $\beta$ are different from the parameter set 
adopted in the models with ``fid". In a similar way, the models with ``alpha1" and ``alpha2" have the values of the parameter, $\alpha \equiv 
v_{\rm up}/v_{\rm down}$, set to be 1.0 and 2.0, respectively. 
The models with ``cos", ``exp", and ``pwr" adopt the types of the initial asphericity of the explosion as ``cos", ``exponential", and ``power", respectively. 
In the model with ``clp0", the value of the parameter $\epsilon$ is 0\%. 
The models with ``ein1.5", ``ein2.0", and ``ein3.0" have the values of the parameter $E_{\rm in}$ as (1.5, 2.0, and 3.0) $\times$ 10$^{51}$ erg, 
respectively. Finally, the models with ``high" have the same parameter sets as in the corresponding models with ``fid" but the simulations are 
performed with the highest resolution in this paper.  

In Paper~I, we explored mildly aspherical explosions with the progenitor model of n16.3. The obtained maximum velocity of $^{56}$Ni, however, 
is at most only $\sim$ 3000 km s$^{-1}$ and the tails ($\sim$ 4000 km s$^{-1}$) of the observed [Fe II] line profiles were not explained in Paper~I. 
It is noted that the corresponding model to the best model in Paper~I, AM2, is the model n16.3-mo13 in this paper (see Table~\ref{table:models}). 
Then, in this paper, we explore a wider range for the asphericity of the explosions. For example, as can be seen in Figure~\ref{fig:shape} in 
Appendix~\ref{sec:app1}, in the best model in Paper~I, the angle dependence of the initial radial velocities is similar to the distribution for $\beta$ 
= 2 shown in the top left panel. In this paper, explosions in which higher initial radial velocities are more concentrated around the polar axis (see 
the distribution for $\beta$ = 16 in the bottom right panel) are also considered. 
As for the types of the initial asphericity of the explosion, the ``elliptical" form is adopted as a fiducial form because we found that 
models with the ``elliptical" form overall better reproduce observational requirements for SN 1987A discussed in Section~\ref{subsec:res-3d-low}. 
The impacts of the types can be investigated by comparing the results among the models b18.3-fid, b18.3-cos, b18.3-exp, and b18.3-pwr, among which 
only the types of the initial asphericity of the explosion are different (see Section~\ref{subsec:res-3d-low}). 
Moreover, in order to investigate the impact of the 
progenitor model dependence, the four progenitor models (the two BSG models and the other two RSG models) are included. 

\begin{deluxetable*}{lccccccc}
\tabletypesize{\footnotesize}
\tablenum{2}
\label{table:models}
\tablecolumns{8}
\tablecaption{Models of 3D simulations and parameters.}
\tablehead
{
\multicolumn{1}{l}{Model} &
\multicolumn{1}{c}{$\beta$ $\equiv$ $v_{\rm pol}/v_{\rm eq}$\tablenotemark{a}} &
\multicolumn{1}{c}{$\alpha$ $\equiv$ $v_{\rm up}/v_{\rm down}$\tablenotemark{b}} &
\multicolumn{1}{c}{Type of asphel.\tablenotemark{c}} &
\multicolumn{1}{c}{$E_{\rm in}$ (10$^{51}$ erg)\tablenotemark{d}} &
\multicolumn{1}{c}{$\epsilon$\tablenotemark{e}} &
\multicolumn{1}{c}{} &
\multicolumn{1}{c}{}
}
\startdata
b18.3-mo13 & 2 & 2.0 & cos & 2.5 & 30\% &  & \\ 
b18.3-beta2 & 2 & 1.5 & elliptical & 2.5 & 30\% &  & \\ 
b18.3-beta4 & 4 & 1.5 & elliptical & 2.5 & 30\% &  & \\ 
b18.3-beta8 & 8 & 1.5 & elliptical & 2.5 & 30\% &  & \\ 
b18.3-fid & 16 & 1.5 & elliptical & 2.5 & 30\% &  & \\ 
\hline
n16.3-mo13 & 2 & 2.0 & elliptical & 2.5 & 30\% &  & \\ 
n16.3-beta2 & 3 & 1.5 & elliptical & 2.5 & 30\% &  & \\ 
n16.3-beta4 & 4 & 1.5 & elliptical & 2.5 & 30\% &  & \\ 
n16.3-beta8 & 8 & 1.5 & elliptical & 2.5 & 30\% &  & \\ 
n16.3-fid & 16 & 1.5 & elliptical & 2.5 & 30\% &  & \\ 
\hline
s18.0-mo13 & 2 & 2.0 & cos & 2.5 & 30\% &  & \\ 
s18.0-beta2 & 2 & 1.5 & elliptical & 2.5 & 30\% &  & \\ 
s18.0-beta4 & 4 & 1.5 & elliptical & 2.5 & 30\% &  & \\ 
s18.0-beta8 & 8 & 1.5 & elliptical & 2.5 & 30\% &  & \\ 
s18.0-fid& 16 & 1.5 & elliptical & 2.5 & 30\% &  & \\ 
\hline
s19.8-mo13 & 2 & 2.0 & cos & 2.5 & 30\% &  & \\ 
s19.8-beta2 & 2 & 1.5 & elliptical & 2.5 & 30\% &  & \\ 
s19.8-beta4 & 4 & 1.5 & elliptical & 2.5 & 30\% &  & \\ 
s19.8-beta8 & 8 & 1.5 & elliptical & 2.5 & 30\% &  & \\ 
s19.8-fid & 16 & 1.5 & elliptical & 2.5 & 30\% &  & \\ 
\hline
%
b18.3-alpha1 & 16 & 1.0 & elliptical & 2.5 & 30\% &  & \\ 
b18.3-alpha2 & 16 & 2.0 & elliptical & 2.5 & 30\% &  & \\ 
b18.3-cos & 16 & 1.5 & cos & 2.5 & 30\% & & \\ 
b18.3-exp & 16 & 1.5 & exponential & 2.5 & 30\% &  & \\ 
b18.3-pwr & 16 & 1.5 & power & 2.5 & 30\% &  & \\ 
b18.3-clp0 & 16 & 1.5 & elliptical & 2.5 & 0\% &  & \\ 
b18.3-ein1.5 & 16 & 1.5 & elliptical & 1.5 & 30\% &  & \\ 
b18.3-ein2.0 & 16 & 1.5 & elliptical & 2.0 & 30\% &  & \\ 
b18.3-ein3.0 & 16 & 1.5 & elliptical & 3.0 & 30\% &  & \\ 
\hline
b18.3-high & 16 & 1.5 & elliptical & 2.5 & 30\% &  & \\ 
n16.3-high & 16 & 1.5 & elliptical & 2.5 & 30\% &  & 
\enddata
\scriptsize{
\tablenotetext{a}{Ratio of the initial radial velocities along the polar to the equatorial ($x$-$y$ plane) directions.}
\tablenotetext{b}{Ratio of the initial radial velocities along the positive to the negative $z$-directions.}
\tablenotetext{c}{Type of the distribution of the initial radial velocities described in Appendix A, i.e., ``cos" or ``exponential" or ``power" or ``elliptical", 
which correspond to the shapes of the functions, $f(\theta)$, in Eqs.,~(\ref{eq:vcos}), (\ref{eq:vexp}), (\ref{eq:vpwr}), and (\ref{eq:vell}), respectively.}
\tablenotetext{d}{Energy initially injected to initiate the explosion.}
\tablenotetext{e}{Amplitude of the fluctuations in the initial radial velocities.}
}
\end{deluxetable*}

\section{Results} \label{sec:results}

In this section, the results of the one-dimensional simulations (Section~\ref{subsec:res-1d}), 3D simulations with lower resolution 
(Section~\ref{subsec:res-3d-low}), and 3D high resolution simulations (Section~\ref{subsec:res-3d-high}) are presented in sequence.

\subsection{Results of One-dimensional Simulations} \label{subsec:res-1d}

In this subsection, the results of stability analyses to the four pre-supernova models are shown. 
The growth factors at the time when the shock wave reaches the radius of about 5 $\times$ 10$^{14}$ cm (after the shock breakout for all cases) 
are shown in Figure~\ref{fig:gr2}. From the top to the bottom, the progenitor models b18.3, n16.3, s18.0, and s19.8 are shown, respectively. 
From the left to the right, the cases of the injected energies $E_{\rm in}$ = (1.5, 2.0, 2.5, and 3.0) $\times$ 10$^{51}$ erg are depicted. 
Red 
solid lines are the growth factors for a compressible fluid and black 
dashed lines are those for an incompressible fluid. Thin vertical lines are the 
composition interfaces. As one can see, growth factors are salient around the composition interfaces of He/H and/or C+O/He as shown in 
previous studies \citep[e.g.,][]{1989ApJ...344L..65E,1990ApJ...348L..17B,1991A&A...251..505M}. Since the condition for the RT instability for an 
incompressible fluid is always more stringent than that for the convective instability (Schwarzschild criterion) for a compressible fluid, the 
development of the growth factors for the convective instability dominates the one for the RT instability. The top two panels are for BSG models 
and the bottom two panels are for RSG models. In BSG models, growth factors are high around both the C+O/He and He/H interfaces. 
On the other hand, in RSG models, growth factors are outstanding only around the He/H interfaces, which is attributed to the fact that the 
gradients of the $\rho \,r^3$ value are overall negative in the helium layer of the two RSG models in contrast to the case of the BSG models 
(see Figure~\ref{fig:prog1}). Focusing on the binary merger model (b18.3), the growth factors seem to be proportional to the injected energies 
$E_{\rm in}$ 
in particular at the He/H composition interface. The growth factors depend on several 
factors, e.g., where and when the conditions, Eqs.~(\ref{eq:RT}) and~(\ref{eq:SC}), are realized, the steepness of the density and pressure 
gradients, the time for instabilities to grow. Then, the situation could change depending on the progenitor models and the explosion energies. 
For the cases of the binary merger model (b18.3), a more energetic explosion probably makes the pressure gradients steeper than those for less 
energetic models. As for the other BSG model (n16.3), the growth factors are not sensitive to the explosion energies. For the case of the model 
s18.0 (RSG), in the less energetic model (the left panel), growth factors are the most prominent around the He/H composition interface. 

\begin{figure*}[htb]
\begin{center}
\hspace*{-1cm}
\includegraphics[width=18cm,keepaspectratio,clip]{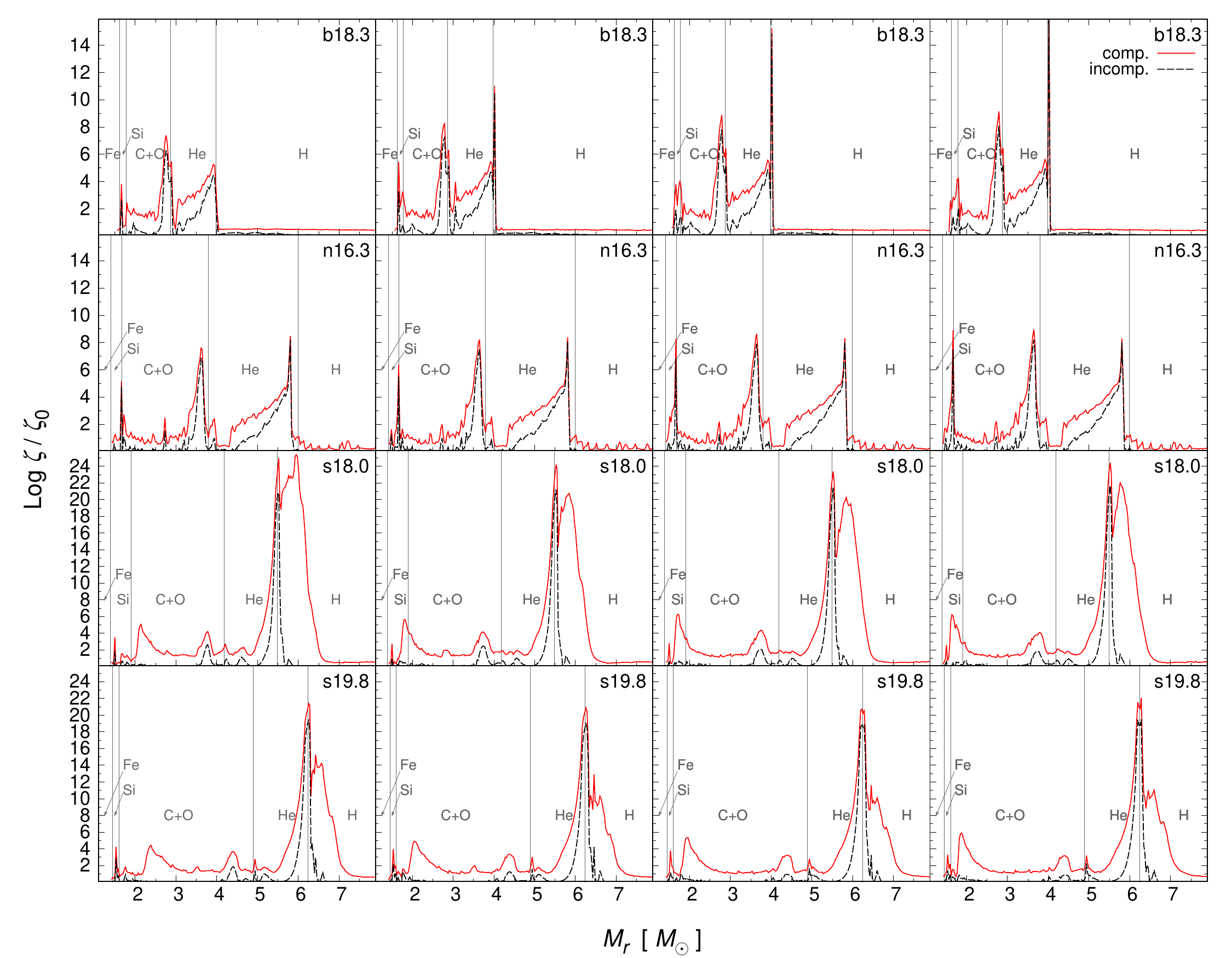}
\end{center}
\caption{Growth factors, $\zeta/\zeta_0$, as a function of the mass coordinate, $M_r$, for the progenitor models b18.3 (top), n16.3 (1st middle), 
s18.0 (2nd middle), and s19.8 (bottom) at the time when the shock wave reaches the radius of about 5 $\times$ 10$^{14}$ cm. From the left to 
the right, the cases of the injected energies $E_{\rm in}$ = (1.5, 2.0, 2.5, and 3.0) $\times$ 10$^{51}$ erg are shown, respectively. 
Red 
solid lines are the growth factors for compressible fluid and black 
dashed lines are ones for incompressible fluid. Thin vertical lines are the composition 
interfaces. See the text for the details.}
\label{fig:gr2}
\end{figure*}

\subsection{Results of Three-dimensional Simulations: Lower Resolution Cases} \label{subsec:res-3d-low}

In this section, the results of the 3D simulation with lower resolution are presented.
The results are summarized in Table~\ref{table:results}, where $E_{\rm exp}$ is the explosion energy (see Eq.~(\ref{eq:exp}) for the definition), 
$M_{\rm ej}$ ($^{56}$Ni) is the ejected total mass of $^{56}$Ni, $M_{\rm ej}$ ($^{44}$Ti) is the ejected total mass of $^{44}$Ti, 
$M_{3.0}$ ($^{56}$Ni), $M_{4.0}$ ($^{56}$Ni), and  $M_{4.7}$ ($^{56}$Ni) are the masses of $^{56}$Ni of which radial velocity is $\geq$ 
3000 km s$^{-1}$, $\geq$ 4000 km s$^{-1}$, and $\geq$ 4700 km s$^{-1}$, respectively, the value in the 8th column, 
$M_{3.0}$/$M_{\rm ej}$ ($^{56}$Ni), is the ratio of the values of 5th to 3rd columns, $v_{\rm NS}$ is the NS kick velocity 
(see Section~\ref{subsec:ns-kick} and Eq.~(\ref{eq:kick})). 
The 9th column, ``No.",  denotes the sequential serial number (model number) for Figure~\ref{fig:table3}. 
The values in Table~\ref{table:results} are obtained at the end of the simulation when the blast wave reaches $\sim$ 2 $\times$ 10$^{14}$ cm (after the shock breakout for all models). 
Explosion energies are defined by the expression: 
\begin{equation}
E_{\rm exp} = \int \int \int_V \left( \frac{1}{2}\,\rho \,\vec{v}^2 
+ \rho\, E +\rho \,\Phi \right) \mathrm{d} x \, \mathrm{d} y \,\mathrm{d} z,
\label{eq:exp}
\end{equation}
where $V$ is the computational domain, $\vec{v}$ is the velocity, $E$ is the internal energy, $\Phi$ is the gravitational potential. 

As noted in Section~\ref{subsec:model-1d}, the explosion energy of SN 1987A has been estimated from the observations in the range of (1--2) 
$\times$ 10$^{51}$ erg. The injected energy $E_{\rm in}$ is 2.5 $\times$ 10$^{51}$ erg except for models b18.3-ein1.5, b18.3-ein2.0, and b18.3-ein3.0 
in which $E_{\rm in}$ = (1.5, 2.0, and 3.0) $\times$ 10$^{51}$ erg, respectively. From the 2nd column of Table~\ref{table:results}, obtained explosion 
energies, $E_{\rm exp}$ from the lower resolution simulations are roughly $\lesssim$ 2 $\times$ 10$^{51}$ erg for the models with $E_{\rm in}$ = 2.5 
$\times$ 10$^{51}$ erg and those values are within the accepted range, i.e., (1--2) $\times$ 10$^{51}$ erg mentioned above. For models b18.3-ein1.5, 
b18.3-ein2.0, and b18.3-ein3.0, the obtained $E_{\rm exp}$ is roughly (1, 1.5, and 2.5) $\times$ 10$^{51}$ erg, respectively. As one can see, the 
$E_{\rm exp}$ values are well approximated as $E_{\rm in} - $ 0.5 $\times$ 10$^{51}$ erg. The $E_{\rm exp}$ value for the model b18.3-ein3.0 is 
outside the accepted range. Then, the model b18.3-ein3.0 is an extreme case. 

From the theoretical modeling of the observed optical light curves \citep[e.g.,][]{1988ApJ...330..218W,1990ApJ...360..242S}, the mass of ejected 
$^{56}$Ni has been deduced as 0.07 $M_{\odot}$. Obtained ejected masses of $^{56}$Ni from the simulations depend on the degrees of the 
asymmetry of bipolar-like explosion ($\beta \equiv v_{\rm pol}/v_{\rm eq}$), asymmetries against the equatorial plane 
($\alpha \equiv v_{\rm up}/v_{\rm down}$), and the progenitor models. For example, looking at the values for models b18.3-beta2, b18.3-beta4, 
b18.3-beta8, and b18.3-fid ($\beta$ = 2, 4, 8, and 16, respectively), the larger the $\beta$ value, the smaller the ejected mass of $^{56}$Ni. 
Comparing models b18.3-alpha1, b18.3-fid, and b18.3-alpha2 ($\alpha$ = 1.0, 1.5, 2.0, respectively), the larger the $\alpha$ value, the smaller 
the ejected mass of $^{56}$Ni. In the order of the models s19.8-fid, b18.3-fid, n16.3-fid, and s18.3-fid, the mass of ejected $^{56}$Ni is 
increasing. The dependence of the mass of ejected $^{56}$Ni on the pre-supernova models may reflect the structure (density and temperature) 
of the innermost regions around the composition interface of Fe/Si (see Figure~\ref{fig:prog1}). Overall, the obtained values of the mass of ejected 
$^{56}$Ni are (0.8--1) $\times$ 10$^{-1}$ $M_{\odot}$ 
but for some models, e.g., s18.0-mo13 and s18.0-beta2, the ejected mass of $^{56}$Ni ($\sim$ 0.13 $M_{\odot}$) is a bit large. 

As mentioned in Section~\ref{sec:intro}, the mass of $^{44}$Ti has been estimated as (3.1 $\pm$ 0.8) $\times$ 10$^{-4}$ $M_{\odot}$ 
\citep{2012Natur.490..373G} or (1.5 $\pm$ 0.3) $\times$ 10$^{-4}$ $M_{\odot}$ \citep{2015Sci...348..670B} from the observations of direct 
$\gamma$-ray lines from the decay of $^{44}$Ti. Overall, the obtained mass of ejected $^{44}$Ti are within the orders of 
10$^{-4}$--10$^{-3}$ $M_{\odot}$. In general, the amount of $^{44}$Ti synthesized by a neutrino-driven supernova is of the order of 
10$^{-5}$ $M_{\odot}$ \citep{2011ApJ...738...61F} and the large values ($\sim$ 10$^{-4}$ $M_{\odot}$) deduced from the observations 
are in some sense a mystery. A jetlike (globally aspherical) explosions could be essential for a strong alpha-rich freezeout to be realized to 
obtain a high mass ratio of $^{44}$Ti to $^{56}$Ni \citep{1997ApJ...486.1026N,1998ApJ...492L..45N}. 

It is noted that the calculations of the explosive nucleosynthesis in this paper are performed with only the small nuclear reaction network 
(19 nuclei are included). Then, the amount of $^{44}$Ti (roughly two orders of magnitude less than that of $^{56}$Ni) is inaccurate compared 
with the value of $^{56}$Ni. Additionally, the innermost regions around the composition interface of Fe/Si where the explosive nucleosynthesis 
occurs are slightly neutron-rich. Then, the synthesis of neutron-rich isotopes, $^{57}$Ni and $^{58}$Ni, is also expected. As demonstrated in 
the Appendix in Paper~II, the calculated masses of $^{56}$Ni and $^{44}$Ti could be overestimated by factors of $\sim$ 1.5 and 3, respectively, 
compared with those calculated by a larger nuclear reaction network (464 nuclei are included). Then, for example, the mass of ejected $^{56}$Ni 
and $^{44}$Ti for the model b18.3-fid, 8.1 $\times$ 10$^{-2}$ 
$M_{\odot}$ and 7.4 $\times$ 10$^{-4}$ $M_{\odot}$, could be translated as 5.4 
$\times$ 10$^{-2}$ $M_{\odot}$ and 2.5 $\times$ 10$^{-4}$ $M_{\odot}$, respectively, although the factors should depend on the inner structure 
of the progenitor models and the explosion asymmetries. Then, the obtained values of the masses of the ejected $^{56}$Ni and $^{44}$Ti are 
roughly consistent with the values suggested by the observations. The values of the masses of high velocity $^{56}$Ni listed in the 5--7th 
columns in Table~\ref{table:results}, could also be overestimated but the high velocity $^{56}$Ni is considered to be synthesized in outer less 
neutron-rich regions. Then, the correction for the values in the 5--7th columns should be much smaller than that for the value in the 3rd column. 
The value in the 8th column could be underestimated depending on the overestimation of the value in the 3rd column. The correction factors 
themselves, however, are rather uncertain and hereafter, we proceed with discussion based on the values listed (directly calculated by the 
numerical code in this paper). 

Hereafter, effects of asymmetries of explosions on the matter mixing are explored. 
The parameters related to the asymmetry of an explosion 
are $\beta$, $\alpha$, and the type of the asphericity of the explosion, i.e., one of ``cos", ``power", ``exponential", and ``elliptical" (see 
Eqs.~(\ref{eq:vcos}), (\ref{eq:vpwr}), (\ref{eq:vexp}), and (\ref{eq:vell}), respectively, and Table~\ref{table:models}). As seen in Figure~\ref{fig:shape} 
(in Appendix~\ref{sec:app1}), the larger the $\beta$ value, the higher the concentration of initial radial velocities along the polar ($z$-axis), if the type of 
the asphericity is fixed. It is noted that the type of ``cos" was adopted in Paper~I and Paper~II. As can be seen, the differences of the initial radial 
velocity distribution among $\beta = 2, 4, 8, 16$ are not so large around the polar axis, if the type is fixed to be ``cos". In the order of ``cos", ``power", 
``exponential", and ``elliptical", the concentration of radial velocities becomes higher. 
In this paper, the type of ``elliptical" is adopted as the fiducial one (see Section~\ref{subsec:model-3d}). 
%
\begin{figure*}[htb]
\begin{tabular}{cc}
\begin{minipage}{0.5\hsize}
\begin{center}
\includegraphics[width=7cm,keepaspectratio,clip]{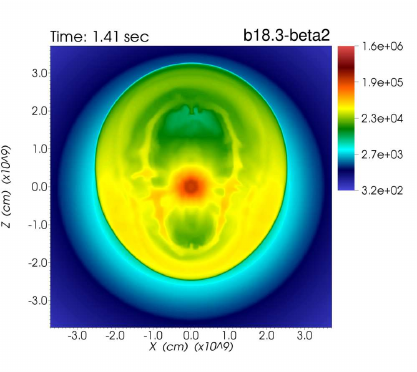}
\end{center}
\end{minipage}
\begin{minipage}{0.5\hsize}
\begin{center}
\includegraphics[width=7cm,keepaspectratio,clip]{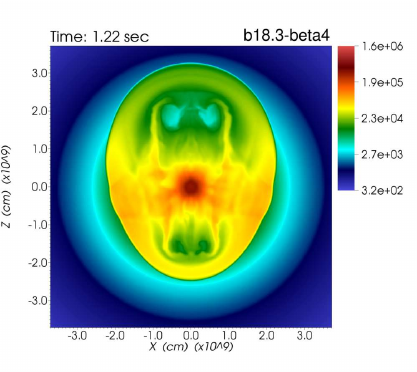}
\end{center}
\end{minipage}
\vspace{-0.4cm}
\\
\begin{minipage}{0.5\hsize}
\begin{center}
\includegraphics[width=7cm,keepaspectratio,clip]{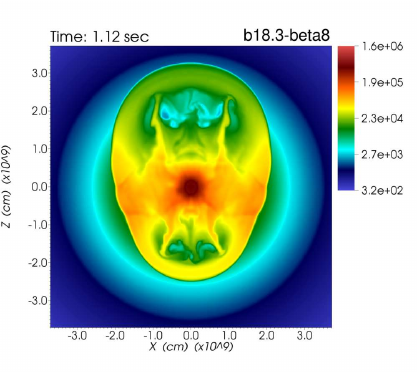}
\end{center}
\end{minipage}
\begin{minipage}{0.5\hsize}
\begin{center}
\includegraphics[width=7cm,keepaspectratio,clip]{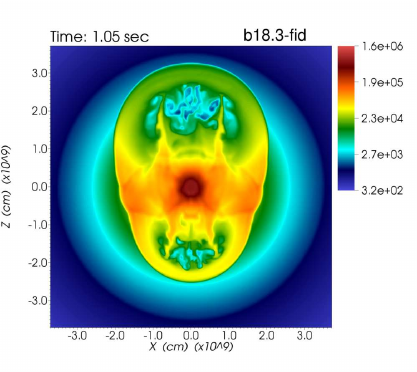}
\end{center}
\end{minipage}
\end{tabular}
\caption{Density color maps (2D slices of the $x$-$z$ plane) in a logarithmic scale at an early phase of the explosion ($\sim$ 1 sec). 
The unit of the values in the color bars is g cm$^{-3}$. 
Top left, top right, bottom left, and bottom right panels are for models b18.3-beta2, b18.3-beta4, b18.3-beta8, and b18.3-fid, respectively.}
\label{fig:dens-low-beta}
\end{figure*}
%
The dependence of the density distribution at an early phase ($\sim$ 1 sec) on each parameter related to the asphericity is discussed below. 
Figure~\ref{fig:dens-low-beta} shows the dependence on the parameter $\beta$ (other parameters are fixed). 
%
The shape of the blast wave (the interface between red and green colors) slightly depends on the parameter $\beta$. As expected, the 
larger the $\beta$ value is, the more elliptical the shape is but the elliptical shape is not so evident soon after the explosion compared with 
one of the initial radial velocity distribution seen in Figure~\ref{fig:shape}~(in Appendix~\ref{sec:app1}). Inside the blast wave, the 
density distribution is more sensitive to $\beta$ than that for the outer part. High density regions (red-colored) for models with 
larger $\beta$ are more concentrated around the equatorial plane ($z = 0$). As can be seen, instabilities are developed in regions inside 
the blast wave (high entropy bubbles: blue-colored regions). The growth of instabilities at such an early phase may be due to 
Kelvin-Helmholtz instability (shear velocity is necessary for its growth) and/or RT instability. The larger the $\beta$ value, the stronger 
the growth of instabilities (see in particular the bottom two panels). 
\begin{figure*}[htb]
\begin{tabular}{cc}
\begin{minipage}{0.5\hsize}
\begin{center}
\includegraphics[width=7cm,keepaspectratio,clip]{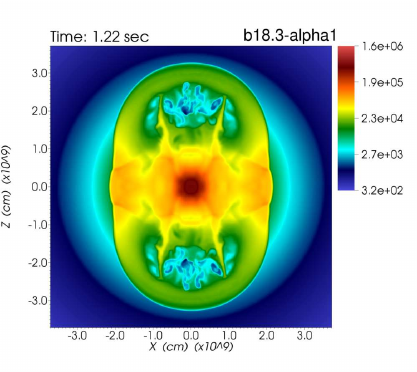}
\end{center}
\end{minipage}
\begin{minipage}{0.5\hsize}
\begin{center}
\includegraphics[width=7cm,keepaspectratio,clip]{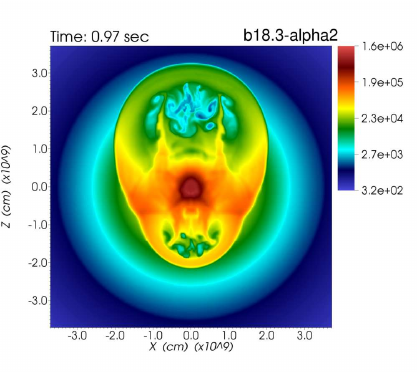}
\end{center}
\end{minipage}
\vspace{-0.4cm}
\end{tabular}
\caption{Same as Figure~\ref{fig:dens-low-beta} but for models b18.3-alpha1 (left) and b18.3-alpha2 (right).}
\label{fig:dens-low-alpha}
\end{figure*}
%
In Figure~\ref{fig:dens-low-alpha} the dependence on the parameter $\alpha$ is presented. 
In the case of $\alpha = 1.0$ (left panel), the shape of the blast wave is almost symmetric against the equatorial plane (as expected). On the other hand, 
in the case of $\alpha = 2.0$ (right) the shape and extension are very different between the upper and lower regions. 
Compared with the case of $\alpha = 1.0$, in the case of $\alpha = 2.0$, there are the following features: the development of hydrodynamic instabilities 
and high entropy bubbles are prominent in the upper regions, the blast wave reaches the radius of $3 \times 10^9$ cm earlier than in the case of 
$\alpha = 1.0$ (see the time for each model), and there are high density regions (red color) 
in equatorial regions. Focusing on the regions disturbed by instabilities in the upper regions, the regions in the case of $\alpha = 2.0$ are a bit larger than those 
in the case of $\alpha = 1.0$, whereas lower density regions (darker blue color) are recognized in the case of $\alpha = 1.0$.   
The features in the case of $\alpha = 1.5$ (bottom left panel in Figure~\ref{fig:dens-low-beta}) are roughly in between the two cases above ($\alpha = 1.0$ and $\alpha = 2.0$). 
%
\begin{figure*}[htb]
\begin{tabular}{cc}
\begin{minipage}{0.5\hsize}
\begin{center}
\includegraphics[width=7cm,keepaspectratio,clip]{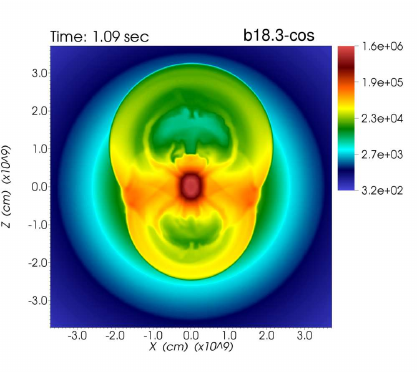}
\end{center}
\end{minipage}
\begin{minipage}{0.5\hsize}
\begin{center}
\includegraphics[width=7cm,keepaspectratio,clip]{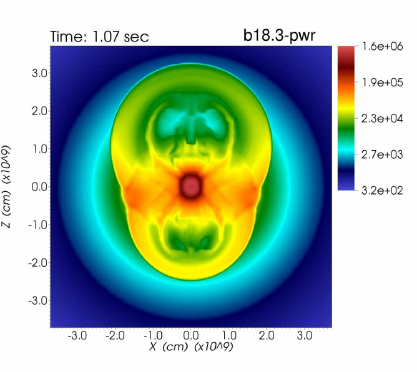}
\end{center}
\end{minipage}
\vspace{-0.4cm}
\\
\begin{minipage}{0.5\hsize}
\begin{center}
\includegraphics[width=7cm,keepaspectratio,clip]{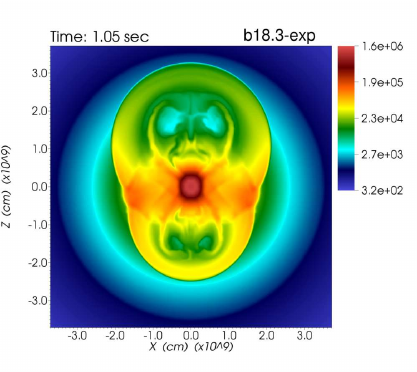}
\end{center}
\end{minipage}
\begin{minipage}{0.5\hsize}
\begin{center}
\includegraphics[width=7cm,keepaspectratio,clip]{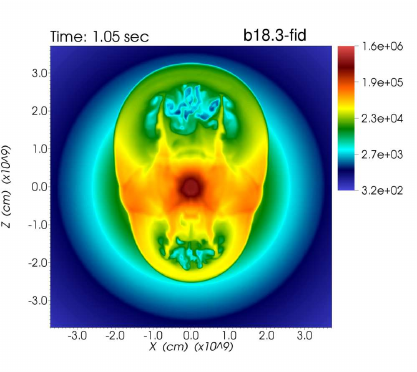}
\end{center}
\end{minipage}
\end{tabular}
\caption{Same as Figure~\ref{fig:dens-low-beta} but for models b18.3-cos (top left), b18.3-power (top right), b18.3-exp (bottom left), 
and b18.3-fid (bottom right).}
\label{fig:dens-low-shape}
\end{figure*}
%
%
Figure~\ref{fig:dens-low-shape} shows the dependence on the type of the aspherical explosion. 
As can be seen, the shape of the blast wave is not so different among the four types but the shape of the type ``elliptical" (bottom right) 
is slightly more elliptical. In the order of the models, b18.3-cos, b18.3-pwr, b18.3-exp, and b18.3-fid, the regions of high entropy bubbles 
inside the blast wave are more pronounced and more disturbed due to instabilities. 
%
\begin{figure*}[htb]
\begin{tabular}{cc}
\begin{minipage}{0.5\hsize}
\begin{center}
\includegraphics[width=7cm,keepaspectratio,clip]{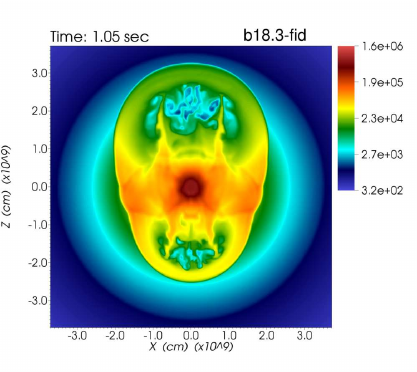}
\end{center}
\end{minipage}
\begin{minipage}{0.5\hsize}
\begin{center}
\includegraphics[width=7cm,keepaspectratio,clip]{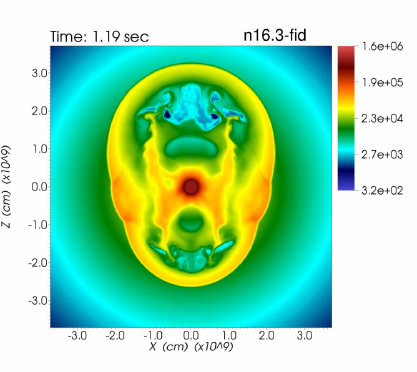}
\end{center}
\end{minipage}
\vspace{-0.4cm}
\\
\begin{minipage}{0.5\hsize}
\begin{center}
\includegraphics[width=7cm,keepaspectratio,clip]{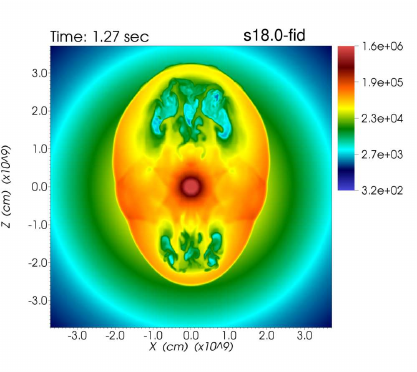}
\end{center}
\end{minipage}
\begin{minipage}{0.5\hsize}
\begin{center}
\includegraphics[width=7cm,keepaspectratio,clip]{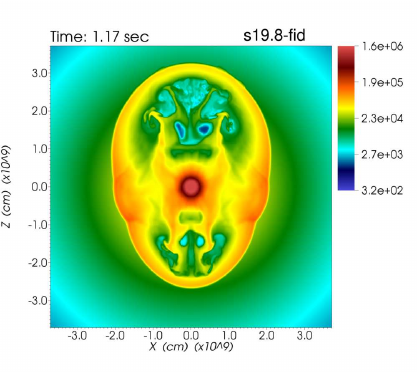}
\end{center}
\end{minipage}
\end{tabular}
\caption{Same as Figure~\ref{fig:dens-low-beta} but for models b18.3-fid (top left), n16.3-fid (top right), s18.0-fid (bottom left), and 
s19.8-fid (bottom right).}
\label{fig:dens-low-model}
\end{figure*}
%
In Figure~\ref{fig:dens-low-model} the dependence of the early morphology of the explosion on the progenitor models is depicted. 
%
The shape of the blast wave is different between the two BSG models 
(b18.3 and n16.3: the top panels) and the other RSG models (s18.0 and s19.8: the bottom panels). At the time presented here, the blast 
wave is inside the C+O layer ($r < 3 \times 10^{9}$ cm). As seen in Figure~\ref{fig:prog1}, the density structure are different among the 
progenitor models. The density structures are relatively similar between the two RSG models, while the sizes (in both the mass and the radius, 
see Figure~\ref{fig:prog1}, \ref{fig:prog2}, and~\ref{fig:gr2}, respectively) of the C+O cores and the density gradients are different 
between the two BSG models. The size of the C+O core of the model b18.3 (binary merger model) is smaller than that of n16.3 (single star 
model). The $\rho\,r^3$ gradient in the C+O core of the model b18.3 is flatter than that of the model n16.3. It is difficult to find a clear 
correlation between the morphology of the explosion at the early phase and the density structure inside the C+O core. 
%
Nevertheless, the bipolar structure for the two RSG models is more prominent (the width of the bipolar structure is narrower) than that for the two BSG models. 
Between the BSG models, the shape of the bipolar 
structure of the model n16.3 is wider than that of b18.3 because of the steeper $\rho\,r^3$ gradient and the larger size of the C+O core, 
which cause rapid deceleration of the shock wave. Among the four pre-supernova models, the model n16.3 has a distinct $\rho\,r^3$ profile 
inside the silicon layer compared with those of the others, i.e., the profile of the silicon layer of n16.3 is rather flat compared with those of 
the others (the gradients of $\rho \,r^3$ for the others are overall negative), which causes the deceleration of the earliest phase. Then, the 
structures of the silicon layer could affect the morphologies of the early phases. 

\begin{figure*}[htb]
\begin{tabular}{cc}
\begin{minipage}{0.5\hsize}
\begin{center}
\includegraphics[width=7cm,keepaspectratio,clip]{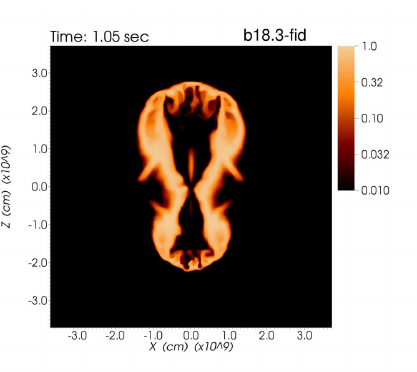}
\end{center}
\end{minipage}
\begin{minipage}{0.5\hsize}
\begin{center}
\includegraphics[width=7cm,keepaspectratio,clip]{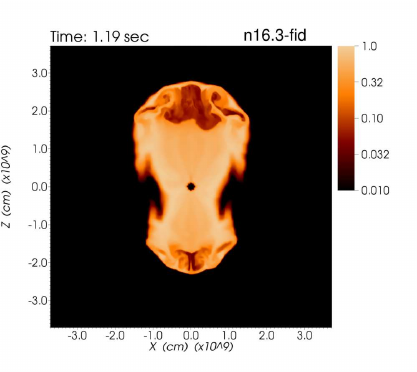}
\end{center}
\end{minipage}
\vspace{-0.4cm}
\\
\begin{minipage}{0.5\hsize}
\begin{center}
\includegraphics[width=7cm,keepaspectratio,clip]{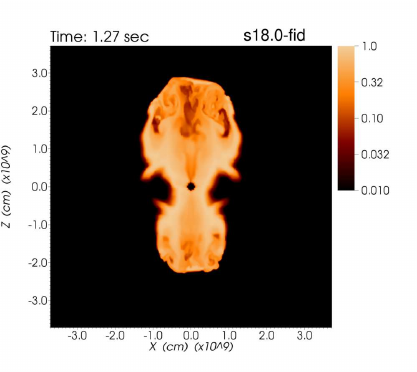}
\end{center}
\end{minipage}
\begin{minipage}{0.5\hsize}
\begin{center}
\includegraphics[width=7cm,keepaspectratio,clip]{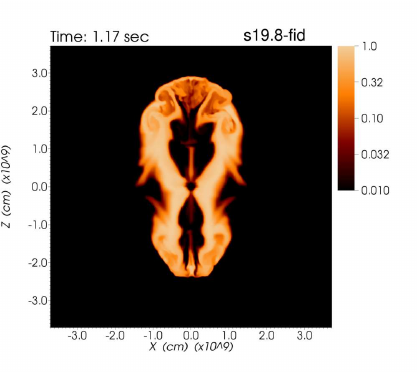}
\end{center}
\end{minipage}
\end{tabular}
\caption{Color maps (2D slices of the $x$-$z$ plane) of the mass fraction of $^{56}$Ni at an early phase of the explosion ($\sim$ 1 sec). 
The colors are logarithmically scaled.  
Top left, top right, bottom left, and bottom right panels are for models b18.3-fid, n16.3-fid, s18.0-fid, and s19.8-fid, respectively.}
\label{fig:ni56-low-model-early}
\end{figure*}
%
Hereafter, spatial distributions of representative elements are presented. 
Figure~\ref{fig:ni56-low-model-early} shows the distributions of $^{56}$Ni at an early phase ($\sim$ 1 sec). 
The dependence on the progenitor model is presented. 
The shapes of the outer edge of the distribution of $^{56}$Ni are 
not so different among the progenitor models, although the widths of the bipolar structure are slightly different reflecting the density 
distribution as seen in Figure~\ref{fig:dens-low-model}. The inner distributions of $^{56}$Ni are rather different among the models. 
%
Hole structures (cavities) of $^{56}$Ni inside the outer edge are found in the models b18.3 and s19.8. 
It is noted that the small spherical holes ($r \lesssim$ 10$^{8}$ cm) at the origin are the regions corresponding to the compact object. The products of the explosive 
nucleosynthesis sensitively depend on the peak temperature and density during the burning process 
\citep[see e.g.,][]{2015ApJ...807..110J}. In a high entropy regime, the synthesis of $^{56}$Ni is limited due to the so-called alpha-rich 
freezeout. The cavities inside the outer edges could correspond to the regions of strong alpha-rich freezeout. 
%
\begin{figure*}[htb]
\begin{tabular}{cc}
\begin{minipage}{0.5\hsize}
\begin{center}
\includegraphics[width=7cm,keepaspectratio,clip]{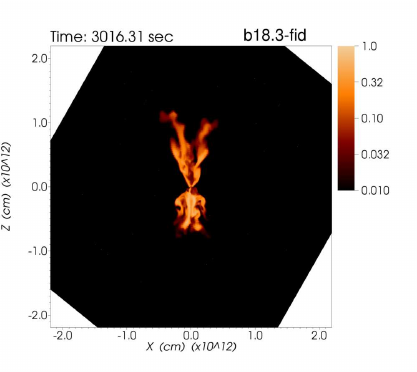}
\end{center}
\end{minipage}
\begin{minipage}{0.5\hsize}
\begin{center}
\includegraphics[width=7cm,keepaspectratio,clip]{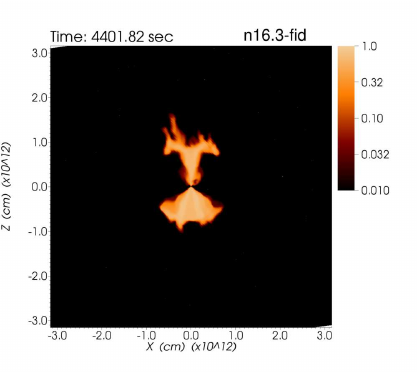}
\end{center}
\end{minipage}
\vspace{-0.4cm}
\\
\begin{minipage}{0.5\hsize}
\begin{center}
\includegraphics[width=7cm,keepaspectratio,clip]{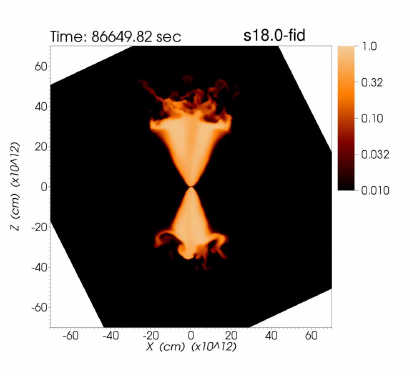}
\end{center}
\end{minipage}
\begin{minipage}{0.5\hsize}
\begin{center}
\includegraphics[width=7cm,keepaspectratio,clip]{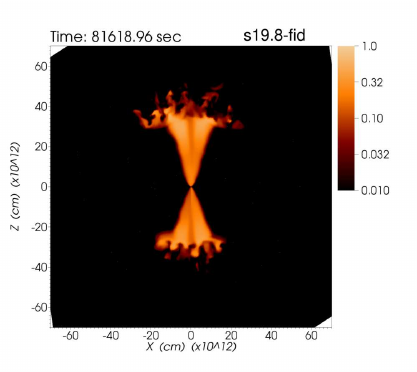}
\end{center}
\end{minipage}
\end{tabular}
\caption{Same as Figure~\ref{fig:ni56-low-model-early} but for just before the shock breakout. The white color regions are outside the 
computational domain. Since the stellar radii are much different between the two BSG models (top two panels) and the two RSG models (bottom two panels), 
spatial scales shown are much different between the top two panels and the bottom two panels.}
\label{fig:ni56-low-model-late}
\end{figure*}
%
%
Figure~\ref{fig:ni56-low-model-late} shows the distributions of $^{56}$Ni just before the shock breakout. 
Depending on the structures of the progenitor 
models, the distributions are rather different. In the two RSG models s18.0 and s19.8, a bi-cone-like structure is clearly seen. On top of the 
bi-cone-like structures, small-scale fingers due to RT instabilities are prominent. Between the two BSG models, b18.3 and n16.3, the distribution 
of $^{56}$Ni in the model of b18.3 is more shrunk and wobbling than that in the model n16.3. Comparing the distributions of $^{56}$Ni in 
Figure~\ref{fig:ni56-low-model-early} and Figure~\ref{fig:ni56-low-model-late}, the initial bipolar-like distributions are roughly kept even just 
before the shock breakout but the shapes are rather modified during the shock propagation.  
%
\begin{figure*}[htb]
\begin{tabular}{cc}
\begin{minipage}{0.5\hsize}
\begin{center}
\includegraphics[width=7cm,keepaspectratio,clip]{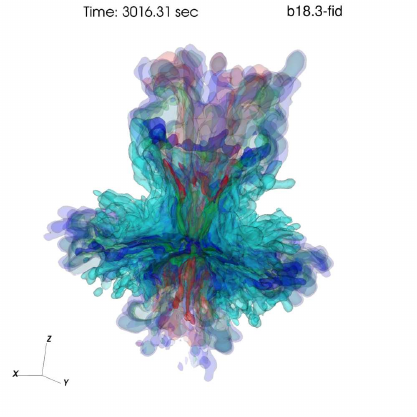}

\end{center}
\end{minipage}
\begin{minipage}{0.5\hsize}
\begin{center}
\includegraphics[width=7cm,keepaspectratio,clip]{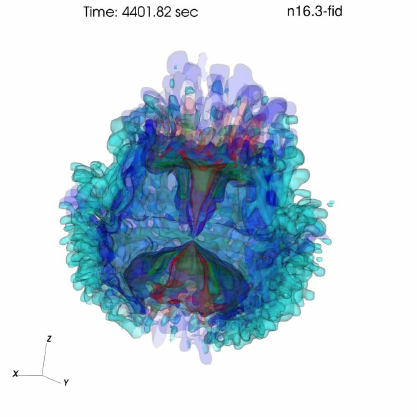}
\end{center}
\end{minipage}
\\
\begin{minipage}{0.5\hsize}
\begin{center}
\includegraphics[width=7cm,keepaspectratio,clip]{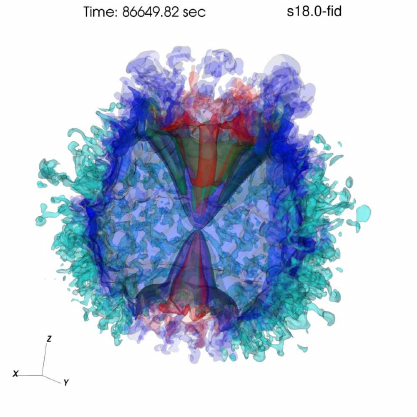}
\end{center}
\end{minipage}
\begin{minipage}{0.5\hsize}
\begin{center}
\includegraphics[width=7cm,keepaspectratio,clip]{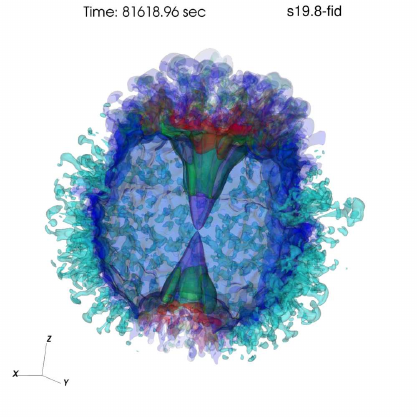}
\end{center}
\end{minipage}
\end{tabular}
\caption{Isosurfaces of the mass fractions of elements, $^{56}$Ni (red), $^{28}$Si (green), $^{16}$O (blue), and $^{4}$He (light blue), 
just before the shock breakout in a 3D view. 
Isosurfaces of the 10 \% (lighter color) and 70 \% (darker color) of the maximum value for each element are shown. 
To see the inner structure, the regions of $x > 0$ and $y > 0$  are clipped. 
Top left, top right, bottom left, and bottom right panels are for models b18.3-fid, n16.3-fid, s18.0-fid, and s19.8-fid, respectively. 
Interactive 3D models on Sketchfab corresponding to top left (https://skfb.ly/6OZDr), top right (https://skfb.ly/6OZD8), 
bottom left (https://skfb.ly/6OZDt), and bottom right (https://skfb.ly/6OZD9) panels are available.}
\label{fig:elem-3d-low-model}
\end{figure*}
%
Figure~\ref{fig:elem-3d-low-model} shows the 3D distribution of elements, $^{56}$Ni, $^{28}$Si, $^{16}$O, and $^{4}$He, 
just before the shock breakout. The dependence on the progenitor model is shown. 
%
The distributions of $^{56}$Ni are different from each other (as also seen in Figure~\ref{fig:ni56-low-model-late}) and other elements, $^{28}$Si, $^{16}$O, $^{4}$He, are also different from each other. 
The distributions in the two RSG models (s18.0-fid and s19.8-fid) are similar to each other but the distributions in the two BSG models (b18.3-fid 
and n16.3-fid) are rather different. Overall, the distributions of heavier two elements, i.e., $^{56}$Ni and $^{28}$Si, are similar to each other 
compared with the other two elements, $^{16}$O, $^{4}$He. In models n16.3, s18.0, and s19.8, bi-cone-like structures of $^{56}$Ni and $^{28}$Si 
are seen. The bi-cone-like structures in the model n16.3 are more asymmetric against the equatorial plan ($x$-$y$ plane). 
A distinct feature of the model b18.3-fid is that the distributions of $^{16}$O and $^4$He are more concentrated around the equatorial plane 
(the fingers are extended from more central regions) than those in the other models. 
On the other hand, the distributions of $^{16}$O and $^4$He in the models n16.3, s18.0, and s19.8  
are more roundly extended around the elements, $^{56}$Ni and $^{28}$Si. 
The reason for the different distributions of $^{16}$O and $^4$He among the progenitor models is discussed in \S~\ref{subsec:res-3d-high}. 

As introduced in Section~\ref{sec:intro}, the observations of SN 1987A have suggested the existence of high velocity $^{56}$Ni in helium and hydrogen layers, which has lead to 
the invocation of the matter mixing to convey the inner most material into outer layers: from the observations of [Fe II] lines \citep{1990ApJ...360..257H}, tails of the lines reach 
$\sim$ 4000 km s$^{-1}$ and at least 4\% of the iron had a velocity of $\gtrsim$ 3000 km s$^{-1}$; from the fine-structure developed in H$_{\alpha}$ line 
\citep[the Bochum event:][]{1988MNRAS.234P..41H}, the existence of a high velocity (4700 $\pm$ 500 km s$^{-1}$) $^{56}$Ni clump of $\sim$ 10$^{-3}$ $M_{\odot}$ 
has been suggested \citep{1995A&A...295..129U}. 
Such observational constrains can be a test for the models in this paper. Then, we consider three conditions to test models as follows: 
%
%
i) the ratio of the mass of $^{56}$Ni that has velocity $\geq$ 3000 km s$^{-1}$ to the total $^{56}$Ni mass is greater than 4\%; 
ii) the mass of $^{56}$Ni that has velocity $\geq$ 4000 km s$^{-1}$ is greater than 10$^{-3}$ $M_{\odot}$; 
iii) the mass of $^{56}$Ni that has velocity $\geq$ 4700 km s$^{-1}$ is greater than 10$^{-3}$ $M_{\odot}$.
The first condition is based on \citet{1990ApJ...360..257H}. For the second condition, there has been no clear constrain on the mass but we take $\sim$ 
10$^{-3}$ $M_{\odot}$ as a minimum requirement based on the fact that the tails of [Fe II] line reach 4000 km s$^{-1}$ and the suggestion from 
\citet{1995A&A...295..129U}. The third condition is directly based on \citet{1995A&A...295..129U} and more stringent than the second one. The derivation 
of the values in \citet{1995A&A...295..129U} was, however, based on a simple modeling of the H$_{\alpha}$ line and errors in the velocity are bit large. 
Then, we regard the third condition as an optional one. From the calculated models, masses of representative elements, in particular $^{56}$Ni, and their 
radial velocities are discussed by comparing with the conditions above. In Table~\ref{table:results}, masses of $^{56}$Ni that have radial velocities greater 
than specific values are listed in the 5, 6, and 7th columns, i.e., $M_{3.0}$ ($^{56}$Ni), $M_{4.0}$ ($^{56}$Ni), and $M_{4.7}$ ($^{56}$Ni), respectively. 
The second and third conditions can be tested by seeing the 6th and 7th columns. The first condition can be tested from the 8th column, 
$M_{3.0}$/$M_{\rm ej}$ ($^{56}$Ni). In Figure~\ref{fig:table3}, the values of 5th--8th columns for all models listed in Table~\ref{table:results} are plotted 
(see Figure~\ref{fig:table3} for the discussion in this Section and Section~\ref{subsec:res-3d-high}, when necessary). 
The models of lower resolution simulations that satisfy both the first and second conditions are b18.3-beta8, b18.3-fid, s18.0-fid, s19.8-beta4, 
s19.8-beta8, s19.8-fid, b18.3-alpha2, b18.3-clp0, and b18.3-ein3.0. It is worth noting that there is no model with the n16.3 progenitor that satisfies 
the two conditions simultaneously. The models that include ``mo13" have the same values for the parameters, $\beta$ and $\alpha$, as in Paper~I but among the 
models there is no model that satisfies the two conditions. Only three models, i.e., s19.8-beta8, b18.3-alpha2, and b18.3-ein3.0, satisfy not only the two 
conditions but also the third one. 
The dependence of the radial velocity of $^{56}$Ni on the parameter $\beta \equiv v_{\rm pol}/v_{\rm eq}$ can be checked by comparing models, e.g., 
b18.3-beta2, b18.3-beta4, b18.3-beta8, and b18.3-fid. The larger the $\beta$ value is, the larger the values, $M_{4.0}$ ($^{56}$Ni) and 
$M_{3.0}$/$M_{\rm ej}$ ($^{56}$Ni), are. It is noted that the larger $\beta$ value is, the stronger the concentration of initial radial velocities around the 
polar axis is (see Figure~\ref{fig:shape} in Appendix~\ref{sec:app1}). The dependence on the type of the asphericity can be seen by comparing models 
b18.3-fid, b18.3-cos, b18.3-pwr, and b18.3-exp. In the order of b18.3-cos, b18.3-pwr, b18.3-exp, and b18.3-fid, the values, $M_{4.0}$ ($^{56}$Ni) and 
$M_{3.0}$/$M_{\rm ej}$ ($^{56}$Ni), increase. As seen in Figure~\ref{fig:shape}, in the type of ``elliptical", the concentration of initial radial velocities 
around the polar axis is the most prominent if the $\beta$ values is fixed. 
Compared with b18.3-fid, the model b18.3-alpha2 has larger $M_{4.0}$ ($^{56}$Ni) and $M_{3.0}$/$M_{\rm ej}$ ($^{56}$Ni) values, which reflects 
a stronger explosion in a certain direction in the model b18.3-alpha2 with $\alpha \equiv v_{\rm up}/v_{\rm down}$ = 2.0. The 
$M_{4.0}$ ($^{56}$Ni) and $M_{3.0}$/$M_{\rm ej}$ ($^{56}$Ni) values in the models b18.3-fid and b18.3-clp0 are not so different from each other, 
the values in the model b18.3-clp0 (no fluctuation in the initial radial velocities: see Appendix~\ref{sec:app2}) are slightly larger than those in the model 
b18.3-fid though. Then, the existence of the initial clumpiness (the fluctuations in the initial radial velocities) has a negative role at least for the b18.3 model. 
The role of initial clumpiness could, 
however, change depending on the structure of the progenitor model. Actually, in Paper~I, the existence of an initial clumpiness has a positive 
role for obtaining high velocity $^{56}$Ni with the b16.3 model (see the results for models AM2 and AM3 in Paper~I). As a summary, if we exclude 
models with RSG progenitor models and/or the highest $E_{\rm in}$ value (3.0 $\times$ 10$^{51}$ erg) model, b18.3-beta8, b18.3-fid, b18.3-alpha2, 
and b18.3-clp0, could be promising for SN 1987A at this time. 
\begin{figure*}[htb]
\begin{tabular}{cc}
\hspace*{-0.4cm}
\begin{minipage}{0.5\hsize}
\begin{center}
\includegraphics[width=7cm,keepaspectratio,clip]{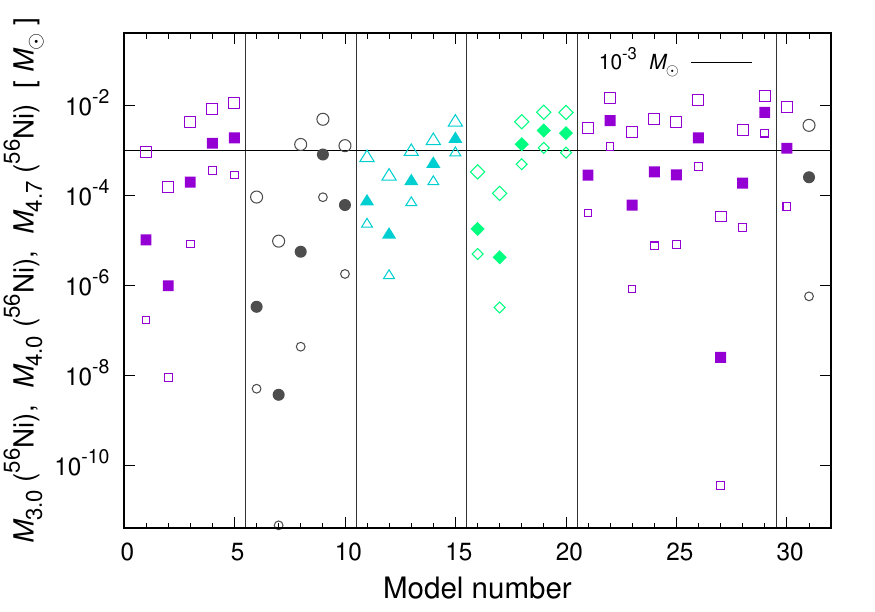}
\end{center}
\end{minipage}
\hspace*{-0.4cm}
\begin{minipage}{0.5\hsize}
\begin{center}
\includegraphics[width=7cm,keepaspectratio,clip]{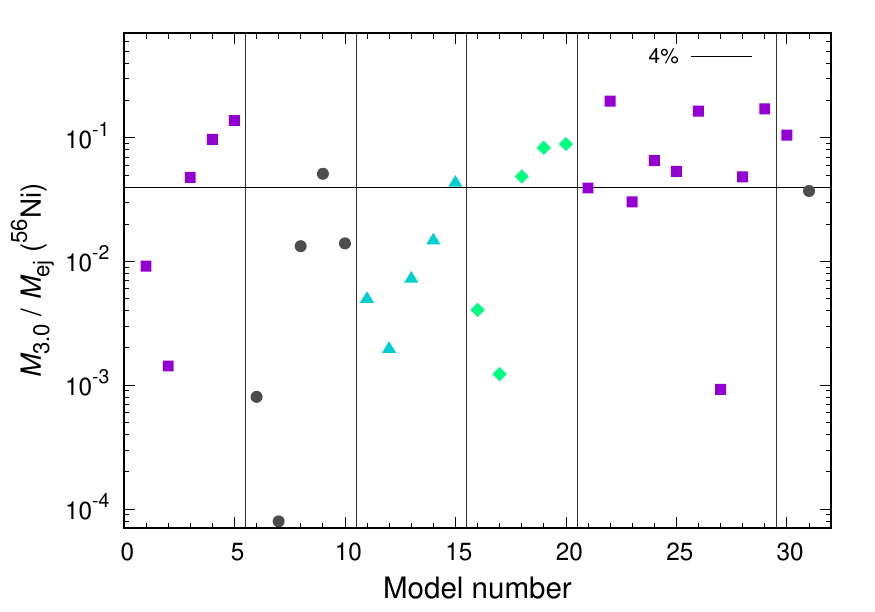}
\end{center}
\end{minipage}
\end{tabular}
\caption{Left panel: masses of $^{56}$Ni that have velocities higher than 3000 km s$^{-1}$, 4000 km s$^{-1}$, and 4700 km s$^{-1}$ at the end of the simulation, 
i.e., $M_{3.0}$ ($^{56}$Ni)~(large open points), $M_{4.0}$ ($^{56}$Ni)~(filled points), and $M_{4.7}$ ($^{56}$Ni)~(small open points), respectively, as a function 
of the model number (see the 10th column in Table~\ref{table:results}). 
Squares, circles, triangles, and diamonds denote the points for the models with the progenitor models b18.3, n16.3, s18.0, and s19.8, respectively. 
The horizontal solid line is the value of 10$^{-3}$ $M_{\odot}$. 
Right panel: ratios of the mass of $^{56}$Ni that has velocity higher than 3000 km s$^{-1}$ to the total ejected $^{56}$Ni mass 
at the end of the simulation, $M_{3.0}$/$M_{\rm ej}$ ($^{56}$Ni), as a function of the model number. The four shapes denote the same as in the left panel. 
The horizontal solid line is the value of 4 $\times$ 10$^{-2}$ (4\%). 
}
\label{fig:table3}
\end{figure*}

Here, mass distributions of representative elements including $^{56}$Ni are discussed. 
%
\begin{figure*}[htb]
\begin{tabular}{cc}
\begin{minipage}{0.5\hsize}
\begin{center}
\includegraphics[width=7cm,keepaspectratio,clip]{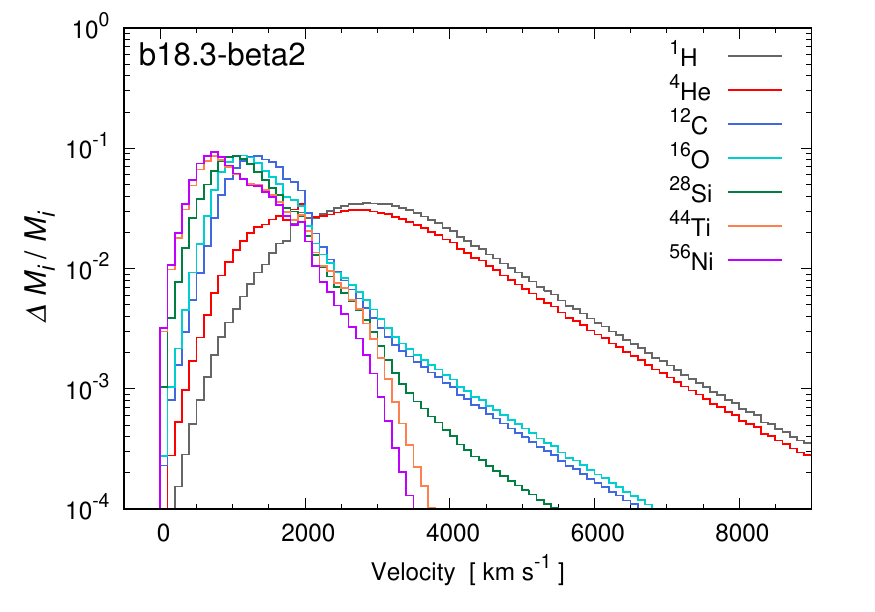}
\end{center}
\end{minipage}
\begin{minipage}{0.5\hsize}
\begin{center}
\includegraphics[width=7cm,keepaspectratio,clip]{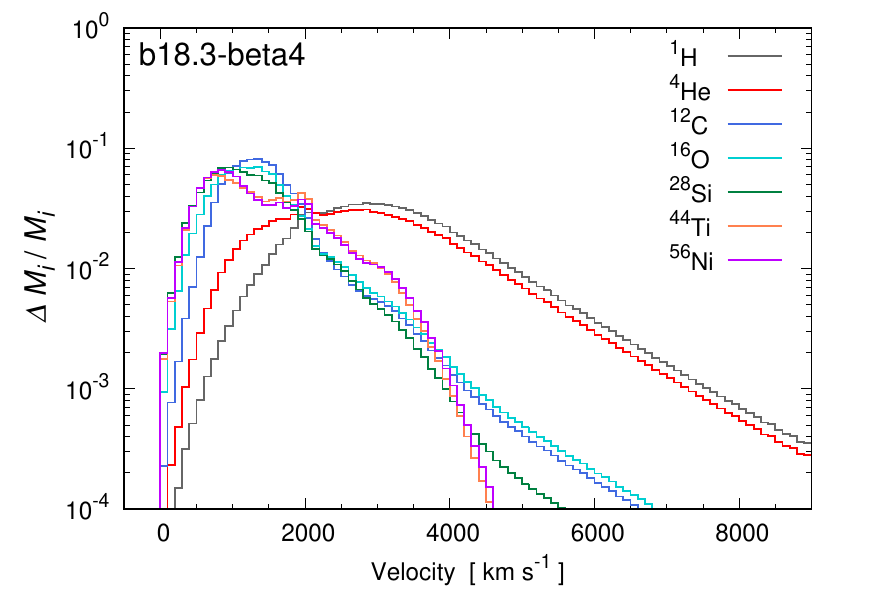}
\end{center}
\end{minipage}
\\
\begin{minipage}{0.5\hsize}
\begin{center}
\includegraphics[width=7cm,keepaspectratio,clip]{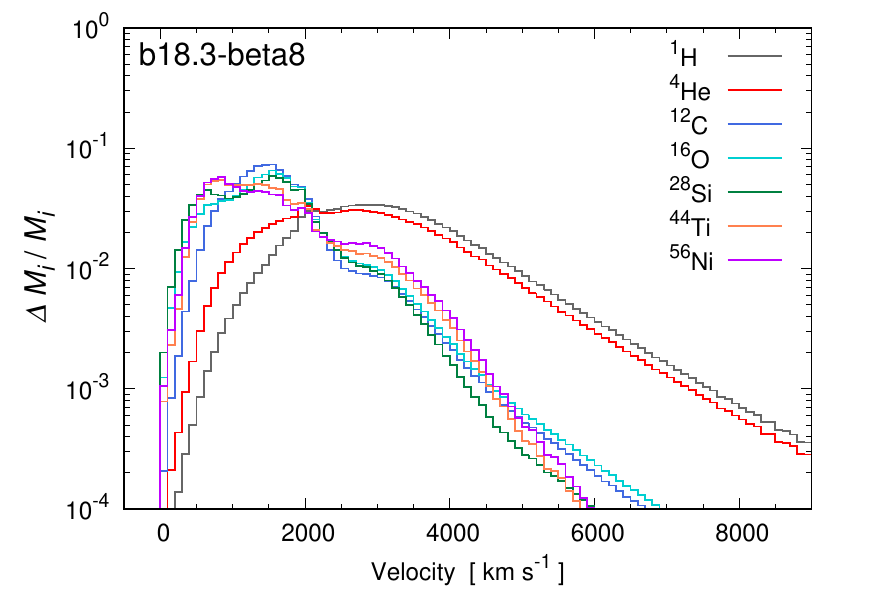}
\end{center}
\end{minipage}
\begin{minipage}{0.5\hsize}
\begin{center}
\includegraphics[width=7cm,keepaspectratio,clip]{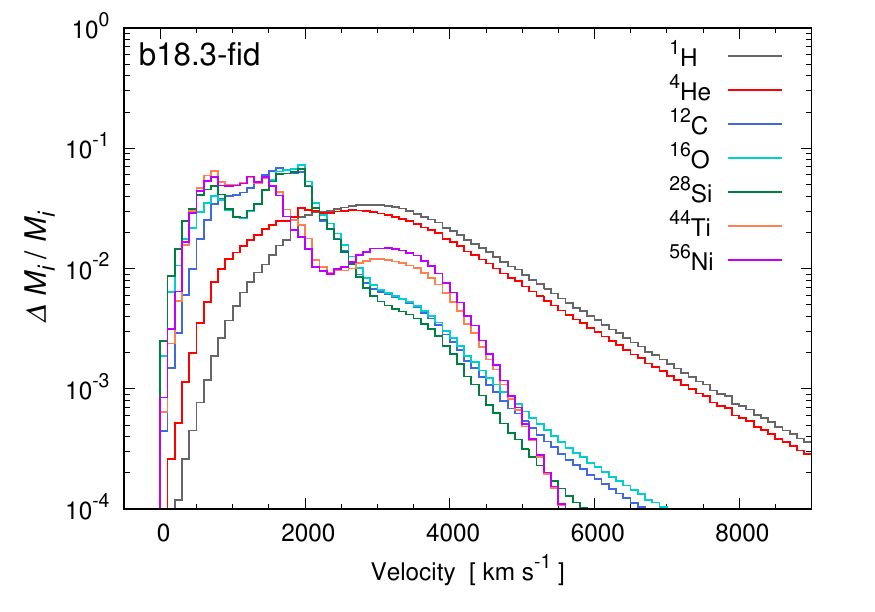}
\end{center}
\end{minipage}
\end{tabular}
\caption{
Normalized masses of elements, $^1$H, $^4$He, $^{12}$C, $^{16}$O, $^{28}$Si, $^{44}$Ti, and $^{56}$Ni, as a function of radial velocity at the end of the simulation 
(hereafter, the times in parentheses after model names denote the simulation time corresponding to each model shown). 
Top left, top right, bottom left, and bottom right panels are for models b18.3-beta2 (76,819 sec), b18.3-beta4 (78,501 sec), b18.3-beta8 (78,128 sec), and b18.3-fid (80,621 sec), respectively. 
$\it{\Delta} M_i$ is the mass of the element, $i$, in the velocity range of $v \sim v + \it{\Delta} v$. $M_i$ is the total ejected mass of 
element, $i$. The size of velocity bins, $\it{\Delta} v$, is 100 km s$^{-1}$.}
\label{fig:vel-low-beta}
\end{figure*}
%
Figure~\ref{fig:vel-low-beta} shows distribution of elements as a function of radial velocity at the end of the simulation 
(see the figure caption for the definitions of several variables). The dependence on the parameter $\beta$ is shown. 
As can be seen, helium and hydrogen in outer layers have very high velocities of $\gtrsim$ 6000 km s$^{-1}$ 
and the distribution of elements, $^1$H, $^4$He, $^{12}$C, $^{16}$O, $^{28}$Si in the velocities of $\gtrsim$ 5000 km s$^{-1}$ hardly depends on the $\beta$ value. 
On the other hand, the distributions of in the velocities of $\lesssim$ 5000 km s$^{-1}$ are different among the four models. 
The most distinct feature is that the larger the $\beta$ value is, the more extended ($\gtrsim$ 4000 km s$^{-1}$) the high velocity tails of $^{56}$Ni 
and $^{44}$Ti are. The amounts of $^{12}$C, $^{16}$O, and $^{28}$Si around 4000 km s$^{-1}$ are also more enhanced than those for models with 
higher $\beta$ values. It is notable that inward mixing of hydrogen down to the velocity of $\sim$ 1000 km s$^{-1}$ are recognized as seen in 
Figure~\ref{fig:vel-low-beta}. The minimum velocities of hydrogen are comparable with the values \citep[e.g., 800 km s$^{-1}$:][]{1990ApJ...360..242S} 
suggested by modeling of the light curves of SN 1987A \citep[][]{1990ApJ...360..242S,2000ApJ...532.1132B} 
and the values are marginally consistent with the value ($\lesssim$ 700 km s$^{-1}$) deduced from the spectral modeling \citep{1998ApJ...497..431K}. 
As mentioned above, among the models displayed in Figure~\ref{fig:vel-low-beta}, only models b18.3-beta8 (bottom left) and b18.3-fid (bottom right) 
have the amount of high velocity 
$^{56}$Ni required from the observations. 
%
\begin{figure*}[htb]
\begin{tabular}{cc}
\begin{minipage}{0.5\hsize}
\begin{center}
\includegraphics[width=7cm,keepaspectratio,clip]{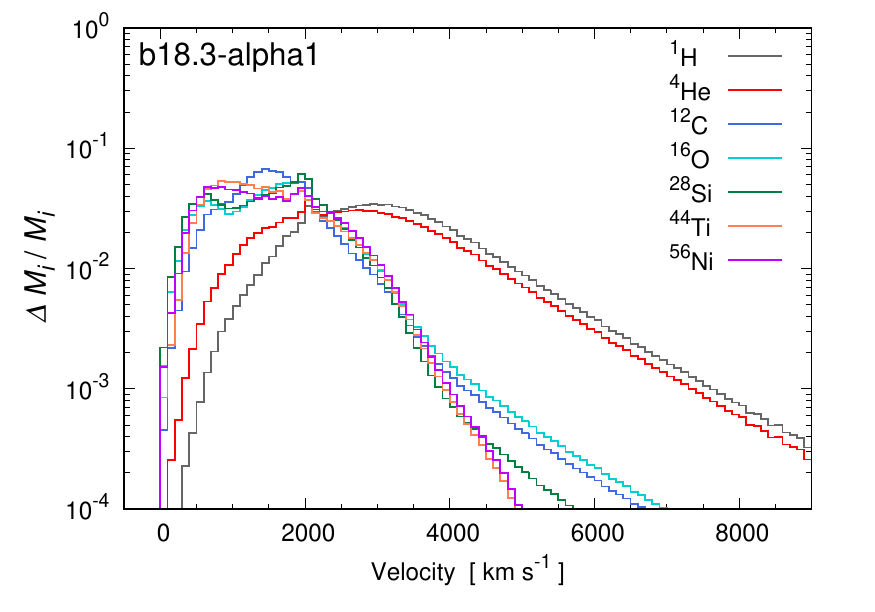}
\end{center}
\end{minipage}
\begin{minipage}{0.5\hsize}
\begin{center}
\includegraphics[width=7cm,keepaspectratio,clip]{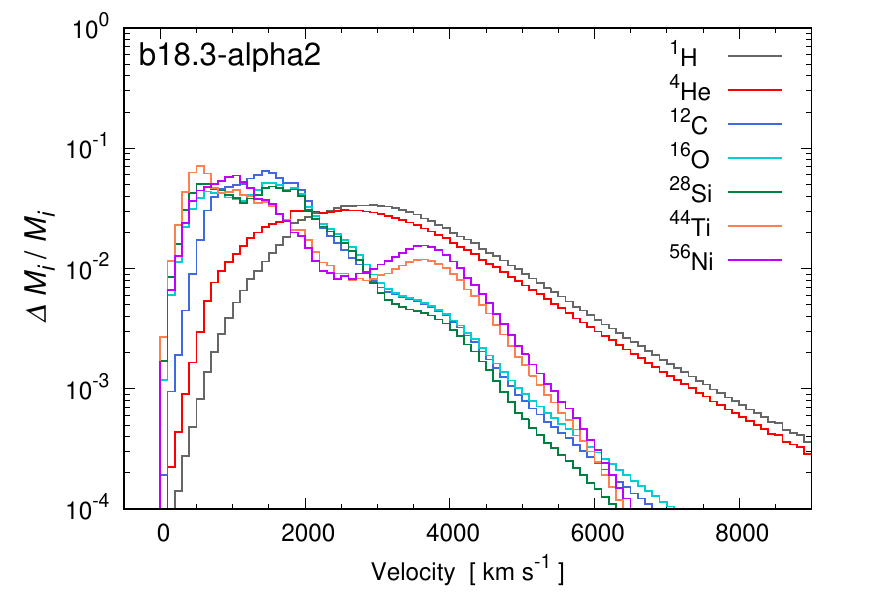}
\end{center}
\end{minipage}
\end{tabular}
\caption{Same as Figure~\ref{fig:vel-low-beta} but for models b18.3-alpha1 (left; 77,425 sec), b18.3-alpha2 (right; 77,261 sec).}
\label{fig:vel-low-alpha}
\end{figure*}
%
In Figure~\ref{fig:vel-low-alpha}, the dependence on the parameter $\alpha$ is presented. As for $^1$H and $^4$He, the distributions are similar 
between the two cases 
($\alpha = 1.0$ and $\alpha = 2.0$). On the other hand, the distributions for other elements, $^{12}$C, $^{16}$O, $^{28}$Si, $^{44}$Ti, and $^{56}$Ni, 
are rather different between the two cases in particular for velocities of $\gtrsim$ 2000 km s$^{-1}$. Compared with the case of $\alpha = 1.0$, 
in the case of $\alpha = 2.0$, the amount of elements, $^{12}$C, $^{16}$O, and $^{28}$Si, around velocities of about 2000 km s$^{-1}$ is slightly reduced, whereas 
the amount around 4000 km s$^{-1}$ is enhanced. Such features are more prominent for elements, $^{44}$Ti and $^{56}$Ni. In the case of $\alpha = 2.0$, a bump around velocities of 2500--3000 km s$^{-1}$ is recognized for $^{44}$Ti and $^{56}$Ni and the second peak appears at about 3500 km s$^{-1}$. 
In the case of $\alpha = 2.0$, the high velocity tail for elements, $^{44}$Ti and $^{56}$Ni, is extended to velocities greater than 4000 km s$^{-1}$. 
In the middle case ($\alpha = 1.5$) between the two cases (see the bottom left panel in Figure~\ref{fig:vel-low-beta}), the peak for elements, 
$^{12}$C, $^{16}$O, and $^{28}$Si, around 2000 km s$^{-1}$ is more prominent than for the case of $\alpha = 2.0$, whereas the second peak for 
elements, $^{44}$Ti and $^{56}$Ni, is shifted to a bit lower velocity regions and is broader than for the case of $\alpha = 2.0$. 
%
\begin{figure*}[htb]
\begin{tabular}{cc}
\begin{minipage}{0.5\hsize}
\begin{center}
\includegraphics[width=7cm,keepaspectratio,clip]{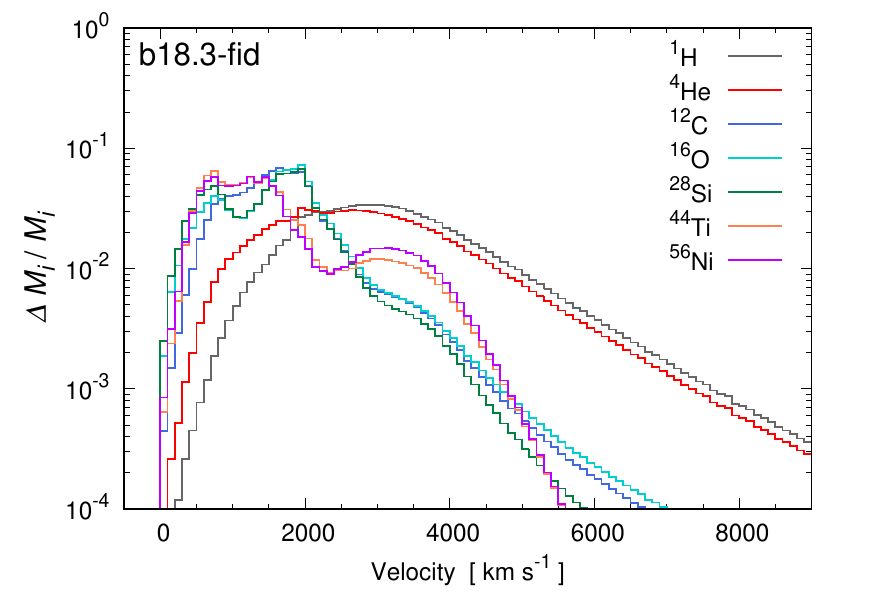}
\end{center}
\end{minipage}
\begin{minipage}{0.5\hsize}
\begin{center}
\includegraphics[width=7cm,keepaspectratio,clip]{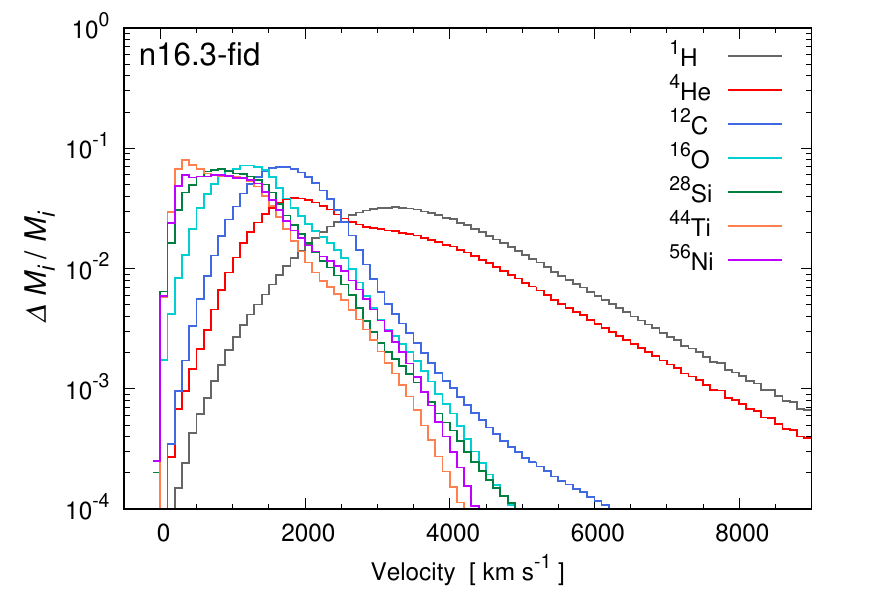}
\end{center}
\end{minipage}
\\
\begin{minipage}{0.5\hsize}
\begin{center}
\includegraphics[width=7cm,keepaspectratio,clip]{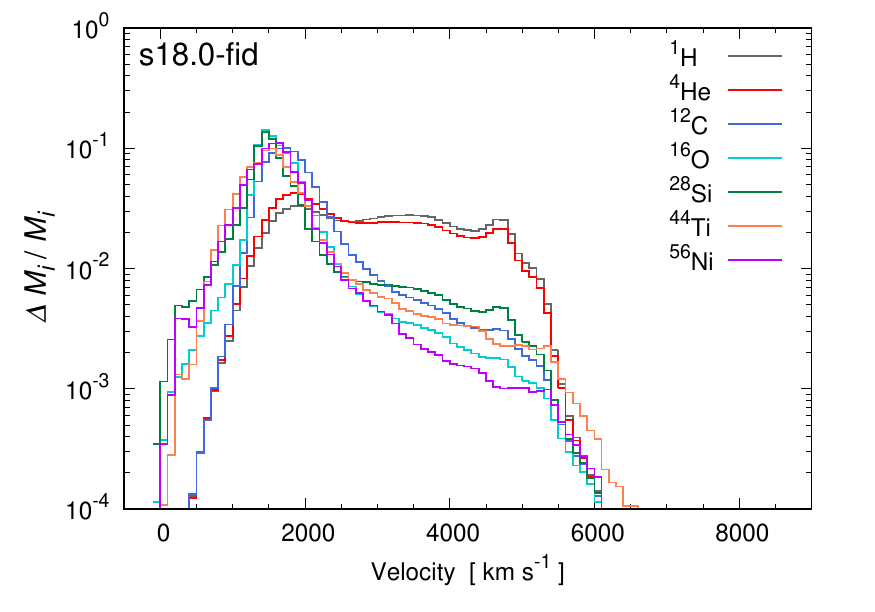}
\end{center}
\end{minipage}
\begin{minipage}{0.5\hsize}
\begin{center}
\includegraphics[width=7cm,keepaspectratio,clip]{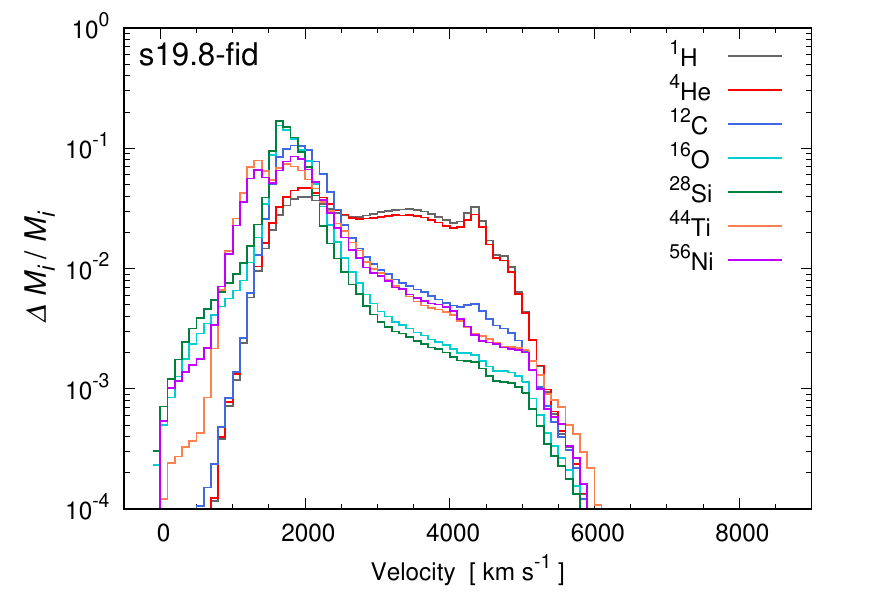}
\end{center}
\end{minipage}
\end{tabular}
\caption{Same as Figure~\ref{fig:vel-low-beta} but for models b18.3-fid (top left; 80,621 sec), n16.3-fid (top right; 85,714 sec), s18.0-fid (bottom left; 279,037 sec), and s19.8-fid (bottom right; 283,161 sec).}
\label{fig:vel-low-model}
\end{figure*}
%
Figure~\ref{fig:vel-low-model} shows the dependence on the progenitor model. 
As seen in the figure, there are significant 
differences between the two BSG models (top two panels) and the two RSG models (bottom two panels). In the two RSG models, the maximum 
velocities of elements are apparently limited to around $\sim$ 5000 km s$^{-1}$, which is in contrast to the BSG models. 
This feature in the RSG models is 
attributed to the structures of the extended ($\gtrsim$ 6 $\times$ 10$^{13}$ cm) hydrogen envelopes. The blast wave is continuously decelerated 
during the propagation in the extended hydrogen envelope in the RSG models. Another feature is that even inner most elements, $^{56}$Ni and 
$^{44}$Ti, finally reach the highest velocity regions ($\sim$ 5000 km s$^{-1}$), although the amounts are not so significant in particular in the model 
s18.0-fid. Among the two BSG models, there is clear difference in the extension of the high velocity tails of $^{56}$Ni and $^{44}$Ti. 
In the model b18.3-fid (top left), a bump is present around $\lesssim$ 4000 km s$^{-1}$ and the tail is more extended than that in the model n16.3-fid. 

So far, only radial velocities of elements are discussed but the observed [Fe II] lines \citep{1990ApJ...360..257H} involved with the line of sight velocity of iron 
(the decay product of $^{56}$Ni) should also be explained from the models. Based on the simulation results, by changing the direction of the axis of the 
bipolar-like explosion ($z$-axis in the simulation box) to observers on the Earth, distributions of the $^{56}$Ni mass in the line of sight velocity are estimated.  
%
\begin{figure}
\begin{center}
\includegraphics[width=7cm,keepaspectratio,clip]{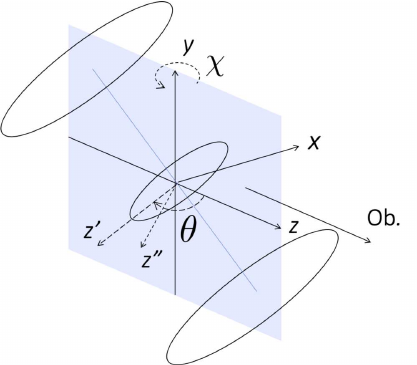}
\end{center}
\caption{Schematic picture for an assumed direction of the bipolar-like explosion axis to observers on the Earth and to the triple ring structure 
\citep[for the configuration of the triple ring structure to observers, see e.g.,][]{2005ApJ...627..888S,2005ApJS..159...60S,2011A&A...527A..35T}. In order to 
see the impact of the direction of the explosion axis on the line of sight velocity of $^{56}$Ni, the simulation box (initially, the explosion axis is directed to 
the $z$-axis) is rotated for the estimation. First, $z$-axis in the simulation box is set to be directed to the observers on Earth. Then, the simulation box is 
rotated around the original $x$-axis by an angle of $\theta$ (the $z$-axis is rotated to be the $z'$-axis). Finally, the box is rotated around the original 
$y$-axis by an angle of $\chi$ (the $z'$-axis is rotated to be the $z''$-axis).}
\label{fig:geo}
\end{figure}
Figure~\ref{fig:geo} shows a schematic picture for an assumed direction of the axis of the bipolar-like explosion to observers on Earth and to the triple ring 
structure. Two rotation angles, $\theta$ and $\chi$ are defined as in Figure~\ref{fig:geo}. For the configuration of the triple ring structure, see 
\citet{2005ApJ...627..888S,2005ApJS..159...60S} and \citet{2011A&A...527A..35T}. The inclination angle of the ER is $\sim$ 43$^{\circ}$ 
\citep{2011A&A...527A..35T}. 
%
\begin{figure*}[htb]
\begin{tabular}{cc}
\begin{minipage}{0.5\hsize}
\begin{center}
\includegraphics[width=7cm,keepaspectratio,clip]{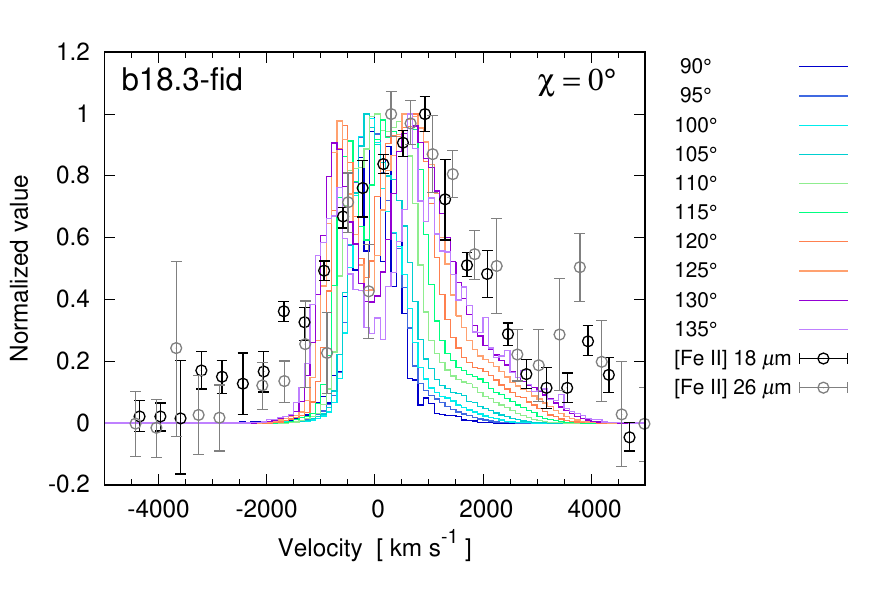}
\end{center}
\end{minipage}
\begin{minipage}{0.5\hsize}
\begin{center}
\includegraphics[width=7cm,keepaspectratio,clip]{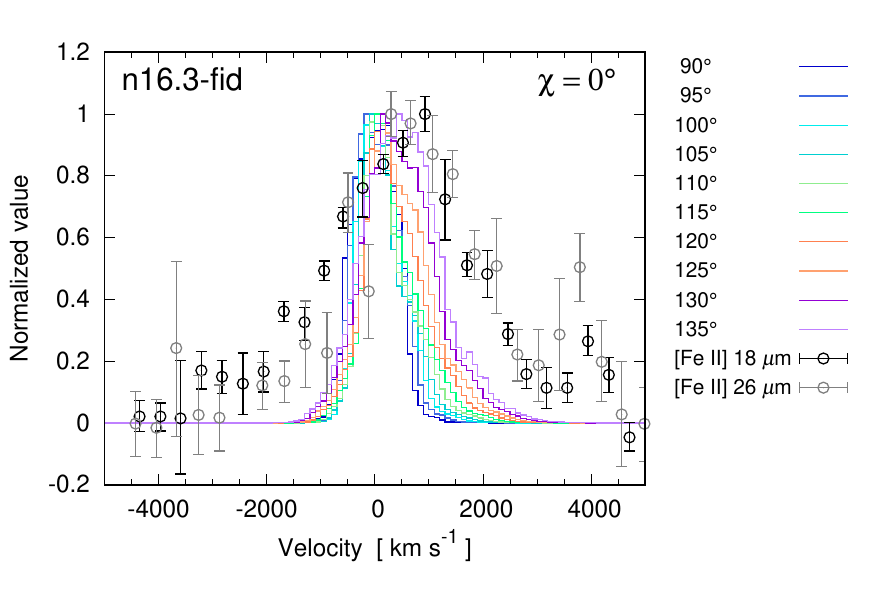}
\end{center}
\end{minipage}
\\
\begin{minipage}{0.5\hsize}
\begin{center}
\includegraphics[width=7cm,keepaspectratio,clip]{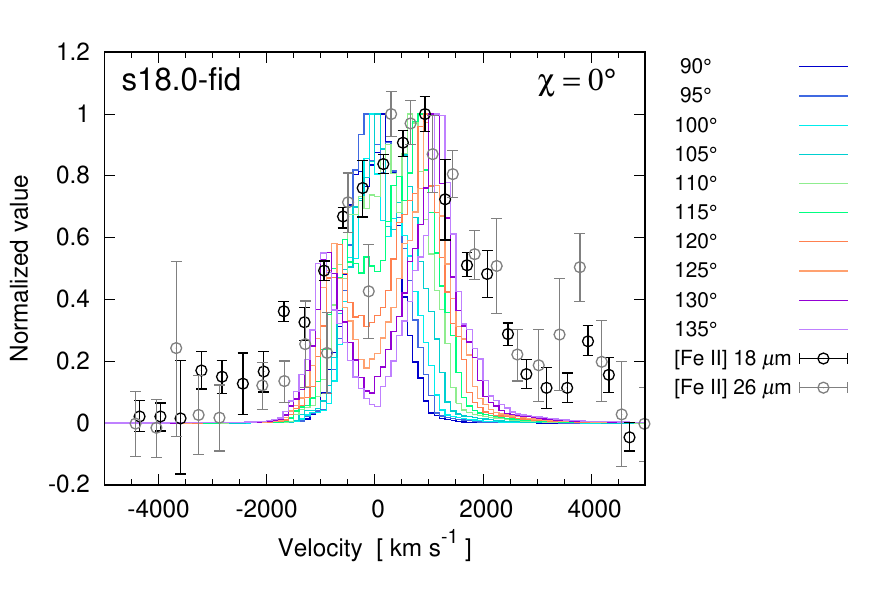}
\end{center}
\end{minipage}
\begin{minipage}{0.5\hsize}
\begin{center}
\includegraphics[width=7cm,keepaspectratio,clip]{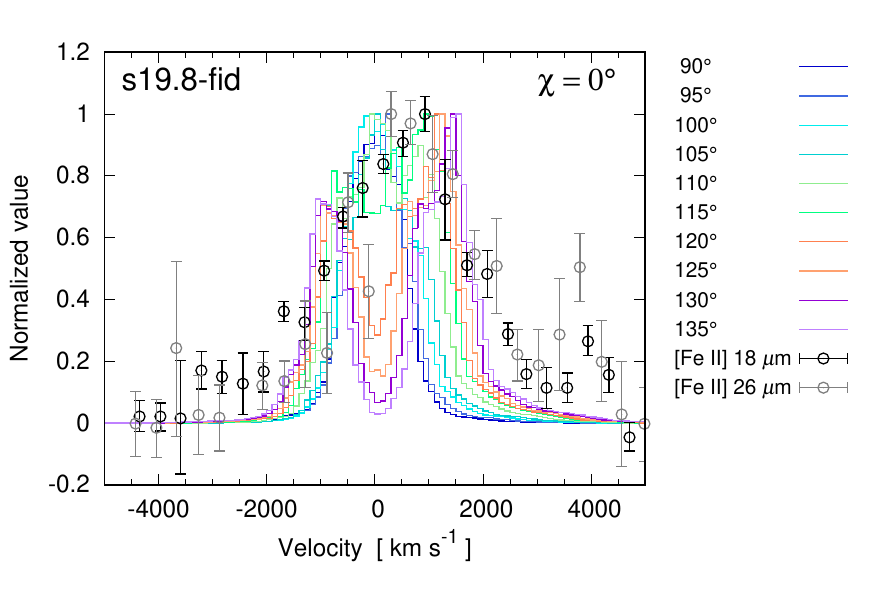}
\end{center}
\end{minipage}
\end{tabular}
\caption{Normalized masses of $^{56}$Ni as a function of the line of sight velocity at the end of the simulation. 
Top left, top right, bottom left, and bottom right panels are for models b18.3-fid (80,621 sec), n16.3-fid (85,714 sec), s18.0-fid (279,037 sec), and s19.8-fid (283,161 sec), respectively. 
The points with error bars (1 $\sigma$) are normalized observed fluxes of the [Fe II] lines, 18 $\mu$m 
and 26 $\mu$m \citep{1990ApJ...360..257H}
, where continuum levels are subtracted. 
Normalizations are carried out in order for the peak value to be unity. Each solid line is the results with 
an angle, $\theta$, defined in Figure~\ref{fig:geo}. Here, the rotation angle $\chi$ is fixed to be 0$^{\circ}$.}
\label{fig:vel-dir-low-model}
\end{figure*}
%
%
Figure~\ref{fig:vel-dir-low-model} shows normalized masses of $^{56}$Ni as a function of the line of sight velocity (solid lines). 
The dependence of the distribution on the angle $\theta$, defined in Figure~\ref{fig:geo} is presented compared with the observed 
[Fe II] lines (points with error bars), 18 $\mu$m and 26 $\mu$m \citep{1990ApJ...360..257H}. Here, the rotation angle $\chi$ is fixed to be 0$^{\circ}$. 
For the rotation angle $\theta$, considering the fact that the bulk of the [Fe II] line is redshifted \citep{1990ApJ...360..257H} and the 3D distribution of the 
inner ejecta seems to be slightly tilted to the ER plane from the observations of [Si I] + [Fe II] lines, $\theta$ is changed in the range between 90$^{\circ}$ 
and 135$^{\circ}$. The dependence on the progenitor models is also presented. 
For all models shown in the figure, the smaller the rotation angle $\theta$ is, the more concentrated around the velocity center the distributions are. Compared with the 
model b18.3, the model n16.3 apparently lacks high velocity component ($\gtrsim$ 2000 km s$^{-1}$). In the model b18.3, the tail ($\lesssim$ 4000 
km s$^{-1}$) and the peak ($\sim$ 1000 km s$^{-1}$) at the redshifted side are best reproduced for the rotation angle $\theta$ of $\gtrsim$ 130$^{\circ}$ 
among the four models, although the distribution of the blueshifted side is insufficient. In models, b18.3-fid, s18.0-fid, and s19.8-fid, in particular the latter 
two RSG models, double-peak structures are seen for the rotation angle $\theta$ of $\gtrsim$ 120$^{\circ}$. The clear double-peak structures in the RSG 
models reflect the bi-cone-like distribution of $^{56}$Ni as seen in the bottom panels in Figure~\ref{fig:ni56-low-model-late}. It is noted that in the points 
of [Fe II] line of 26 $\mu$m, a valley around the velocity center (the bottom of the valley is only one point) is recognized but clear double peaks as seen 
in the RSG models are inconsistent with the overall distributions from the observations. 
%
\begin{figure*}[htb]
\begin{tabular}{cc}
\begin{minipage}{0.5\hsize}
\begin{center}
\includegraphics[width=7cm,keepaspectratio,clip]{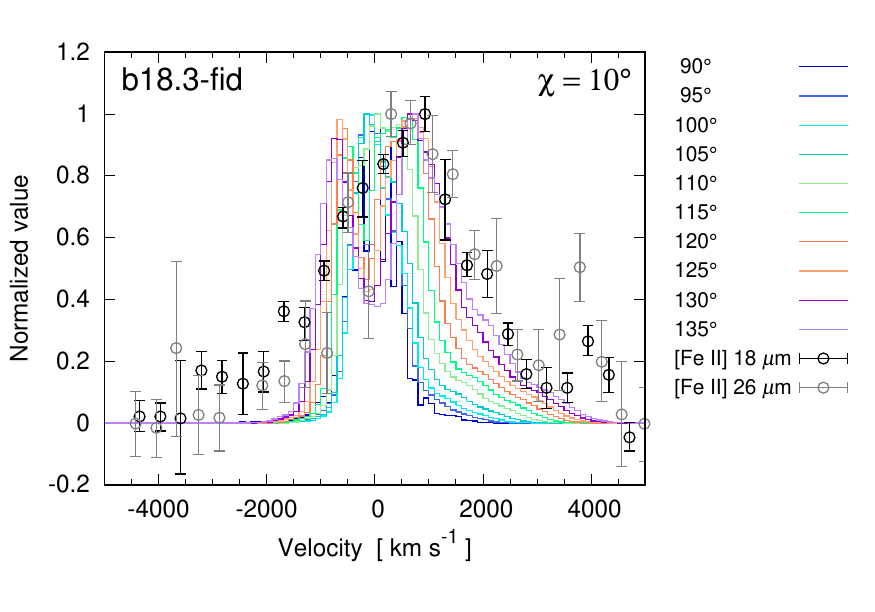}
\end{center}
\end{minipage}
\begin{minipage}{0.5\hsize}
\begin{center}
\includegraphics[width=7cm,keepaspectratio,clip]{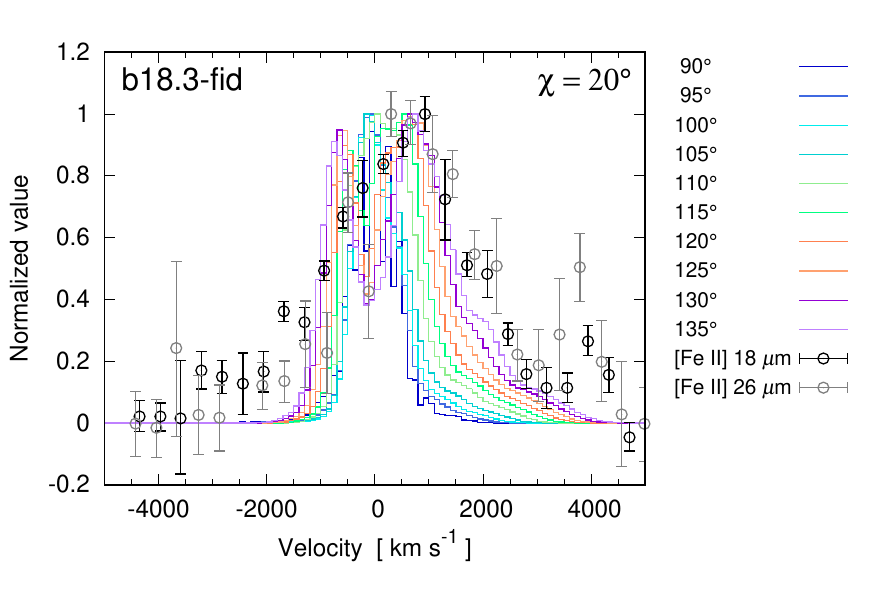}
\end{center}
\end{minipage}
\\
\begin{minipage}{0.5\hsize}
\begin{center}
\includegraphics[width=7cm,keepaspectratio,clip]{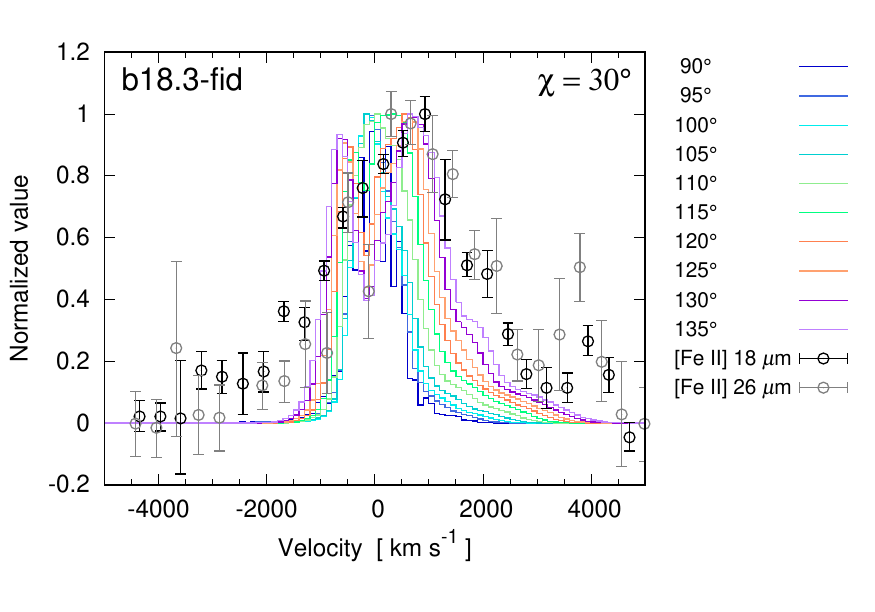}
\end{center}
\end{minipage}
\begin{minipage}{0.5\hsize}
\begin{center}
\includegraphics[width=7cm,keepaspectratio,clip]{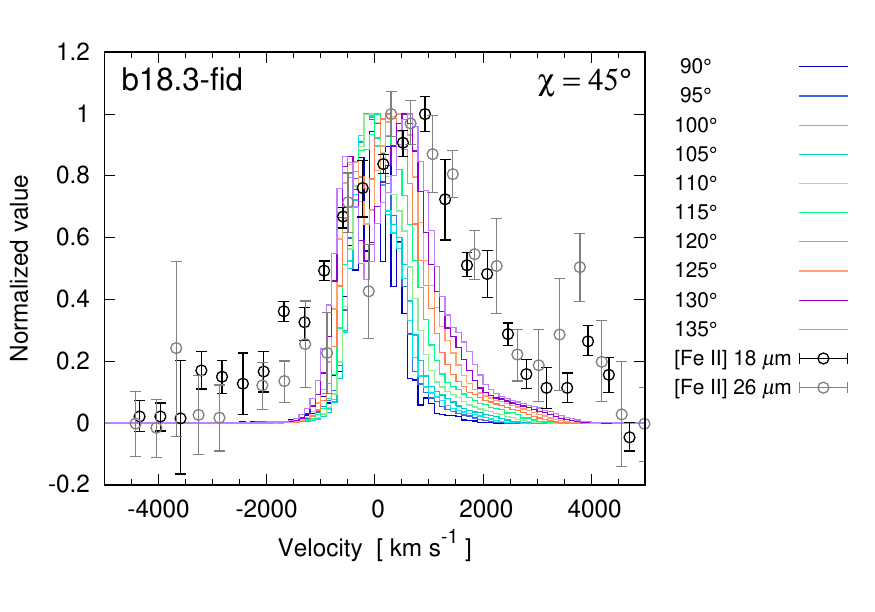}
\end{center}
\end{minipage}
\end{tabular}
\caption{Same as Figure~\ref{fig:vel-dir-low-model} but for the model b18.3-fid (80,621 sec) with the parameter $\chi$ of 
10$^{\circ}$ (top left), 20$^{\circ}$ (top right), 30$^{\circ}$ (bottom left), and 45$^{\circ}$ (bottom right).}
\label{fig:vel-dir-low-chi}
\end{figure*}
%
Figure~\ref{fig:vel-dir-low-chi} shows the dependence on the rotation angle $\chi$ (the model is fixed as b18.3-fid). 
The case of $\chi$ = 0$^{\circ}$ is shown in the top left panel in Figure~\ref{fig:vel-dir-low-model}. As can be 
seen, among the models of the cases of $\chi$ = 0$^{\circ}$, 10$^{\circ}$, and 20$^{\circ}$, there is no distinct differences in the overall 
distributions but the tail at the redshifted side is slightly better explained in the case of $\chi$ = 10$^{\circ}$ than the other two cases. In the cases 
of $\chi$ = 30$^{\circ}$ and 45$^{\circ}$, the tails ($\gtrsim$ 1500 km s$^{-1}$) are apparently reduced compared with the cases of smaller $\chi$. 
%
\begin{figure*}[htb]
\begin{tabular}{cc}
\begin{minipage}{0.5\hsize}
\begin{center}
\includegraphics[width=7cm,keepaspectratio,clip]{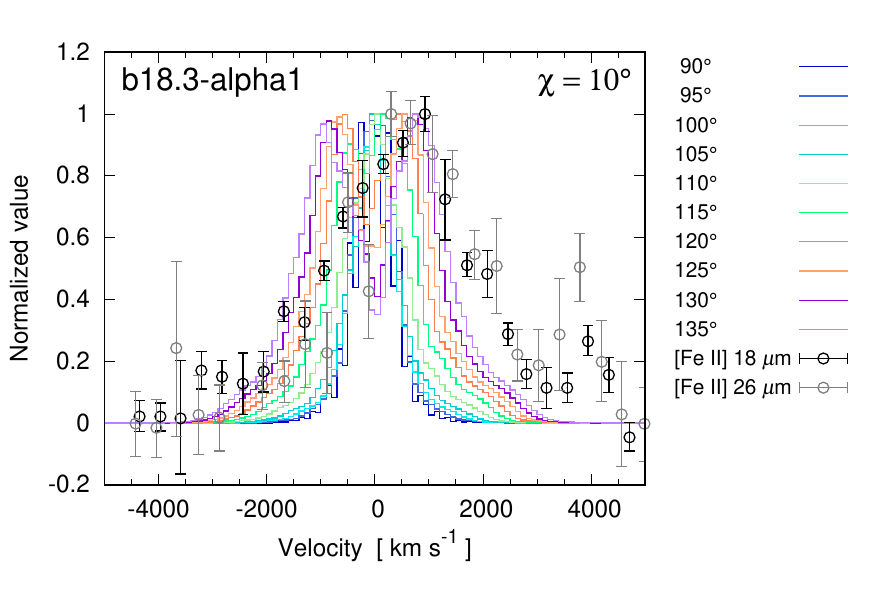}
\end{center}
\end{minipage}
\begin{minipage}{0.5\hsize}
\begin{center}
\includegraphics[width=7cm,keepaspectratio,clip]{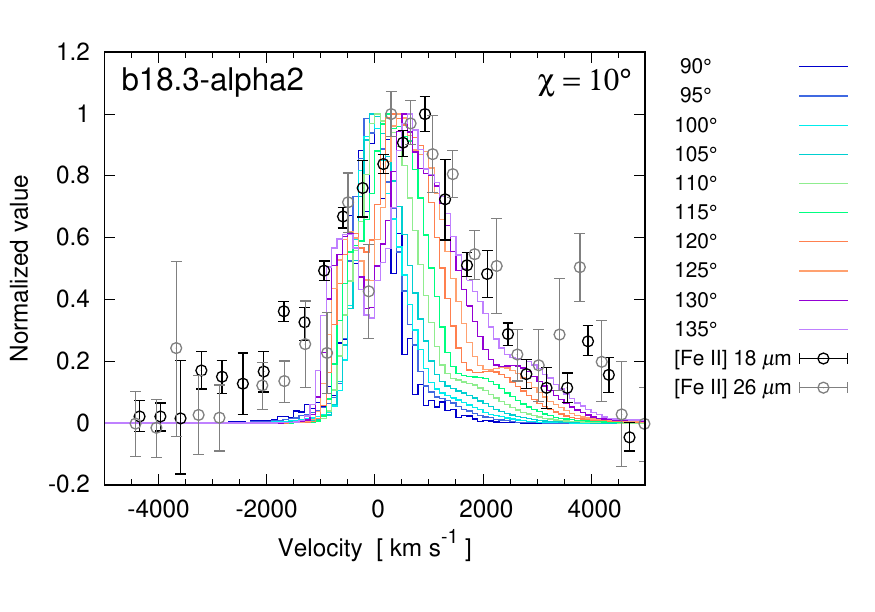}
\end{center}
\end{minipage}
\end{tabular}
\caption{Same as Figure~\ref{fig:vel-dir-low-model} but for models b18.3-alpha1 (left; 77,425 sec) and b18.3-alpha2 (right; 77,261 sec) and for the parameter 
$\chi$ = 10$^{\circ}$.}
\label{fig:vel-dir-low-ud}
\end{figure*}
%
Figure~\ref{fig:vel-dir-low-ud} shows the dependence on the parameter $\alpha \equiv v_{\rm up}/v_{\rm down}$. 
The values of $\alpha$ in the models, b18.3-alpha1 and b18.3-alpha2, are 1.0 and 2.0, respectively. 
The case of $\alpha$ = 1.5 is shown in top left panel in Figure~\ref{fig:vel-dir-low-chi}. For the case of $\alpha$ = 1.0, in which there is no global asymmetry 
in the explosion against the equatorial plane, the distributions are symmetric against the velocity center, as expected, and symmetric double peaks 
are seen for the rotation angle $\theta$ of $\gtrsim$ 120$^{\circ}$. For the case of $\alpha$ = 2.0, compared with the case of $\alpha$ = 1.5, the 
tail at the redshifted side is slightly enhanced and the peak at the blueshifted side is reduced for the rotation angle $\theta$ of $\gtrsim$ 
120$^{\circ}$. In the models b18.3-fid and b18.3-alpha2, the sharp cut-offs at the blueshifted side are seen at the velocities around 
2000 km s$^{-1}$ and 1500 km s$^{-1}$, respectively. Then, the observed tails at the blueshifted side are more difficult to reproduce in the model 
b18.3-alpha2 than the model b18.3-fid. 

Based on the results of lower resolution simulations and arguments on the constraints from the observations of SN 1987A on the mass of high velocity 
$^{56}$Ni and the mass distributions of the line of sight velocity of $^{56}$Ni, favored values of parameters related to the asymmetric explosions and 
progenitor models are presented by comparing representative models. As mentioned in Section~\ref{sec:intro}, the progenitor of SN 1987A, 
Sk $-$69$^{\circ}$ 202, was a compact BSG at the time of the explosion. Then, models with one of the BSG pre-supernova models b18.3 and n16.3 
are appropriate. From the constraints on the mass of high velocity $^{56}$Ni, i.e., 
i) the ratio of the mass of $^{56}$Ni that has $\geq$ 3000 km s$^{-1}$ to the total $^{56}$Ni mass is greater than 4\%; 
ii) the mass of $^{56}$Ni that has $\geq$ 4000 km s$^{-1}$ is greater than 10$^{-3}$ $M_{\odot}$, 
models with large $\beta \equiv v_{\rm pol}/v_{\rm eq}$ values (8 or 16), b18.3-beta8, b18.3-fid, b18.3-alpha2, and b18.3-clp0 are selected as candidates 
for SN 1987A. As mentioned before, there is no explosion model with the BSG model n16.3 (single star evolution) that satisfies the two conditions on the mass of 
high velocity $^{56}$Ni simultaneously. Among the models b18.3-beta8, b18.3-fid, and b18.3-clp0, the mass of high velocity $^{56}$Ni in the model b18.3-beta8 is a bit 
smaller than that in the other two models (see Table~\ref{table:results}). The difference in the initial setup between the models b18.3-fid and b18.3-clp0 
is the existence of the fluctuations in the initial radial velocities. 
It is a bit arbitrary but motivated by the recent observations of the CO and SiO molecules, and dust in 
the inner ejecta of SN 1987A \citep{2017ApJ...842L..24A,2019arXiv191002960C}, which have 
indicated that the inner ejecta is 
clumpy 
\citep[the first observational evidence of clumpiness of the ejecta of SN 1987A was found from narrow features in emission lines e.g. {$\left[ \right.$}O I{$\left. \right]$}:][]{1991supe.conf...95S}
, we thus prefer the 
model with the initial fluctuations, i.e., b18.3-fid. Finally, the model b18.3-alpha2 is also a candidate as well as the model b18.3-fid. 
As discussed before, considering the deficiency in the tail at the blueshifted side in the mass distribution of $^{56}$Ni as a function of the line of sight velocity, we 
select the model with a moderate value for the parameter $\alpha \equiv v_{\rm up}/v_{\rm down}$ = 1.5, i.e., b18.3-fid, as a fiducial model in the 
models of lower resolution simulations. 

\begin{deluxetable*}{lccccccccr}
\tabletypesize{\footnotesize}
\tablewidth{16cm}
\tablenum{3}
\label{table:results}
\tablecolumns{10}
\tablecaption{Results of 3D simulation models.}
\tablehead
{
\multicolumn{1}{l}{\fontsize{9pt}{9pt}\selectfont Model} & 
\multicolumn{1}{c}{\fontsize{9pt}{9pt}\selectfont $E_{\rm exp}$\tablenotemark{a}} &
\multicolumn{1}{c}{\fontsize{9pt}{9pt}\selectfont $M_{\rm ej}$ ($^{56}$Ni)\tablenotemark{b}} &
\multicolumn{1}{c}{\fontsize{9pt}{9pt}\selectfont $M_{\rm ej}$ ($^{44}$Ti)\tablenotemark{c}} &
\multicolumn{1}{c}{\fontsize{9pt}{9pt}\selectfont $M_{3.0}$ ($^{56}$Ni)\tablenotemark{d}} &
\multicolumn{1}{c}{\fontsize{9pt}{9pt}\selectfont $M_{4.0}$ ($^{56}$Ni)\tablenotemark{e}} &
\multicolumn{1}{c}{\fontsize{9pt}{9pt}\selectfont $M_{4.7}$ ($^{56}$Ni)\tablenotemark{f}} &
\multicolumn{1}{c}{\fontsize{9pt}{9pt}\selectfont $M_{3.0}$/$M_{\rm ej}$ ($^{56}$Ni)\tablenotemark{g}} &
\multicolumn{1}{c}{\fontsize{9pt}{9pt}\selectfont $v_{\rm NS}$\tablenotemark{h}} &
\multicolumn{1}{l}{\fontsize{9pt}{9pt}\selectfont No.\tablenotemark{i}}
\\
&
\multicolumn{1}{c}{\fontsize{8pt}{8pt}\selectfont (erg)} & 
\multicolumn{1}{c}{\fontsize{8pt}{8pt}\selectfont ($M_{\odot}$)} & 
\multicolumn{1}{c}{\fontsize{8pt}{8pt}\selectfont ($M_{\odot}$)} & 
\multicolumn{1}{c}{\fontsize{8pt}{8pt}\selectfont ($M_{\odot}$)} &
\multicolumn{1}{c}{\fontsize{8pt}{8pt}\selectfont ($M_{\odot}$)} & 
\multicolumn{1}{c}{\fontsize{8pt}{8pt}\selectfont ($M_{\odot}$)} & 
\multicolumn{1}{c}{\fontsize{8pt}{8pt}\selectfont ( -- )} &
\multicolumn{1}{c}{\fontsize{8pt}{8pt}\selectfont (km s$^{-1}$)} &
}
\startdata
b18.3-mo13 & \,\,1.95 (51)\tablenotemark{j} & \,\,9.83 (-2)\tablenotemark{k} & \,\,7.79 (-4)\tablenotemark{k} & 9.02 (-4) & 1.02 (-5) & 1.69 (-7) & \,\,9.17 (-3)\tablenotemark{k} & 5.05 (2) & 1 \\
b18.3-beta2 & 1.95 (51) & 1.06 (-1) & 7.76 (-4) & 1.51 (-4) & 9.75 (-7) & 8.86 (-9) & 1.43 (-3) & 3.35 (2) & 2 \\
b18.3-beta4 & 1.95 (51) & 8.93 (-2) & 6.81 (-4) & 4.26 (-3) & 1.95 (-4) & 8.27 (-6) & 4.77 (-2) & 2.98 (2) & 3 \\
b18.3-beta8 & 1.97 (51) & 8.39 (-2) & 7.12 (-4) & 8.12 (-3) & 1.43 (-3) & 3.59 (-4) & 9.68 (-2) & 2.83 (2) & 4 \\
b18.3-fid & 1.99 (51) & 8.10 (-2) & 7.39 (-4) & 1.12 (-2) & 1.89 (-3) & 2.81 (-4) & 1.38 (-1) & 2.75 (2) & 5 \\
\hline
%
n16.3-mo13 & 1.88 (51) & 1.13 (-1) & 8.11 (-4) & 9.08 (-5) & 3.33 (-7) & 5.00 (-9) & 8.05 (-4) & 5.56 (2) & 6 \\
n16.3-beta2 & 1.90 (51) & 1.20 (-1) & 8.23 (-4) & 9.53 (-6) & 3.69 (-9) & 4.63 (-12) & 7.94 (-5) & 3.74 (2) & 7 \\
n16.3-beta4 & 1.89 (51) & 1.01 (-1) & 6.04 (-4) & 1.35 (-3) & 5.55 (-6) & 4.32 (-8) & 1.33 (-2) & 3.43 (2) & 8 \\
n16.3-beta8 & 1.90 (51) & 9.53 (-2) & 6.58 (-4) & 4.87 (-3) & 8.07 (-4) & 9.03 (-5) & 5.11 (-2) & 3.34 (2) & 9 \\
n16.3-fid & 1.91 (51) & 9.04 (-2) & 7.17 (-4) & 1.27 (-3) & 6.08 (-5) & 1.78 (-6) & 1.40 (-2) & 3.20 (2) & 10 \\
\hline
%
s18.0-mo13 & 1.85 (51) & 1.36 (-1) & 1.15 (-3) & 6.74 (-4) & 7.12 (-5) & 2.24 (-5) & 4.95 (-3) & 5.67 (2) & 11\\
s18.0-beta2 & 1.87 (51) & 1.33 (-1) & 1.03 (-3) & 2.60 (-4) & 1.30 (-5) & 1.62 (-6) & 1.95 (-3) & 3.99 (2) & 12 \\
s18.0-beta4 & 1.85 (51) & 1.27 (-1) & 1.11 (-3) & 9.19 (-4) & 2.01 (-4) & 6.76 (-5) & 7.21 (-3) & 3.28 (2) & 13 \\
s18.0-beta8 & 1.86 (51) & 1.09 (-1) & 1.04 (-3) & 1.61 (-3) & 4.89 (-4) & 1.98 (-4) & 1.47 (-2) & 2.86 (2) & 14\\
s18.0-fid & 1.87 (51) & 9.63 (-2) & 1.22 (-3) & 4.13 (-3) & 1.74 (-3) & 8.65 (-4) & 4.29 (-2) & 2.67 (2) & 15 \\
\hline
%
s19.8-mo13 & 1.87 (51) & 8.18 (-2) & 8.36 (-4) & 3.32 (-4) & 1.78 (-5) & 4.96 (-6) & 4.06 (-3) & 5.81 (2) & 16 \\
s19.8-beta2 & 1.89 (51) & 8.96 (-2) & 8.47 (-4) & 1.10 (-4) & 4.18 (-6) & 3.20 (-7) & 1.23 (-3) & 3.89 (2) & 17 \\
s19.8-beta4 & 1.88 (51) & 8.83 (-2) & 7.61 (-4) & 4.29 (-3) & 1.36 (-3) & 4.90 (-4) & 4.86 (-2) & 3.51 (2) & 18 \\
s19.8-beta8 & 1.88 (51) & 8.36 (-2) & 8.25 (-4) & 6.95 (-3) & 2.73 (-3) & 1.13 (-3) & 8.31 (-2) & 3.31 (2) & 19 \\
s19.8-fid & 1.89 (51) & 7.80 (-2) & 8.93 (-4) & 6.93 (-3) & 2.40 (-3) & 8.87 (-4) & 8.88 (-2) & 2.99 (2) & 20 \\
\hline
%
b18.3-alpha1 & 1.98 (51) & 8.09 (-2) & 6.61 (-4) & 3.17 (-3) & 2.79 (-4) & 4.03 (-5) & 3.92 (-2) & 2.13 (0) & 21 \\
b18.3-alpha2 & 1.99 (51) & 7.45 (-2) & 7.70 (-4) & 1.47 (-2) & 4.55 (-3) & 1.22 (-3) & 1.98 (-1) & 4.23 (2) & 22 \\
b18.3-cos & 2.00 (51) & 8.53 (-2) & 8.61 (-4) & 2.59 (-3) & 6.00 (-5) & 8.30 (-7) & 3.03 (-2) & 2.59 (2) & 23 \\
b18.3-exp & 2.00 (51) & 7.57 (-2) & 7.71 (-4) & 4.94 (-3) & 3.32 (-4) & 7.53 (-6) & 6.53 (-2) & 2.62 (2) & 24 \\
b18.3-pwr & 2.00 (51) & 7.91 (-2) & 7.75 (-4) & 4.24 (-3) & 2.85 (-4) & 8.08 (-6) & 5.35 (-2) & 2.62 (2) & 25 \\
b18.3-clp0 & 1.99 (51) & 8.07 (-2) & 7.39 (-4) & 1.32 (-2) & 1.89 (-3) & 4.36 (-4) & 1.64 (-1) & 2.78 (2) & 26 \\
b18.3-ein1.5 & 9.87 (50) & 4.15 (-2) & 3.53 (-4) & 3.83 (-5) & 2.51 (-8) & 3.53 (-11) & 9.25 (-4) & 1.59 (2) & 27 \\
b18.3-ein2.0 & 1.49 (51) & 6.45 (-2) & 5.89 (-4) & 2.82 (-3) & 1.85 (-4) & 1.94 (-5) & 4.38 (-2) & 2.42 (2) & 28 \\
b18.3-ein3.0 & 2.49 (51) & 9.40 (-2) & 8.63 (-4) & 1.61 (-2) & 6.99 (-3) & 2.40 (-3) & 1.71 (-1) & 3.05 (2) & 29 \\
\hline
b18.3-high & 2.01 (51) & 8.64 (-2) & 5.73 (-4) & 9.06 (-3) & 1.11 (-3) & 5.63 (-5) & 1.05 (-1) & 2.85 (2) & 30 \\
n16.3-high & 1.93 (51) & 9.67 (-2) & 5.38 (-4) & 3.59 (-3) & 2.50 (-4) & 5.66 (-7) & 3.71 (-2) & 3.03 (2) & 31
\enddata
{\scriptsize
\tablenotetext{a}{Explosion energy estimated by Eq.~(\ref{eq:exp})
at the end of the simulation.}
\tablenotetext{b}{Mass of total ejected $^{56}$Ni which has positive radial velocity at the end of the simulation.}
\tablenotetext{c}{Mass of total ejected $^{44}$Ti which has positive radial velocity at the end of the simulation.}
\tablenotetext{d}{Mass of $^{56}$Ni which has velocity higher than 3000 km s$^{-1}$ at the end of the simulation.}
\tablenotetext{e}{Mass of $^{56}$Ni which has velocity higher than 4000 km s$^{-1}$ at the end of the simulation.}
\tablenotetext{f}{Mass of $^{56}$Ni which has velocity higher than 4700 km s$^{-1}$ at the end of the simulation.}
\tablenotetext{g}{Ratio of the values in 5th to 3rd columns.}
\tablenotetext{h}{$v_{\rm NS}$ is the NS kick velocity estimated by Eq.~(\ref{eq:kick}).}
\tablenotetext{i}{Sequential serial number (model number) for Figure~\ref{fig:table3}. The values of 5th--8th columns are plotted in Figure~\ref{fig:table3}.}
%
\tablenotetext{j}{Number in parenthesis denotes the power of ten.}
\tablenotetext{k}{The values in the 3rd and 4th columns (the values in the 8th column) could be overestimated (underestimated). See Section~\ref{subsec:res-3d-low} for the details.}
}
\end{deluxetable*}

\subsection{Results of Three-dimensional Simulations: High Resolution Cases} \label{subsec:res-3d-high}

%
Based on the arguments on the exploration of lower resolution simulations in Section~\ref{subsec:res-3d-low}, two high resolution simulations with the two BSG 
progenitor models b18.3 and n16.3 are performed, where the high resolution models are denoted as b18.3-high and n16.3-high and the corresponding lower 
resolution models are b18.3-fid and n16.3-fid, respectively. The parameters for the aspherical explosion are fixed to be same as the model b18.3-fid 
(see Table~\ref{table:models} for the values of the parameters). First, the results listed in Table~\ref{table:results} are discussed by comparing with those of 
the corresponding lower resolution models. Obtained explosion energies, $E_{\rm exp}$, of higher (lower) resolution models b18.3-high 
(b18.3-fid) and n16.3-high (n16.3-fid) are 2.01 $\times$ 10$^{51}$ erg (1.99 $\times$ 10$^{51}$ erg) and 1.93 $\times$ 10$^{51}$ erg 
(1.91 $\times$ 10$^{51}$ erg), respectively. The values in the high resolution models are slightly higher compared with those of lower resolution models 
but the values are consistent enough with those of lower resolution models. The ejected masses of $^{56}$Ni in the models b18.3-high (b18.3-fid) and n16.3-high 
(n16.3-fid) are 8.64 $\times$ 10$^{-2}$ $M_{\odot}$ (8.10 $\times$ 10$^{-2}$ $M_{\odot}$) and 9.67 $\times$ 10$^{-2}$ $M_{\odot}$ (9.04 $\times$ 10$^{-2}$ 
$M_{\odot}$), respectively. The ejected masses of $^{44}$Ti in the models b18.3-high (b18.3-fid) and n16.3-high (n16.3-fid) are 5.73 $\times$ 10$^{-4}$ 
$M_{\odot}$ (7.39 $\times$ 10$^{-4}$ $M_{\odot}$) and 5.38 $\times$ 10$^{-4}$ $M_{\odot}$ (7.17 $\times$ 10$^{-4}$ $M_{\odot}$), respectively. Therefore, the 
masses of $^{56}$Ni in the lower resolution models are underestimated by $\sim$ 5\% compared with those of high resolution models. On the other hand, the 
masses of $^{44}$Ti in the lower resolution models are overestimated by 20--30\% compared with those of high resolution models. 
As mentioned in Section~\ref{subsec:res-3d-low}, the nuclear reaction network in the simulations in this paper includes only 19 nuclei and the mass fractions of 
$^{56}$Ni and $^{44}$Ti could be overestimated by factor of $\sim$ 1.5 and 3, respectively, compared with those calculated with larger nuclear reaction 
network. If we correct for the overestimation, the masses of $^{56}$Ni in the models b18.3-high and n16.3-high could be $\lesssim$ 0.06 $M_{\odot}$, which is 
roughly consistent with the value suggested by the observations, 0.07 $M_{\odot}$ \citep[e.g.,][]{1990ApJ...360..242S}. The masses of $^{44}$Ti in the models 
b18.3-high and n16.3-high could be $\lesssim$ 2 $\times$ 10$^{-4}$ $M_{\odot}$, which is also consistent with the values deduced from the observations, 
(3.1 $\pm$ 0.8) $\times$ 10$^{-4}$ $M_{\odot}$ and (1.5 $\pm$ 0.3) $\times$ 10$^{-4}$ $M_{\odot}$ 
\citep[][respectively]{2012Natur.490..373G,2015Sci...348..670B}. 

The values related to the observational constraints on the mass of high velocity $^{56}$Ni (see Table~\ref{table:results} and Figure~\ref{fig:table3}), 
i.e., $M_{4.0}$ ($^{56}$Ni) and $M_{3.0}$/$M_{\rm ej}$ ($^{56}$Ni), in the model b18.3-high (b18.3-fid) are 1.11 $\times$ 10$^{-3}$ $M_{\odot}$ 
(1.89 $\times$ 10$^{-3}$ $M_{\odot}$) and 1.05 $\times$ 10$^{-1}$ (1.38 $\times$ 10$^{-1}$), respectively. The values of $M_{4.0}$ ($^{56}$Ni) 
and $M_{3.0}$/$M_{\rm ej}$ ($^{56}$Ni) in the model n16.3-high (n16.3-fid) are 2.50 $\times$ 10$^{-4}$ $M_{\odot}$ (6.08 $\times$ 10$^{-5}$ $M_{\odot}$) 
and 3.71 $\times$ 10$^{-2}$ (1.40 $\times$ 10$^{-2}$), respectively. 
Then, the values 
in the model b18.3-fid tend to be overestimated compared with those of the model b18.3-high. On the other hand, the values in the model n16.3-fid 
tend to be underestimated compared with those of the model n16.3-high. The opposite responses to the increase in resolution of the simulations between the two 
progenitor models may be attributed to the difference of the significance of RT instabilities. Since the progenitor model n16.3 has larger C+O and helium cores 
with steep gradients in the $\rho \,r^3$ profile, RT instability may play a more significant role in order to convey the innermost $^{56}$Ni into outer high velocity 
layers than that in the model b18.3. In the model n16.3-high, by capturing smaller scale perturbations, the growth of RT instabilities could be faster than in the 
model n16.3-fid. Hence, the mass of high velocity $^{56}$Ni in the model n16.3-high could be large due to the efficient growth of the instabilities compared 
with that in the model b16.3-fid. While the role of RT instabilities is less important in the model b18.3, the situation could be opposite to the model n16.3. 
Although the obtained masses of the high velocity $^{56}$Ni are slightly different from those of lower resolution models, it is not changed between the high 
resolution and lower resolution models whether the model satisfies the observational constraints or not. The model b18.3-high satisfies the two conditions, 
i.e., 
i) the ratio of the mass of $^{56}$Ni that has $\geq$ 3000 km s$^{-1}$ to the total $^{56}$Ni mass is greater than 4\%; 
ii) the mass of $^{56}$Ni that has $\geq$ 4000 km s$^{-1}$ is greater than 10$^{-3}$ $M_{\odot}$. 
On the other hand, n16.3-high fails to satisfy the two conditions.

\begin{figure*}[htb]
\begin{tabular}{cc}
\begin{minipage}{0.5\hsize}
\begin{center}
\includegraphics[width=7cm,keepaspectratio,clip]{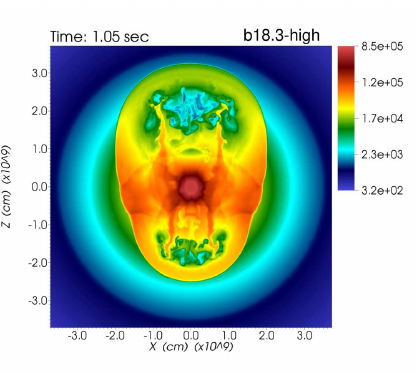}
\end{center}
\end{minipage}
\begin{minipage}{0.5\hsize}
\begin{center}
\includegraphics[width=7cm,keepaspectratio,clip]{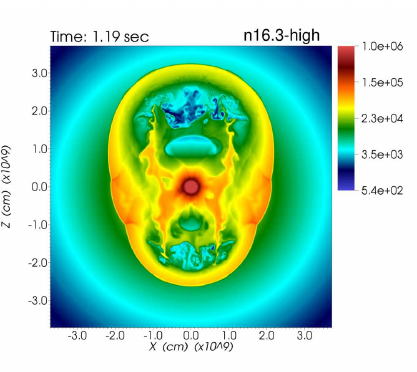}
\end{center}
\end{minipage}
\vspace{-0.3cm}
\\
\begin{minipage}{0.5\hsize}
\begin{center}
\includegraphics[width=7cm,keepaspectratio,clip]{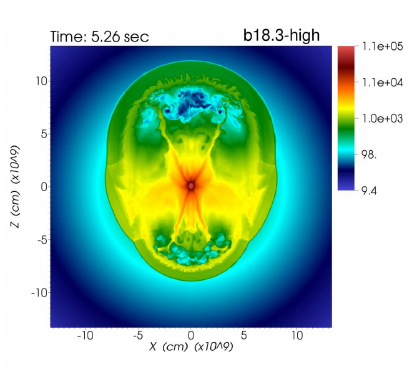}
\end{center}
\end{minipage}
\begin{minipage}{0.5\hsize}
\begin{center}
\includegraphics[width=7cm,keepaspectratio,clip]{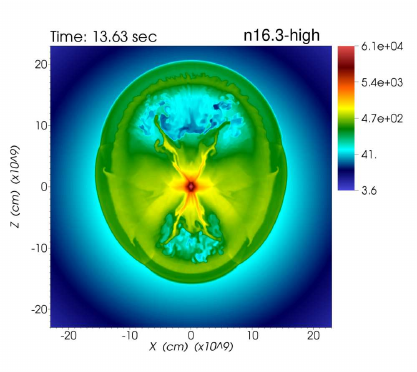}
\end{center}
\end{minipage}
\vspace{-0.3cm}
\\
\begin{minipage}{0.5\hsize}
\begin{center}
\includegraphics[width=7cm,keepaspectratio,clip]{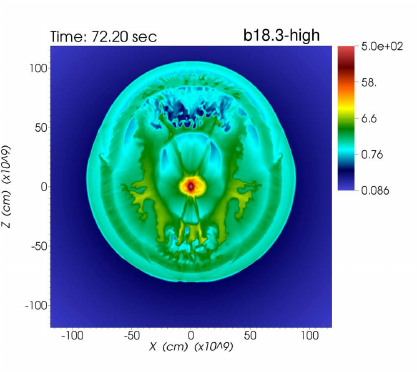}
\end{center}
\end{minipage}
\begin{minipage}{0.5\hsize}
\begin{center}
\includegraphics[width=7cm,keepaspectratio,clip]{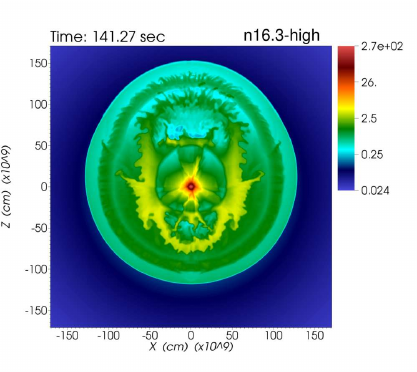}
\end{center}
\end{minipage}
\end{tabular}
\caption{Density color maps (2D slices in the $x$-$z$ plane) in a logarithmic scale. 
The unit of the values in the color bars is g cm$^{-3}$. Left (right) panels are for the model b18.3-high (n16.3-high). 
In the top, the middle, and the bottom panels, maps at the times when the blast wave is inside the C+O layer, the helium layer, and the 
hydrogen layer are shown, respectively. 
%
An animation (density color maps over time) for this figure and Figure~\ref{fig:dens-high-time2} is available. 
In the animation embedded in this figure, snapshots only for the model b18.3-high (left panels) are shown. 
The video starts at $t=0$ s and ends at $t=68,357.48$ s. The realtime duration of the video is 16 seconds.}
\label{fig:dens-high-time1}
\end{figure*}
%
\begin{figure*}[htb]
\begin{tabular}{cc}
\begin{minipage}{0.5\hsize}
\begin{center}
\includegraphics[width=7cm,keepaspectratio,clip]{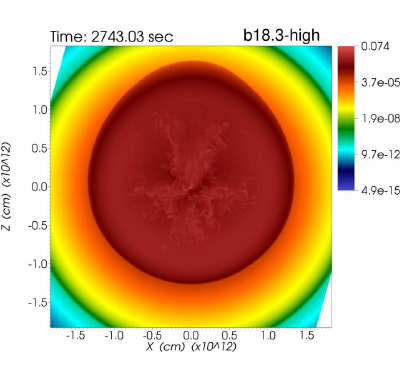}
\end{center}
\end{minipage}
\begin{minipage}{0.5\hsize}
\begin{center}
\includegraphics[width=7cm,keepaspectratio,clip]{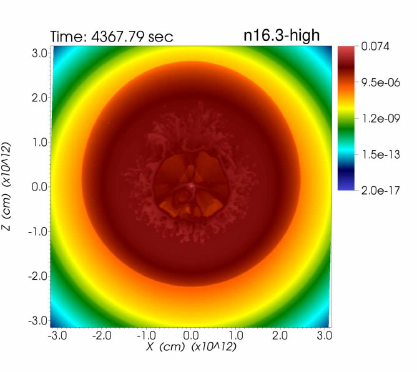}
\end{center}
\end{minipage}
\vspace{-0.3cm}
\\
\begin{minipage}{0.5\hsize}
\begin{center}
\includegraphics[width=7cm,keepaspectratio,clip]{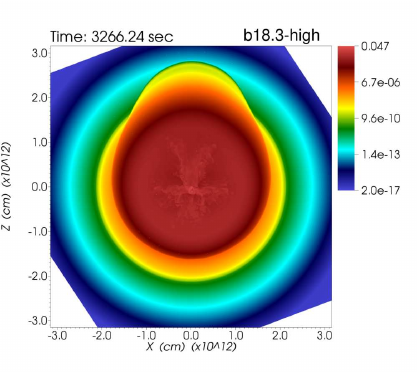}
\end{center}
\end{minipage}
\begin{minipage}{0.5\hsize}
\begin{center}
\includegraphics[width=7cm,keepaspectratio,clip]{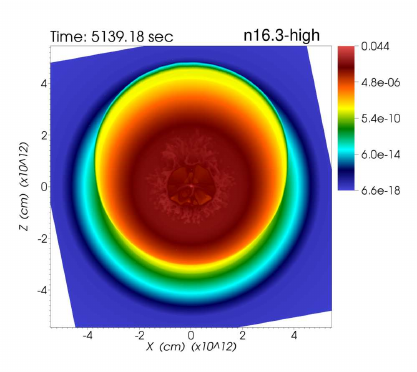}
\end{center}
\end{minipage}
\vspace{-0.3cm}
\\
\begin{minipage}{0.5\hsize}
\begin{center}
\includegraphics[width=7cm,keepaspectratio,clip]{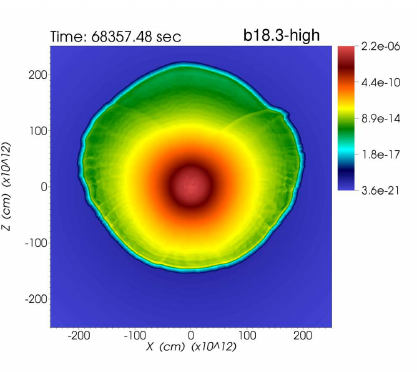}
\end{center}
\end{minipage}
\begin{minipage}{0.5\hsize}
\begin{center}
\includegraphics[width=7cm,keepaspectratio,clip]{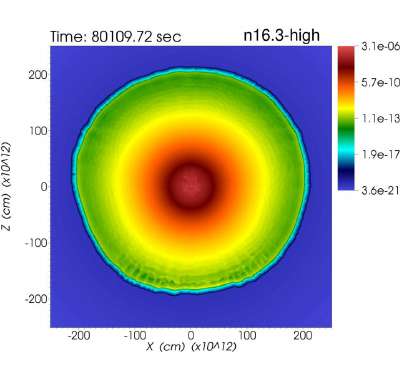}
\end{center}
\end{minipage}
\end{tabular}
\caption{Density color maps (2D slices in the $x$-$z$ plane) in a logarithmic scale. 
The unit of the values in the color bars is g cm$^{-3}$. 
Left (right) panels are for the model b18.3-high (n16.3-high). In the top, the middle, and the bottom panels, 
maps just before the shock breakout, just after the shock breakout, and at the end of the simulation 
are shown, respectively. 
An animation (density color maps over time) for this figure and Figure~\ref{fig:dens-high-time2} is available. 
In the animation embedded in this figure, snapshots only for the model n16.3-high (right panels) are shown. 
The video starts at $t=0$ s and ends at $t=80,109.72$ s. The realtime duration of the video is 16 seconds.}
\label{fig:dens-high-time2}
\end{figure*}
%
Figures~\ref{fig:dens-high-time1} and~\ref{fig:dens-high-time2} show the time evolution of the density distributions for models 
b18.3-high (left) and n16.3-high (right). 
The distributions at the early phase ($\sim$ 1 sec: top panels in Figure~\ref{fig:dens-high-time1}) can be compared with the corresponding density distributions 
of lower resolution simulations in the top panels in Figure~\ref{fig:dens-low-model}. At that time, the blast wave ($r$ $\sim$ 3 $\times$ 10$^{9}$ cm) is inside 
the C+O layer. The shape of the blast wave is almost the same as in the corresponding lower resolution simulation. As seen in Figures~\ref{fig:dens-low-model} 
and~\ref{fig:dens-high-time1}, in the high resolution models, smaller scale structures due to instabilities are recognized than those in the 
lower resolution models. As is the case with the lower resolution simulations, the bipolar structure in the model n16.3-high is wider than that of b18.3-high 
due to stronger deceleration inside the C+O layer with steeper gradient in the $\rho \,r^3$ profile (see Figures~\ref{fig:prog1} and~\ref{fig:prog2}) 
than in the case of b18.3-high. Inside the bipolar structure, fingers due to RT and/or Kelvin-Helmholtz instabilities develop along the bipolar axis. 
After the blast wave goes through the C+O/He interface, fingers of RT instability start to grow on top of the reverse shock developed by the deceleration during 
the shock propagation inside the helium layer (see middle panels in Figure~\ref{fig:dens-high-time1}). The radii of the composition interface of 
the C+O/He of the progenitor models b18.3 and n16.3 are 3.5 $\times$ 10$^{9}$ and 6.1 $\times$ 10$^{9}$ cm, respectively. 
Part of the tips (terminal ends) of the fingers developed at an early phase (fingers seen in the top panels) seems to touch the reverse shock. 
After the blast wave passes the He/H interface, another reverse shock develops outside the previous one caused by the strong deceleration during the shock 
propagation inside the hydrogen layer (see bottom panels in Figure~\ref{fig:dens-high-time1}). On top of the newly developed (outer) reverse shock, 
fingers of RT instability start to grow. Then, a nested double shell structure with fingers develops. The radii of the He/H composition interface in 
the models b18.3 and n16.3 are 2.9 $\times$ 10$^{10}$ and 5.2 $\times$ 10$^{10}$ cm, respectively. It is noted that RT fingers start to grow from the 
inner shell not only along the polar direction but also along near the equatorial plane, where denser material (yellow color) exists than in polar regions inside 
the inner shell. 
Before the shock breakout, the inner shell (inner reverse shock) with RT fingers is swept up by the outer inward shell (reverse shock) during the 
propagation of the blast wave into the hydrogen envelope. Just before the shock breakout, the outer reverse shock has swept up almost all inner ejecta 
in the model b18.3-high (see the top left panel in Figure~\ref{fig:dens-high-time2}). On the other hand, in the model n16.3-high, the last formed 
reverse shock is still propagating inward even after the shock breakout (see the top right and middle right panels in Figure~\ref{fig:dens-high-time2}). 
In the top right panel (just before the shock breakout), the reverse shock is around $r$ $\sim$ 1 $\times$ 10$^{12}$ cm. After the shock breakout 
(middle panels in Figure~\ref{fig:dens-high-time2}), the blast wave is accelerating due to the steep pressure gradient around the original stellar 
surface, leaving the inner ejecta far behind. Here, the radii of the stellar surface in the models b18.3 and n16.3, are 2.1 $\times$ 10$^{12}$ cm 
and 3.4 $\times$ 10$^{12}$ cm, respectively. 
Depending on the density and pressure gradients around the stellar surface, the shock breakout in 
the model b18.3-high takes place in a more aspherical way than in the model n16.3-high (middle panels in Figure~\ref{fig:dens-high-time2}). 
The times of the shock breakout in the models b18.3-high and n16.3-high are $\sim$ 3000 sec and $\sim$ 5000 sec, respectively, which should reflect 
the acceleration/deceleration during the shock propagation depending on the density structure of the progenitor model but in the end
it is roughly proportional to the stellar radius. 
Finally, at the end of the simulation (bottom panels in Figure~\ref{fig:dens-high-time2}), the inner ejecta consisting of the material originally inside the 
helium core is far behind the blast wave. The shape of the blast wave in the model b18.3-high is more aspherical than that in the model n16.3-high. 
%
\begin{figure*}[htb]
\begin{tabular}{cc}
\begin{minipage}{0.5\hsize}
\begin{center}
\includegraphics[width=7cm,keepaspectratio,clip]{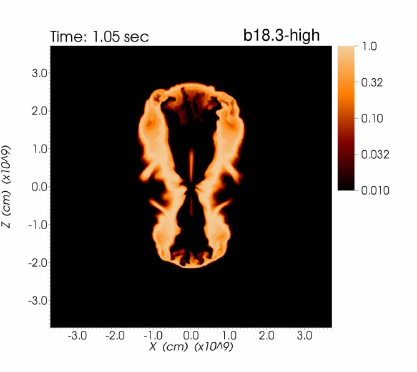}
\end{center}
\end{minipage}
\begin{minipage}{0.5\hsize}
\begin{center}
\includegraphics[width=7cm,keepaspectratio,clip]{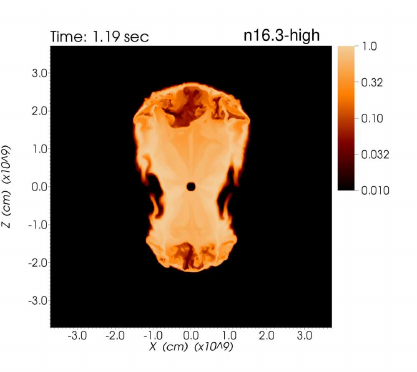}
\end{center}
\end{minipage}
\vspace{-0.3cm}
\\
\begin{minipage}{0.5\hsize}
\begin{center}
\includegraphics[width=7cm,keepaspectratio,clip]{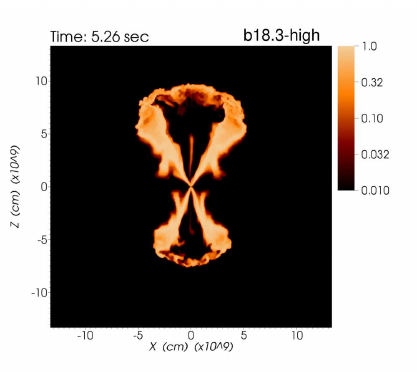}
\end{center}
\end{minipage}
\begin{minipage}{0.5\hsize}
\begin{center}
\includegraphics[width=7cm,keepaspectratio,clip]{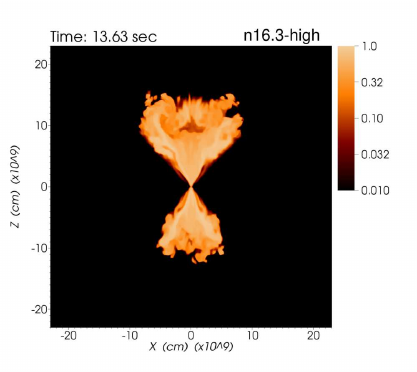}
\end{center}
\end{minipage}
\vspace{-0.3cm}
\\
\begin{minipage}{0.5\hsize}
\begin{center}
\includegraphics[width=7cm,keepaspectratio,clip]{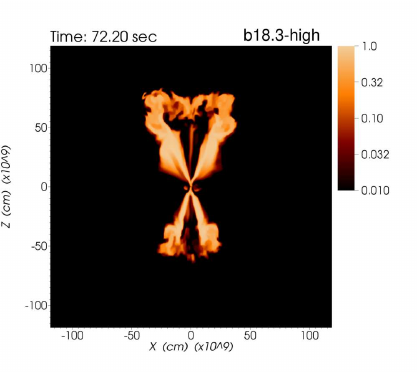}
\end{center}
\end{minipage}
\begin{minipage}{0.5\hsize}
\begin{center}
\includegraphics[width=7cm,keepaspectratio,clip]{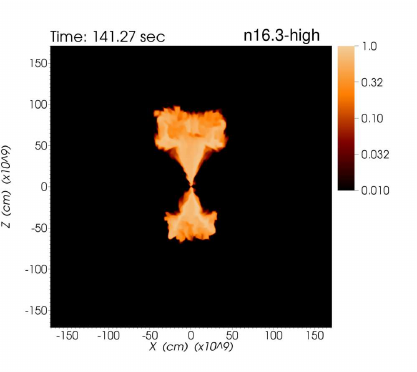}
\end{center}
\end{minipage}
\end{tabular}
\caption{Same as Figure~\ref{fig:dens-high-time2} but for color maps of the mass fraction of $^{56}$Ni.
}
\label{fig:ni56-high-time1}
\end{figure*}
%
\begin{figure*}[htb]
\begin{tabular}{cc}
\begin{minipage}{0.5\hsize}
\begin{center}
\includegraphics[width=7cm,keepaspectratio,clip]{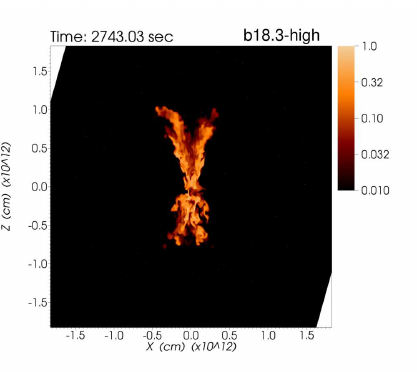}
\end{center}
\end{minipage}
\begin{minipage}{0.5\hsize}
\begin{center}
\includegraphics[width=7cm,keepaspectratio,clip]{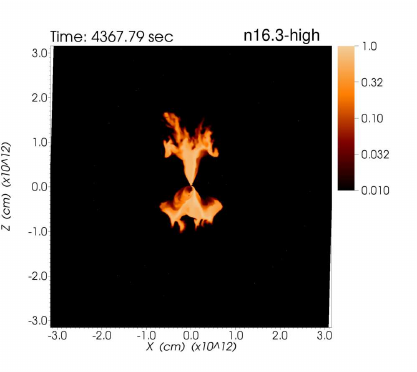}
\end{center}
\end{minipage}
\vspace{-0.3cm}
\\
\begin{minipage}{0.5\hsize}
\begin{center}
\includegraphics[width=7cm,keepaspectratio,clip]{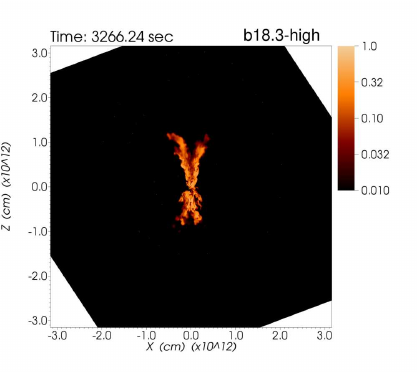}
\end{center}
\end{minipage}
\begin{minipage}{0.5\hsize}
\begin{center}
\includegraphics[width=7cm,keepaspectratio,clip]{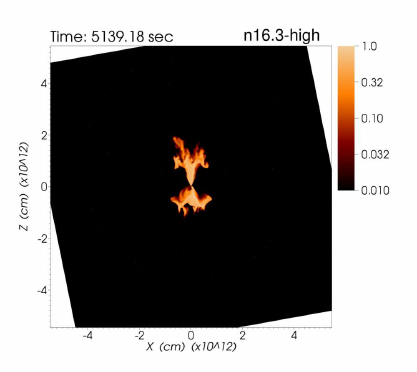}
\end{center}
\end{minipage}
\vspace{-0.3cm}
\\
\begin{minipage}{0.5\hsize}
\begin{center}
\includegraphics[width=7cm,keepaspectratio,clip]{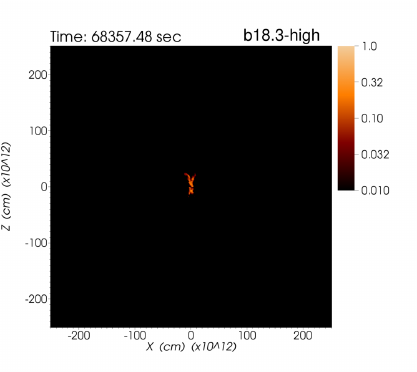}
\end{center}
\end{minipage}
\begin{minipage}{0.5\hsize}
\begin{center}
\includegraphics[width=7cm,keepaspectratio,clip]{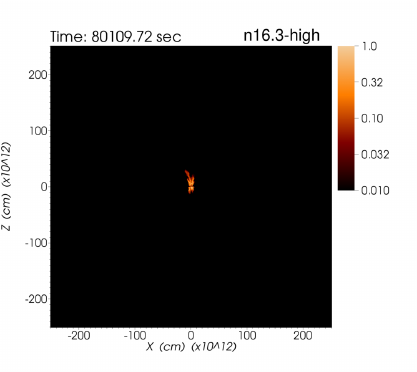}
\end{center}
\end{minipage}
\end{tabular}
\caption{Same as Figure~\ref{fig:dens-high-time2} but for color maps of the mass fraction of $^{56}$Ni.
}
\label{fig:ni56-high-time2}
\end{figure*}
Figures~\ref{fig:ni56-high-time1} and~\ref{fig:ni56-high-time2} show the time evolution of the $^{56}$Ni distribution for models b18.3-high (left) and 
n16.3-high (right), which correspond to the time evolutions of the density distribution in Figure~\ref{fig:dens-high-time1} and~\ref{fig:dens-high-time2}, 
respectively.
The early ($\sim$ 1 sec) distributions of $^{56}$Ni in the high resolution models can be compared with the corresponding lower resolution 
models b18.3-fid and n16.3-fid (top panels in Figure~\ref{fig:ni56-low-model-early}). 
Although smaller scale structures are seen in the models 
b18.3-high and n16.3-high, than in the lower resolution models, the overall distributions of $^{56}$Ni are the same as in the lower resolution models. At the early 
phase, $^{56}$Ni exists inside the fingers on the tips of the bipolar structure (top panels in Figure~\ref{fig:ni56-high-time1}). During the propagation of the 
blast wave into the helium layer (middle panels Figure~\ref{fig:ni56-high-time1}), part of $^{56}$Ni falls back into the compact object along the equatorial 
plane and the equatorial regions of the bipolar distribution are shrunk. At this phase, the tips of the bipolar distribution of $^{56}$Ni reach the shell (reverse 
shock) with RT fingers seen in the corresponding density distribution (middle panels in Figure~\ref{fig:dens-high-time1}). After the blast wave passes the 
He/H interface, as mentioned before, the nested double shell structure forms (bottom panels in Figure~\ref{fig:dens-high-time1}). An interesting 
difference between the two progenitor models is whether the tips of the $^{56}$Ni distribution reach the outer shell (the newly formed reverse shock during 
the shock propagation into the hydrogen layer) or not. In the model b18.3-high, the tips of $^{56}$Ni penetrate the inner shell and touch the outer shell (see 
bottom left panels in Figures~\ref{fig:dens-high-time1} and~\ref{fig:ni56-high-time1}). On the other hand, in the model n16.3-high, $^{56}$Ni remains 
confined to inside the inner shell (bottom right panels in Figures~\ref{fig:dens-high-time1} and~\ref{fig:ni56-high-time1}). As seen in the top two 
panels in Figure~\ref{fig:gr2} in Section~\ref{subsec:res-1d}, for both progenitor models, instabilities grow around the composition interfaces of the 
C+O/He and He/H. Then, the inner and outer shells (bottom panels in Figure~\ref{fig:dens-high-time1}) can approximately be regarded as the C+O/He and He/H composition 
interfaces, respectively. Hence, the bulk of $^{56}$Ni in the model n16.3-high is confined to the C+O layer, whereas part of $^{56}$Ni 
in the model b18.3-high penetrates the helium layer to reach the hydrogen layer. After the inward outer shell sweeps up the inner shell, in the model 
n16.3-high, part of $^{56}$Ni penetrates into the outer high velocity layers consisting of helium and hydrogen but the bulk of $^{56}$Ni remains confined 
to the helium core (top right panels in Figures~\ref{fig:dens-high-time2} and~\ref{fig:ni56-high-time2}). In the model b18.3-high, part of $^{56}$Ni 
reaches the tips of extended RT fingers (top left panels in Figures~\ref{fig:dens-high-time2} and~\ref{fig:ni56-high-time2}). After the shock breakout 
(middle and bottom panels in Figure~\ref{fig:ni56-high-time2}), the inner ejecta, including $^{56}$Ni, is left far behind the blast wave.

\begin{figure*}[htb]
\begin{tabular}{cc}
\begin{minipage}{0.5\hsize}
\begin{center}
\includegraphics[width=7cm,keepaspectratio,clip]{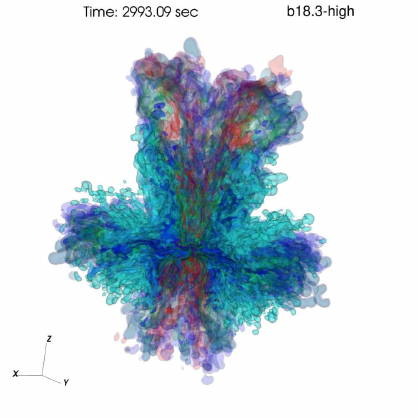}
\end{center}
\end{minipage}
\begin{minipage}{0.5\hsize}
\begin{center}
\includegraphics[width=7cm,keepaspectratio,clip]{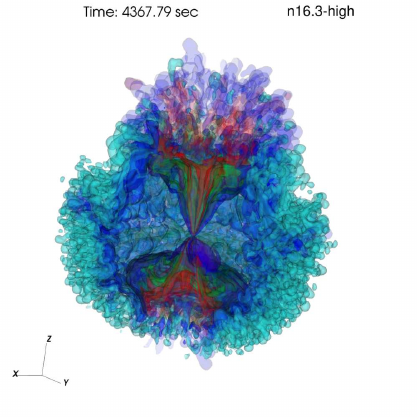}
\end{center}
\end{minipage}
\end{tabular}
\caption{Same as Figure~\ref{fig:elem-3d-low-model} but for models b18.3-high (left) and n16.3-high (right). 
%
An animation (distributions from different view angles and ones with and without clipping) for this figure is available. 
The realtime duration of the video is 12 seconds. 
Interactive 3D models on Sketchfab corresponding to left (https://skfb.ly/6OZDu) and right (https://skfb.ly/6OZDv) 
panels are also available.}
\label{fig:elem-3d-high}
\end{figure*}
%
Figure~\ref{fig:elem-3d-high} shows the distributions of elements, $^{56}$Ni, $^{28}$Si, $^{16}$O, and $^{4}$He, just before 
the shock breakout for models b18.3-high (left) and n16.3-high (right). 
Compared with the distributions in the corresponding lower resolution models (top 
panels in Figure~\ref{fig:elem-3d-low-model}), the global morphologies of the distributions are consistent with the lower resolution models, 
although smaller-scale structures are resolved. In the model b18.3-high, the fingers are extended from the central region, which reflects the fact that 
the reverse shock developed during the shock propagation into the hydrogen envelope has already swept up the inner ejecta before the shock breakout. 
While in the model n16.3-high, the reverse shock is still propagating into the inner ejecta even after the shock breakout and the fingers consisting of 
$^{16}$O and $^{4}$He are extended from the reverse shock surface. Then, a diluted space inside the shell with the $^{16}$O and $^{4}$He fingers is 
visible in the model n16.3-high (right panel). As mentioned above, in the model n16.3-high, a small fraction of $^{56}$Ni penetrates into the extended 
fingers along the polar direction but the bulk of $^{56}$Ni is confined to the helium core. In the model b18.3-high, the penetration of $^{56}$Ni into the tips 
of the extended fingers is observed. 

\begin{figure*}[htb]
\begin{tabular}{cc}
\hspace*{-1cm}
\begin{minipage}{0.5\hsize}
\begin{center}
\includegraphics[width=7cm,keepaspectratio,clip]{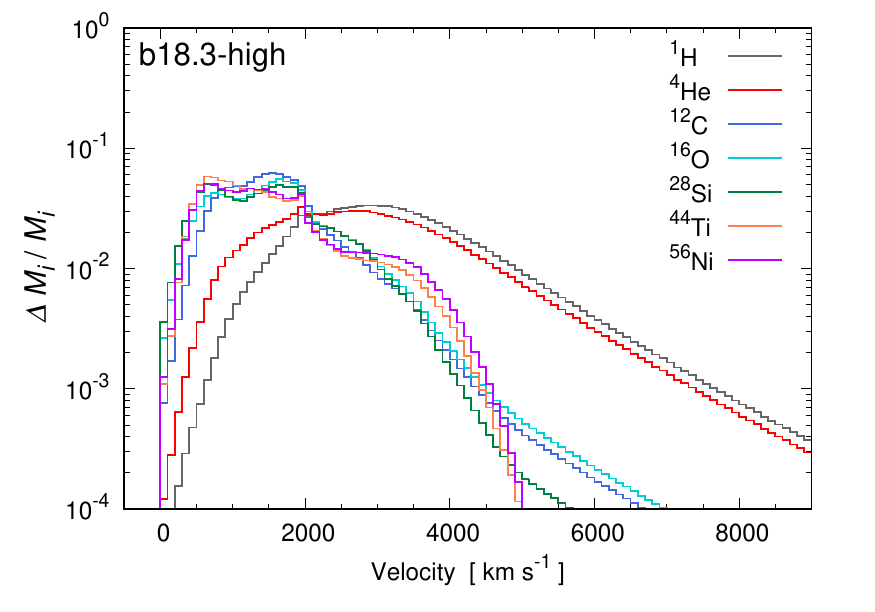}
\end{center}
\end{minipage}
\begin{minipage}{0.5\hsize}
\begin{center}
\includegraphics[width=7cm,keepaspectratio,clip]{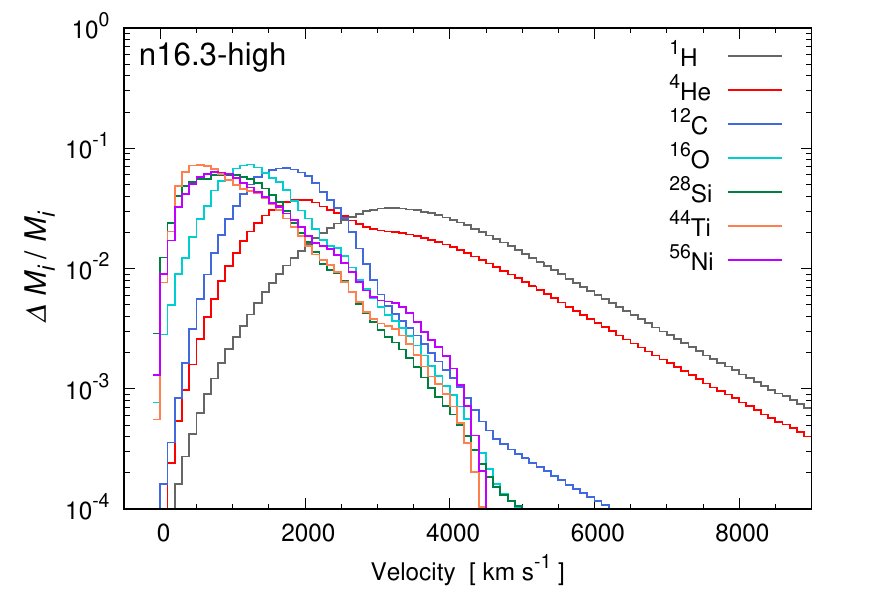}
\end{center}
\end{minipage}
\end{tabular}
\label{fig:vel-high}
\caption{Same as Figure~\ref{fig:vel-low-model} but for models b18.3-high (left; 68,357 sec) and n16.3-high (right; 80,110 sec).}
\label{fig:vel-high}
\end{figure*}
Figure~\ref{fig:vel-high} shows the distributions of radial velocity of $^{56}$Ni for models b18.3-high (left) and n16.3-high (right). 
The distributions can be compared 
with the lower resolution model, b18.3-fid and n16.3-fid (top panels in Figure~\ref{fig:vel-low-model}). Compared with the lower resolution model b18.3-fid, the high 
velocity $^{56}$Ni is slightly reduced in the model b18.3-high. A bump seen at around $\sim$ 3000 km s$^{-1}$ in the b18.3-fid (Figure~\ref{fig:vel-low-model}) is 
flattened in the model b18.3-high 
and instead lower velocity $^{56}$Ni ($\sim$ 2000 km s$^{-1}$) is slightly enhanced. On the other hand, compared with the lower 
resolution model n16.3-fid, in the model n16.3-high, $^{56}$Ni of velocity around 2500 and 3500 km s$^{-1}$ is slightly enhanced. As mentioned in 
Section~\ref{subsec:res-3d-low}, opposite responses to the increase in resolution of the simulations are seen, which may be attributed to the difference of the significance 
of the RT instability between the two progenitor models. Although there are slight differences between the lower and high resolution models, the superiority of the 
model b18.3 in terms of the amount of high velocity $^{56}$Ni is not changed between the lower and high resolution models. The mass of the high velocity $^{56}$Ni 
($\gtrsim$ 3000 km s$^{-1}$) in the model b18.3-high is larger than that of the model n16.3-high by a factor of $\sim$ 3 (see also Table~\ref{table:results}). 

\begin{figure*}[htb]
\begin{tabular}{cc}
\hspace*{-0.5cm}
\begin{minipage}{0.5\hsize}
\begin{center}
\includegraphics[width=7.5cm,keepaspectratio,clip]{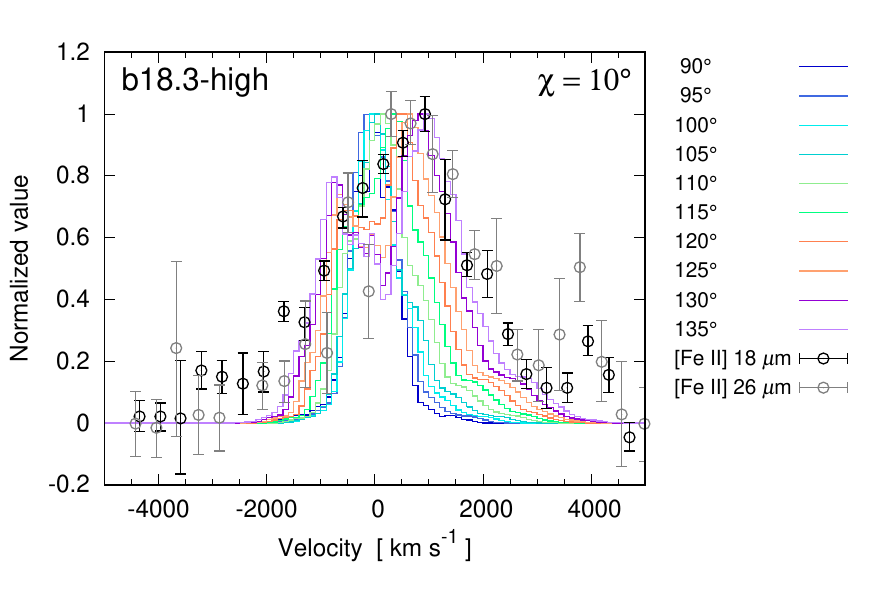}
\end{center}
\end{minipage}
\begin{minipage}{0.5\hsize}
\begin{center}
\includegraphics[width=7.5cm,keepaspectratio,clip]{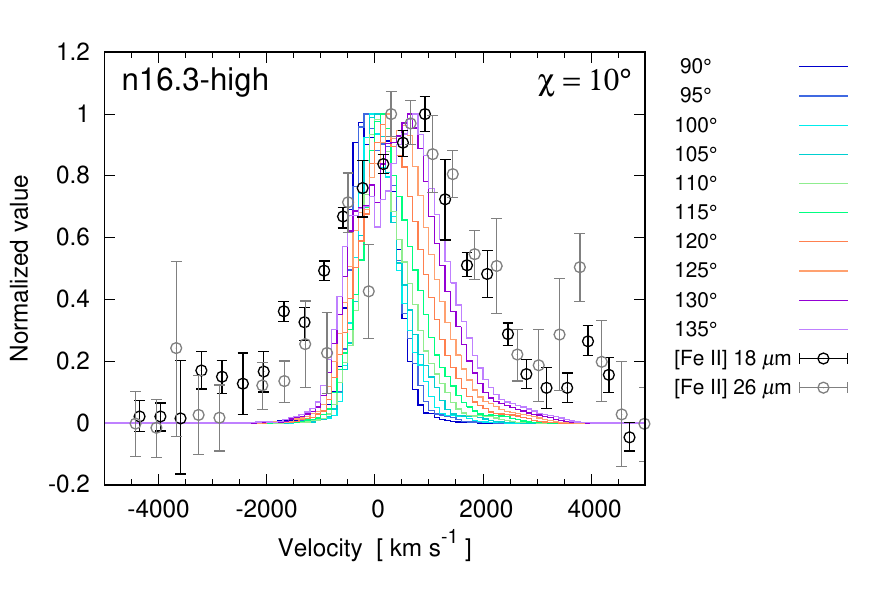}
\end{center}
\end{minipage}
\end{tabular}
\caption{Same as Figure~\ref{fig:vel-dir-low-model} but for models b18.3-high (left; 68,357 sec) and n16.3-high (right; 80,110 sec), and for the parameter $\chi$ = 10$^{\circ}$.}
\label{fig:vel-dir-high}
\end{figure*}
Figure~\ref{fig:vel-dir-high} shows the distributions of the line of sight velocity of $^{56}$Ni for models, b18.3-high (left) and n16.3-high (right). Here, based on the 
discussion on the rotation angle $\chi$ in Section~\ref{subsec:res-3d-low}, the value of $\chi$ is set to be 10$^{\circ}$. The distribution in the b18.3-high can be 
directly compared with the corresponding model b18.3-fid with the same value for $\chi$ (top left panel in Figure~\ref{fig:vel-dir-low-chi}). 
Compared with the 
distribution in the model b18.3-fid, the one of the double peaks for the $\theta$ $\gtrsim$ 120$^{\circ}$ at the blueshifted side around $-700$ km s$^{-1}$ is 
reduced and the value of the peak approaches the values of nearby observed points. The tail around $-2000$ km s$^{-1}$ are slightly enhanced instead the 
tail around $2000$ km s$^{-1}$ is slightly reduced. The distribution in the model n16.3-high can be compared with that in the model n16.3-fid with 
$\chi$ = 0$^{\circ}$ (top right panel in Figure~\ref{fig:vel-dir-low-model}). 
Although the value of $\chi$ in Figure~\ref{fig:vel-dir-high}~($\chi$ = 10$^{\circ}$) is 
different from the one in Figure~\ref{fig:vel-dir-low-model}~($\chi$ = 0$^{\circ}$)
, the distribution in the n16.3-fid with $\chi$ = 10$^{\circ}$ is very similar to the case 
of $\chi$ = 0$^{\circ}$. Compared with the distribution in the model n16.3-fid with $\chi$ = 0$^{\circ}$, the tails around $-1000$ km s$^{-1}$ and 1500 km s$^{-1}$ 
are slightly enhanced. As a summary, the distribution in the model b18.3-high better reproduces the high velocity tails of the observed fluxes of [Fe II] lines, in 
particular the tail at the redshifted side, than that in the model n16.3-high. As for the rotation angle $\theta$, the value of $\theta$ = 130$^{\circ}$--135$^{\circ}$ 
is preferred to fit the observed points.  Motivated by the observed 3D distributions of inner ejecta of SN 1987A 
\citep[see e.g., Figure~11 in][]{2016ApJ...833..147L}~(the ejecta of the redshifted side seems to be closer to us than the ER plane), 
a smaller $\theta$ value may 
be preferred. For the value of $\chi$, $\chi$ = 0$^{\circ}$--20$^{\circ}$ is a possible range as discussed in Section~\ref{subsec:res-3d-low} but in the case of 
$\chi$ = 10$^{\circ}$, the distribution better explains the observed fluxes. Therefore, we propose the parameter set of 
($\theta$, $\chi$) = (130$^{\circ}$, 10$^{\circ}$) as fiducial values for the rotation angles. 


\section{Discussion} \label{sec:discussion}

In this section, based on the results presented in Section~\ref{sec:results}, several related topics are discussed in more detail.  

\subsection{Key Properties of the Progenitor Models, and Their Impact on Matter Mixing} \label{subsec:depend-prog}

As presented in Section~\ref{sec:results}~(see e.g., Table~\ref{table:results}), the matter mixing, in particular how the innermost $^{56}$Ni can be conveyed into 
outer high velocity layers consisting of helium and hydrogen, depends on the pre-supernova model. 
%
As discussed in Sections~\ref{subsec:res-3d-low} and~\ref{subsec:res-3d-high}, 
among the investigated models in 
this paper, there is no explosion model with the BSG model n16.3 that satisfies the two observational constraints on the mass of high velocity $^{56}$Ni simultaneously. 
On the other hand, with the BSG model b18.3 
\citep[binary merger model:][]{2018MNRAS.473L.101U} and two other RSG models s18.0 and s19.8 \citep{2002RvMP...74.1015W,2016ApJ...821...38S}, we 
found several explosion models that satisfy the constraints. Thus, important questions are which properties of the pre-supernova models are essential to the 
matter mixing and why the properties have an advantage in the matter mixing. As presented by \citet{2015A&A...577A..48W}, once the blast wave enters the 
dense helium layer, the reverse shock developed due to the strong deceleration prohibits the inner ejecta from penetrating the helium layer to reach the 
composition interface of He/H where RT instability becomes active. If the inner ejecta consisting of $^{56}$Ni can successfully penetrate the helium layer before 
the development of the reverse shock, $^{56}$Ni can be conveyed into the high velocity hydrogen layer with the help of the RT instability. Actually, high velocity 
$^{56}$Ni ($\sim$ 3500 km s$^{-1}$) is obtained with the BSG model B15 
\citep{2015A&A...577A..48W,2015A&A...581A..40U,2019A&A...624A.116U}. 
As presented in Section~\ref{subsec:res-3d-high}, in 
the model b18.3-high in this paper, such successful penetration of $^{56}$Ni is also demonstrated. On the other hand, in the model n16.3-high, $^{56}$Ni fails 
to penetrate the helium layer before the development of the reverse shock. In \citet{2015A&A...577A..48W}, the same pre-supernova model N20 (n16.3 in this paper) 
with a more realistic explosion model based on the neutrino heating also fails to reproduce the high velocity $^{56}$Ni. A distinct property different between 
the successful BSG models b18.3 and B15 and the unsuccessful model n16.3 (N20) is the helium core mass, $M_{\rm He,c}$. The masses of the helium core 
of the models b18.3 and n16.3 are $\sim$ 4 $M_{\odot}$ and $\sim$ 6 $M_{\odot}$, respectively (see Table~\ref{table:models}). The helium core mass of the 
model B15 is 4.05 $M_{\odot}$, which is very similar to the value of the model b18.3. In terms of the ratio of the helium core mass to the stellar mass, 
$q \equiv M_{\rm He,c}/M$, such difference between the successful and unsuccessful models is also recognized. The $q$ values of the model b18.3 and B15 
are 0.22 and 0.27, respectively, whereas the $q$ value of the model n16.3 (N20) is 0.37. The small masses of the helium core of the models b18.3 and B15 may 
enable the inner ejecta more easily to penetrate the helium core before the development of the reverse shock than the case of the model n16.3. 
On the other hand, as we observed in the model n16.3, the matter mixing in other BSG models with larger helium core mass of 
$\gtrsim$ 5 $M_{\odot}$ 
\citep{2015A&A...577A..48W,2015A&A...581A..40U,2019A&A...624A.116U} 
has revealed that the maximum velocities of $^{56}$Ni are insufficient to explain 
the observations of SN 1987A.
It is worth noting that in several explosion models with the two RSG models, high velocity $^{56}$Ni is also obtained despite the large helium core masses ($q$ values). 
In the case of the two RSG models, the gradient of the $\rho \,r^3$ profile in the helium layer is overall negative in contrast to that of the model b18.3 
(see Figures~\ref{fig:prog1} and \ref{fig:prog2} ) and the hydrogen envelope is very extended ($r$ $\sim$ 10$^{13}$ cm) compared with those of b18.3 and B15 
($r$ $\sim$ 10$^{12}$ cm). Then, during the propagation of the blast wave in the helium layer, the development of a distinct reverse shock is restricted and the inner 
ejecta can reach around the He/H interface without the deceleration by the reverse shock. 
Additionally, thanks to the extended hydrogen envelope, there is enough time for RT instabilities to grow as seen in bottom two panels in Figure~\ref{fig:gr2}~(see the 
highly developed peaks around the composition interfaces of He/H), which enables the inner ejecta to be mixed up into the high velocity hydrogen layer. In this way, 
despite the large helium core mass, high velocity $^{56}$Ni is obtained in the RSG models. 

The BSG model B15 \citep{1988ApJ...324..466W} is one of successful models for the high velocity $^{56}$Ni, however, the helium core mass, $\sim$ 4 $M_{\odot}$, is 
less than the suggested value \citep[6 $\pm$ 1 $M_{\odot}$:][]{1989ARA&A..27..629A} from the observed luminosity of Sk $-$69$^{\circ}$ 202. 
The helium core mass of the model n16.3 (N20), $\sim$ 6 $M_{\odot}$, is appropriate in terms of the luminosity of Sk $-$69$^{\circ}$ 202, 
but the model is made by artificially combining an evolved helium core with a hydrogen envelope obtained from an independent stellar evolution calculation. 
Additionally, as we have seen above, the matter mixing in the model n16.3 fails to obtain the high velocity $^{56}$Ni. As mentioned in Sections~\ref{sec:intro} and~\ref{subsec:pre-sn}, 
for both BSG models based on the single star evolution scenario, several assumptions, reduced metallicity, restricted convection, 
enhanced mass loss, and enhancement of the helium abundance in the hydrogen envelope (the first two are for the model B15 and the latter two are for the envelope 
of the model n16.3), have been implemented to obtain the red-to-blue evolution for Sk $-$69$^{\circ}$ 202. 
Thus, the BSG progenitor models based on the single star evolution have both pros and cons. 
On the other hand, the binary merger model b18.3 
\citep{2018MNRAS.473L.101U}, which satisfies all observational constraints including the luminosity of Sk $-$69$^{\circ}$ 202, has a smaller helium core mass ($q$ value) 
than those of models based on the single star evolution that satisfy the luminosity of Sk $-$69$^{\circ}$ 202, which may reflect the nature of merging processes, the 
penetration of the secondary into the envelope of the primary and the dredge up of the primary's core material into the envelope. Actually, the recent other binary merger 
models \citep{2017MNRAS.469.4649M} that satisfy the observational constraints all have a small helium core mass of $\sim$ 3--4 $M_{\odot}$ ($q$ $\sim$ 0.1--0.2) 
\citep[see Table~4 in ][]{2017MNRAS.469.4649M}. As a summary, from the both aspects of the matter mixing and the observational constraints on Sk $-$69$^{\circ}$ 202, 
the binary merger scenario is preferred for SN 1987A.


\subsection{Morphology of the Supernova Ejecta and the Explosion Mechanism of SN 1987A} \label{subsec:exp-mechanism}

As mentioned in Section~\ref{sec:intro}, the 3D morphology of the inner ejecta of SN 1987A is globally elliptical/elongated \citep{2010A&A...517A..51K,2013ApJ...768...89L,
2016ApJ...833..147L}. The distributions of the observed [Fe II] lines in the Doppler velocity are biased toward the redshifted side. Motivated by the observations, in this 
paper, bipolar-like explosions with asymmetry against the equatorial plane are explored. Theoretically, the shock revival of a canonical core-collapse supernova explosion 
could be triggered by neutrino heating aided by SASI and/or convection and such asymmetric bipolar explosions could be realized if a low unstable mode ($l$ = 1) of SASI 
is dominating as seen in 2D hydrodynamical simulations \citep[e.g.,][]{2006A&A...457..963S,2010PASJ...62L..49S,2016ApJ...817...72P}. 
Since multi-dimensional ab initio hydrodynamical simulations of core-collapse supernovae are practically impossible and the adopted physical effects and the 
approximations in particular for the neutrino transport have been rather varied among the models, a consensus on the explosion mechanism has not been reached. 
Several three-dimensional simulations of core-collapse supernovae have revealed that strong sloshing motion introduced by the low unstable mode of SASI is not evident at 
least at later phases of the shock revival \citep[e.g.,][]{2010ApJ...720..694N,2013ApJ...770...66H,2013ApJ...765..110D}. Not all but some recent 3D hydrodynamical 
models have shown an asymmetric dipolar-like morphology (asymmetric two lobe structure) depending on the progenitor models \citep{2017MNRAS.472..491M,
2019MNRAS.482..351V,2019MNRAS.485.3153B} \citep[see e.g., Figure~1 in][]{2019MNRAS.482..351V}. 
In this paper, in the fiducial model (b18.3-high), the parameter set of ($\alpha \equiv v_{\rm up}/v_{\rm down}$, $\beta \equiv v_{\rm pol}/v_{\rm eq}$) = (1.5, 16) is 
adopted. As seen in Figure~\ref{fig:shape}~(bottom right panel and the case of $\beta = 16$) in Appendix~\ref{sec:app1}, the distribution of initial radial velocities is rather 
concentrated in the polar direction, which invokes a bipolar but more narrowly collimated (jetlike) explosion than those seen in the models mentioned above 
\citep{2017MNRAS.472..491M,2019MNRAS.482..351V,2019MNRAS.485.3153B}. 
Such jetlike explosions could be realized by magnetorotationally-induced core-collapse supernova explosions \citep[e.g.,][]{2009ApJ...691.1360T,2013ApJ...764...10S,
2014ApJ...785L..29M}. 
In MHD simulations, generally both strong magnetic-field and rapid rotation before the core-collapse are necessary for a successful magnetorotationally-induced explosion; however, 
it has not been revealed yet from which evolutionary paths the both conditions are realized simultaneously in the progenitor star just before the core-collapse \citep[see e.g.,][]{2005ApJ...626..350H}. 
To assess whether such magnetorotationally-induced explosions can be realized or not, the understanding of the role of the magnetorotational instability \citep{1998RvMP...70....1B} is important. 
Additionally, depending on the circumstances, the combination of magnetorotational and neutrino heating effects could also trigger a jetlike explosion 
\citep{2016ApJ...817..153S}. 

It is worth noting that even if the explosion is initially narrowly collimated, the morphology is soon modified to be wider at an early phase ($\sim$ 1 sec) as seen in 
Figure~\ref{fig:dens-high-time1}~(top left panel) depending on the structure of the C+O core of the progenitor star and the morphology is continuously changed due 
to the deceleration/acceleration of the blast wave and the growth of instabilities (see Figures~\ref{fig:dens-high-time1} and \ref{fig:dens-high-time2}). This situation 
could be much different from that for Type Ia supernovae, where the vestige of the explosion morphology can be survived at even an early phase of the supernova 
remnant ($\sim$ 100 yr) \citep{2019ApJ...877..136F}. 
As seen in Figures~\ref{fig:ni56-high-time1} and \ref{fig:ni56-high-time2}, the distribution of $^{56}$Ni is time-dependent during the shock propagation in the progenitor star 
but the bipolar structure globally survives even after the shock breakout. In the models, e.g., b18.3-high and n16.3-high, the axisymmetric bipolar structures are identified 
for each element and heavier elements are more concentrated along the bipolar axis as seen in Figure~\ref{fig:elem-3d-high}. Therefore, if such clear axisymmetric 
structures are identified from the future observations in the spatial distributions of emission lines of elements, in particular ones from iron or direct $\gamma$-ray lines from 
the decay of $^{44}$Ti, it would be a clue to deduce the explosion mechanism. Recent observations of 3D distributions of CO and SiO molecules in the inner ejecta of 
SN 1987A \citep{2017ApJ...842L..24A} are an eligible target for the test. Actually, we plan to compare the results (approximate CO and SiO distributions) of 3D MHD simulations 
of further evolutions of the models b18.3-high and n16.3-high with the observed CO and SiO distributions (Orlando et al. 2019, in preparation). Additionally, 
in order to estimate the CO and SiO distributions more accurately, we also plan to calculate (Ono et al., in preparation) the molecule formation in the ejecta with a molecule 
formation network using a post-processing method based on the 3D MHD simulation results above. 
 
\subsection{Neutron Star Kick Velocity} \label{subsec:ns-kick}

The compact object of SN 1987A has not been detected yet but it could be a thermally emitting NS obscured by dust \citep{2015ApJ...810..168O,2018ApJ...864..174A}. If the explosion of SN 1987A 
was an asymmetric one as demonstrated in the models, e.g., b18.3-high, the compact object (probably NS) could have been kicked to the opposite direction to the motion of the 
bulk of the supernova ejecta. Actually, NS kicks are expected from 2D and 3D hydrodynamical simulations of neutrino-driven explosions aided by SASI and/or convection thanks 
to their aspherical nature \citep{2004PhRvL..92a1103S,2006A&A...457..963S,2010PhRvD..82j3016N,2012MNRAS.423.1805N,2010ApJ...725L.106W, 2013A&A...552A.126W}. 
In the context of the neutrino-driven explosion, first, a NS is kicked in the opposite direction to the strongest explosion and later the motion is mediated by the interaction between 
gravitationally combined denser slowly moving clumps left behind the shock and the compact object \citep[gravitational tug-boat mechanism:][]{2013A&A...552A.126W}. 
An asymmetric neutrino emission has been proposed as another mechanism of NS kicks \citep{1987IAUS..125..255W,1993A&AT....3..287B,2005ApJ...632..531S}. 
From the X-ray observations of six core-collapse supernova remnants, Cas A, G292.0+1.8, Puppis A, Kes 73, RCW 103, and N49, it has been revealed that the direction of the NS 
kick relative to the explosion center is opposite to the center of mass of gaseous intermediate elements in the ejecta \citep{2018ApJ...856...18K}, which supports a hydrodynamical 
origin of NS kicks such as the gravitational tug-boat mechanism.
Recent analysis of the observations of Cas A has indicated that heavier elements are more oppositely distributed than lighter ones \citep{2019arXiv190406357H}. From the observed 
proper motions of young pulsars, the 3D NS kick velocities of young pulsars have been deduced typically as 300--500 km s$^{-1}$ \citep{2002ApJ...568..289A,2003AJ....126.3090B,
2005MNRAS.360..974H,2006ApJ...643..332F} but some pulsars have a velocity over 1000 km s$^{-1}$ \citep[e.g.,][]{2005ApJ...630L..61C}. 
%
Motivated by the situation above, we estimate the NS kick velocity by assuming simply momentum conservation (the initial total momentum is zero) as in 
\citet{2013A&A...552A.126W}. The NS kick velocity, $\vec{v}_{\rm NS}$ is estimated as follows:
\begin{equation}
\vec{v}_{\rm \, NS} = - \vec{P}_{\rm gas}/M_{\rm NS} = - \int_{V} \, \rho \, \vec{v} \ \mathrm{d} x \, \mathrm{d} y \,\mathrm{d} z/ M_{\rm NS},
\label{eq:kick}
\end{equation}
where $\vec{P}_{\rm gas}$ is the total momentum inside the computational domain, $V$, except for the innermost regions corresponding to the compact object (NS), $M_{\rm NS}$ 
is the mass inside the regions of the NS, and $\vec{v}$ is the fluid velocity. In the 9th column in Table~\ref{table:results}, the absolute values of the estimated NS kick velocity at the 
end of the simulation, $v_{\rm NS}$, are listed. The values of $v_{\rm NS}$ dominantly depend on the parameter $\alpha \equiv v_{\rm up}/v_{\rm down}$. The values of 
$v_{\rm NS}$ for the models with $\alpha$ = 2.0, b18.3-mo13, n16.3-mo13, s18.3-mo13, s19.8-mo13, and b18.3-alpha2, are $\sim$ 420--580 km s$^{-1}$. The value of 
$v_{\rm NS}$ for the model with $\alpha$ = 1.0, b18.3-alpha1, is $\mathcal{O}$(1) km s$^{-1}$. The values of $v_{\rm NS}$ for the models with $\alpha$ = 1.5 are $\sim$ 250--400 
km s$^{-1}$ except for the model b18.3-ein1.5, for which $E_{\rm in}$ = 1.5 $\times$ 10$^{51}$ erg and $v_{\rm NS}$ $\sim$ 150 km s$^{-1}$. Therefore, overall the larger the 
$\alpha$ value, the larger the NS kick velocity. The NS kick velocities also depend on the parameter, $\beta \equiv v_{\rm pol}/v_{\rm eq}$. For example, as seen in the models 
s18.0-beta2, s18.0-beta4, s18.0-beta8, and s18.0-fid, the larger the $\beta$ value, the smaller the value of $v_{\rm NS}$, which reflects the fact that if the total kinetic energy is 
fixed (here uniform density is considered), the wider the bipolar explosion is, the larger the net momentum is\footnote{Consider a 1 cm$^3$ cube with a kinetic energy of $\frac{1}{2} \rho \,v^2$. 
If the kinetic energy is divided into two 1 cm$^3$ cubes, the total kinetic energy is $\frac{1}{2} \rho \,v^2$ = $\frac{1}{2} \rho \,(\frac{v}{\sqrt{2}})^2 + \frac{1}{2} \rho \,(\frac{v}{\sqrt{2}})^2$, 
whereas the net momentum before dividing is $\rho \,v$ but the momentum after dividing is $\sqrt{2} \,\rho \,v > \rho \,v$.}. 
Overall, the obtained values of $v_{\rm NS}$ are roughly within the range of the observed NS kick velocities. The vector values of the NS kick velocities for the models b18.3-high 
and n16.3-high are $\vec{v}_{\rm \, NS}$ = ($v_x$, $v_y$, $v_z$) = ($-$7.23, 1.28, $-$2.85 
$\times$ 10$^2$) km s$^{-1}$ and (0.103, $-$0.504, $-$3.03 $\times$ 
10$^{2}$) km s$^{-1}$, respectively. Then, the values of $v_x$ and $v_y$ are $\mathcal{O} (1)$ km s$^{-1}$ and the NS kick velocities are directed almost opposite to the $z$-axis 
(the strongest explosion direction). The absolute values for the models b18.3-high and n16.3-high are 2.85 $\times$ 10$^{2}$ and 3.03 $\times$ 10$^{2}$ km s$^{-1}$, respectively, 
which are consistent with the observed values. As seen in Figure~11 in \citet{2018ApJ...856...18K}, the relative positions of the compact objects are not perfectly opposed to the 
positions of the center of mass of the ejecta. Even if the overall features of the observed NS kick velocities can be explained by the hydrodynamical effects demonstrated in this 
paper, such deviations have not been well explained yet. 
Rotation, which is not included in this paper, may play an important role. 
As mentioned in Section~\ref{subsec:res-3d-high}, from the comparison of the line of sight velocities of $^{56}$Ni with the observed [Fe II] line profiles, the parameter set of 
($\theta$, $\chi$) = (130$^{\circ}$, 10$^{\circ}$)~(see Figure~\ref{fig:geo}), which determines the direction of the bipolar explosion axis (the strongest explosion direction) to 
observers on Earth, is preferred for the fiducial (best) model b18.3-high. 
As seen in Figure~\ref{fig:geo}, if the explosion of the redshifted side (the strongest explosion side) is directed to the south side (negative $x$ direction) to us as the case of 
($\theta$, $\chi$) = (130$^{\circ}$, 10$^{\circ}$), the NS kick velocity is directed to the north side. Then, if the best model, b18.3-high with 
($\theta$, $\chi$) = (130$^{\circ}$, 10$^{\circ}$), is correct, we predict that the compact object of SN 1987A will be found in the north part of the inner ejecta. 
Recent observations of dust emission from the inner ejecta of SN 1987A by ALMA have suggested that a dust peak found at the northeast of the center of the remnant could be an indirect detection of the compact object \citep{2019arXiv191002960C}, which is very roughly consistent with our prediction. 

\subsection{Issues in Stellar Evolution Models and Impacts of Possible Large Density Perturbations in the Progenitor Star} \label{subsec:large-dens-perturb}

As presented in Section~\ref{subsec:res-3d-high}, even in the fiducial (best) model b18.3-high, observed fluxes of [Fe II] lines (points with the normalized values of $\gtrsim$ 0.1) 
around the high velocity tails (the absolute Doppler velocity higher than 3000 km s$^{-1}$) can not be reproduced well (see the left panel in Figure~\ref{fig:vel-dir-high}). Another 
possible ingredient that is not included in this paper is large perturbations in density of the pre-supernova models, which was previously investigated in Paper~II. Here, we discuss 
the current status of stellar evolution models and the impact of such large density perturbations. 

Pre-supernova models obtained from stellar evolution calculations are basically spherically symmetric, where one-dimensional (spherical) hydrostatic equations with a 
mixing-length theory \citep[MLT:][]{1958ZA.....46..108B} for convection are solved \citep[see e.g.,][]{1990sse..book.....K}. Since convection is inherently involved in the turbulent 
motion of elements in 3D, the MLT itself has had long standing issues. In the MLT, the length-scale of the mixing of an element (eddy) into surroundings (mixing length: $l$) is 
``assumed" as $l = \alpha \,H_{P}$, where $\alpha$ is a free parameter and $H_{P}$ is the local pressure scale height. Related uncertainties on the treatments of so-called 
semiconvection and overshooting have also been problematic. Semiconvection is a slow mixing process in the region dynamically stable due to the existence of a non-zero 
gradient of the mean molecular weight but vibrationally unstable (the so-called Ledoux criterion is fulfilled but the Schwarzschild criterion is not). The treatments of 
semiconvection and the observational constraints have been investigated for a few decades \citep[e.g.,][]{1985A&A...145..179L,1992A&A...253..131S,2011A&A...529A..63S,
2013A&A...552A..76S,2013A&A...554A.119Z,2012ApJ...756...37L}. Overshooting is the penetration of elements over a convective zone into a dynamically stable region, 
which may be the most uncertain process in the context of the MLT and it has intensively studied in several aspects \citep[e.g.,][]{2006ApJ...653..765R,2007A&A...475.1019C,
2008MNRAS.386.1979D,2013ApJS..205...18Z,2013ApJ...766..118M,2015A&A...580A..61V}. For both the semiconvection and overshooting, non-locality is essential and 
self-consistent non-local convection theories beyond the local MLT have been proposed \citep{1977AcASn..18...86X,1993ApJ...407..284G,1997ApJS..108..529X,
1998ApJ...493..834C,2006ApJ...643..426D,2007MNRAS.375..388L,2016ApJ...818..146Z}, which has partly been motivated by the implications from multi-dimensional 
hydrodynamical simulations mentioned below. 
In general the time scale of the stellar evolution is determined by the nuclear burning which is much longer than the dynamical time scale of the turbulent motion of fluids in a 
convective layer and the crossing time of sound waves. 
Then, it is impossible to cover the whole evolution of a star by multi-dimensional hydrodynamical simulations (for compressible fluids) in which the time step is limited by the 
maximum fluid velocity or the maximum sound speed inside the computational domain from the Courant-Friedrichs-Lewy (CFL) condition. Nevertheless, there have been attempts 
at such multi-dimensional hydrodynamical simulations of the evolution of massive stars which cover one or a few burning shells (for up to a few convection turnover times 
in the case of 3D simulations)~\citep{1994ApJ...433L..41B,1998ApJ...496..316B,2006ApJ...637L..53M,2007ApJ...665..690M,2007ApJ...667..448M,2009ApJ...690.1715A,
2011ApJ...733...78A,2013ApJ...769....1V,2015ApJ...808L..21C,2016ApJ...822...61C,2016ApJ...833..124M,2017MNRAS.471..279C,2018MNRAS.481.2918M,
2019ApJ...881...16Y}. As seen in e.g., Figures~3 and~4 in \citet{2016ApJ...833..124M}, the distributions of $^{28}$Si and radial velocities are very fluctuating 
with large-scale anisotropies. From the investigations above, for example, \citet{2007ApJ...667..448M} argued for turbulent convection where a turbulent layer adjacent to a 
stably stratified layer diffuses into the stable layer over time (turbulent entrainment), which is generally ignored in the stellar evolution models based on local MLTs 
(the authors also suggested that overshooting is best described as an elastic response by convective boundary). 
\citet{2011ApJ...741...33A} 
pointed out the ``$\tau$-mechanism" as a new source of luminosity fluctuations associated with turbulent convective cells based on 3D hydrodynamical simulations of shell 
oxygen burning, which exhibit recurrent fluctuations in turbulent kinetic energy. Recently, a new method to replace the MLT in one-dimensional stellar evolutionary 
computations based on 3D hydrodynamic simulations (``321D" approach) has been presented \citep{2015ApJ...809...30A}. 
Keeping in mind the impact of the density fluctuations in the progenitors on the matter mixing, it is interesting to see how large amplitude of fluctuations could be 
introduced. Overall, among the multi-dimensional hydrodynamical simulations, density fluctuations up to $\sim$ 10\% could be introduced around the edges of the 
convective zone of oxygen burning shell in $\sim$ 20 $M_{\odot}$ stars \citep{1998ApJ...496..316B,2006ApJ...637L..53M,2007ApJ...665..690M} and in the 
envelope of a 5 $M_{\odot}$ of red giant \citep{2013ApJ...769....1V}.
For lower mass stars, high resolution 3D global hydrodynamical simulations of the solar convection \citep{2014ApJ...786...24H,2015ApJ...798...51H} and He shell 
flash in a post-AGB star \citep{2014ApJ...792L...3H} have been performed but the amplitudes of the density fluctuations introduced seem to be small. 
\citet{2014ApJ...785...82S} discussed the discordance between the predictions from stellar evolution models and the last stages of massive stars, some of which 
(progenitors of Type IIn supernovae) exhibit eruptive mass ejection a decade before the core-collapse. The authors suggested that the major reason of the 
discordance may lie in the treatments of turbulent convection, i.e., stellar evolution models with MLTs generally ignore 
i) finite amplitude fluctuations in velocity and temperature and 
ii) their nonlinear interaction with nuclear burning. 
Such mass ejection invokes more violent eruptions from luminous blue variables (LBVs) such as $\eta$ Carinae. 
Actually, the candidate of a LBV, HD 168625, is a nearby twin of Sk $-$69$^{\circ}$ 202, which has a similar triple ring structure. 
From the similarity with HD 168625, \citet{2007AJ....133.1034S} proposed a scenario that Sk $-$69$^{\circ}$ 202 was a LBV evolved as a single star, although 
the single star evolution scenario contradicts the binary merger model b18.3 proposed as the pre-supernova model for the fiducial model (b18.3-high) in this paper. 
It was theoretically demonstrated that some binary mergers are capable of producing LBVs \citep{2014ApJ...796..121J}. 
Actually, it has been proposed that $\eta$ Carinae and a LVB candidate, R4, currently in a binary system were derived from a binary merger of two stars originally in a triple star system 
\citep[for the former and the latter, see][respectively]{2016MNRAS.456.3401P,2000AJ....119.1352P}. 
%
Whatever the evolution scenario is, violent dynamical eruptions from the envelope of a LBV or mass ejection from a rapidly rotating BSG would cause large-scale 
fluctuations in the envelope. 
Herschel observations of the closest RSG, Betelgeuse, have revealed that the observed clumpy structure in the inner part of the circumstellar medium could stem 
from giant convection cells of the outer atmosphere \citep{2012A&A...548A.113D}. The observed close molecular layer and the intensity map computed based on 
3D radiative hydrodynamic simulations of RSGs has also invoked large-scale fluctuations in the envelope of Betelgeuse \citep[][]{2010A&A...515A..12C,
2014A&A...572A..17M} \citep[see Fig.~10 in][]{2014A&A...572A..17M}. 
Hitherto, despite the intensive attempts at the multi-dimensional hydrodynamical simulations mentioned above, the theoretical understanding of stellar envelopes 
of massive stars in particular at the last stage before the core-collapse is far from conclusive and it has not been unveiled how large fluctuations can be introduced 
in the envelopes of pre-supernova stars. 
In addition, recent 3D hydrodynamical simulations of core-collapse supernova explosions have revealed that pre-collapse asphericities around Si/O layers due to 
turbulence could alter the postbounce evolution and enhance the explodability of core-collapse supernovae in the context of the neutrino-driven mechanism 
\citep{2013ApJ...778L...7C,2015ApJ...799....5C,2017MNRAS.472..491M}. 

Motivated by the theoretical and observational situations mentioned above, in Paper~II \citep{2015ApJ...808..164M}, we investigated the influence of large density 
perturbations of the amplitude of up to 50\% in the density of the pre-supernova model (same as the model n16.3 in this paper) based on 2D hydrodynamical 
simulations focusing on the matter mixing.  Among the investigated models, if there are non-radial perturbations (50\%) with radially coherent structures (see the 
top left panel in Figure~2 in Paper~II), high velocity clumps of $^{56}$Ni ($\lesssim$ 4000 km s$^{-1}$) can be obtained at the tails of the highest Doppler velocity 
(see e.g., Figure~15 in Paper~II), even if the explosion is only mildly aspherical (approximately the same as seen in the top left panel in Figure~\ref{fig:shape} 
(the case of $\beta$ = 4) in this paper). The high velocity clumps of $^{56}$Ni obtained correspond to the tips of giant RT fingers. Therefore, by introducing such 
large density perturbations in the model b18.3-high, a better fit to the observed [Fe II] line profiles may be obtained, although such investigation is beyond the scope 
of this paper. 


\section{Summary} \label{sec:summary}

In this paper, we perform 3D hydrodynamic simulations of non-spherical core-collapse supernovae focusing on the matter mixing in SN 1987A. The impact of the 
four pre-supernova models and parameterized aspherical explosions on the matter mixing are investigated. For the aspherical explosions, we explore asymmetric 
bipolar explosions characterized by the parameters, $\alpha \equiv v_{\rm up}/v_{\rm down}$ and $\beta \equiv v_{\rm pol}/v_{\rm eq}$. As one of the progenitor 
models, the BSG pre-supernova model for Sk $-$69$^{\circ}$ 202 based on the slow-merger scenario (b18.3) is adopted in addition to existing single star models, one BSG 
model (n16.3) and the other two RSG models (s18.0 and s19.8). From the simulations results, the radial velocity distribution 
of elements, in particular $^{56}$Ni, and the distribution of the line of sight velocity of $^{56}$Ni are mainly discussed by comparing with the constraints on the 
mass of high velocity $^{56}$Ni and observed [Fe II] line profiles for SN 1987A. First we perform one-dimensional simulations in order to see the pre-supernova 
model dependence of the matter mixing, where the growth factors of instabilities are presented (Section~\ref{subsec:res-1d}). Next, we explore the dependence on 
the parameters of the aspherical explosion and the pre-supernova models based on many lower resolution simulations 
(Section~\ref{subsec:res-3d-low}). Then, with the fiducial (best) parameter set for the explosion, two high resolution simulations, one with the binary merger progenitor model 
b18.3 and the other with the single star progenitor model n16.3, are performed (Section~\ref{subsec:res-3d-high}). Finally, some implications from the results, the key 
properties of the pre-supernova models for the matter mixing (Section~\ref{subsec:depend-prog}), explosion asymmetries and possible explosion mechanisms for 
SN 1987A (Section~\ref{subsec:exp-mechanism}), NS kick velocities (Section~\ref{subsec:ns-kick}), and the impacts of possible large density perturbations in the 
pre-supernova models, are presented (Section~\ref{subsec:large-dens-perturb}). 
Here, the findings and main points in this paper are summarized. 

1. From the analysis of growth factors of instabilities based on one-dimensional simulations, instabilities grow around both the C+O/He and He/H interfaces for the 
two BSG progenitor models (b18.3 and n16.3). On the other hand, instabilities are developed only around the He/H interfaces for the two RSG progenitor models (s18.0 and s19.8), 
which is attributed to the fact that the gradients of $\rho \,r^3$ profile in the helium layer are overall negative for the RSG models in contrast to the case of the BSG 
models. 

2. Initial asphericities of explosions affect early ($\sim$ 1 sec) morphologies of the inner ejecta. As expected, the larger $\beta$ value is, the narrower the bipolar 
structure is. However, compared with the initial radial velocity distributions, the morphologies of the bipolar structure at around $\sim$ 1 sec become wider and 
less distinct due to the deceleration during the shock propagation inside the C+O core.   

3. The early morphologies of the expanding ejecta depend on the structures in the pre-supernova models. At an early phase ($\sim$ 1 sec), depending on the 
gradients of the $\rho \,r^3$ profiles of the C+O and/or the silicon layers, pre-supernova models with steeper $\rho \,r^3$ gradients results in a wider 
bipolar structure in the early morphology of the explosion due to stronger decelerations than those for progenitor models with flatter or negative gradients. 
The BSG progenitor model based on the single star evolution (n16.3) results in the widest bipolar structure in the ejecta. 

4. Later morphologies of the expanding ejecta and the distributions of elements also depend on the structures of the helium and the hydrogen layers of the pre-supernova models. 
The distributions of lower mass elements, e.g., $^{16}$O and $^{4}$He, depend on whether the reverse shock developed during the 
shock propagation in the hydrogen layer sweeps up the inner ejecta or not. In the BSG model based on the binary merger evolution (b18.3), the reverse 
shock last developed sweeps up the inner ejecta completely by the time of the shock breakout. Consequently, the distributions of $^{16}$O and $^{4}$He are 
more concentrated around the equatorial plane. 

5. Among the investigated explosion models, the models with the pre-supernova model n16.3 fail to fulfil simultaneously the two observational constraints on the 
mass of the high velocity $^{56}$Ni, i.e., i)  $M_{3.0}/M_{\rm ej}$ ($^{56}$Ni) $\geq$ 4\% and ii) $M_{4.0}$ ($^{56}$Ni) $\geq$ 10$^{-3}$ $M_{\odot}$ 
(see Table~\ref{table:results} and Figure~\ref{fig:table3}). 
On the other hand, some explosion models with the other pre-supernova models succeed to fulfil the observational constraints for the case of larger $\beta$ values 
(8 or 16).  

6. If the explosion models with RSG models and with extreme explosion energies are excluded from the point of view of the observational constraints on the progenitor of 
SN 1987A, Sk $-$69$^{\circ}$ 202, and its explosion, the best model in this paper is the model b18.3-high in which the binary merger progenitor model b18.3 
\citep{2018MNRAS.473L.101U} and the parameter set of ($\alpha$, $\beta$) = (1.5, 16) are adopted. In the best model, the obtained explosion energy, 
$E_{\rm exp}$, is $\sim$ 2 $\times$ 10$^{51}$ erg. 

7. The obtained values related to the observational constraints on the mass of the high 
velocity $^{56}$Ni, $M_{3.0}/M_{\rm ej}$ ($^{56}$Ni) and $M_{4.0}$ ($^{56}$Ni), for the best explosion model b18.3-high are 10.5\% and 1.1 $\times$ 10$^{-3}$ $M_{\odot}$, 
respectively. The values for the counterpart model n16.3-high, in which the single star progenitor model n16.3 is adopted, are 3.7\% and 2.5 $\times$ 10$^{-4}$ 
$M_{\odot}$, respectively. 

8. The distribution of the line of sight velocity of $^{56}$Ni for the model b18.3-high best reproduces the high velocity tails of the observed [Fe II] line profiles in 
particular at the redshifted side with the angles of ($\theta$, $\chi$) = (130$^{\circ}$, 10$^{\circ}$)~(see Figure~\ref{fig:geo} for the definition of the angles and 
Figure~\ref{fig:vel-dir-high} for the distribution). The distribution for the counterpart model n16.3 apparently lacks the tail at the redshifted side. 

9. The key to obtain such high velocity $^{56}$Ni is the penetration of $^{56}$Ni through the helium layer to reach the hydrogen envelope before the development 
of the strong reverse shock during the shock propagation in the helium layer, which is consistent with the findings in \citet{2015A&A...577A..48W}. 

10. To realize the penetration of $^{56}$Ni through the helium layer, the structures of the C+O and the helium layers are important. At least among the existing BSG 
progenitor models including the models b18.3 and n16.3, the helium core mass $M_{\rm He,c}$ or the mass ratio of the helium core to the stellar mass $q \equiv 
M_{\rm He,c}/M$ appears to be a useful indicator for the successful penetration of $^{56}$Ni 
as seen in previous studies \citep{2015A&A...577A..48W,2015A&A...581A..40U,2019A&A...624A.116U}. 
The value of $M_{\rm He,c}$ of the pre-supernova model for the best explosion model 
b18.3-high is about $\sim$ 4 $M_{\odot}$ ($q$ $\sim$ 0.2). On the other hand, the value for the other BSG model n16.3 is $\sim$ 6 $M_{\odot}$ ($q$ $\sim$ 0.37). 

11. It seems difficult to find such a small $M_{\rm He,c}$ value among the existing BSG progenitor models based on the single star evolution that satisfy both the observed 
luminosity and the effective temperature (the final position in the HR diagram) of Sk $-$69$^{\circ}$ 202. On the other hand, the existing BSG 
models based on the binary merger scenario that satisfy these values naturally have small $M_{\rm He,c}$ values of $\lesssim$ 4 $M_{\odot}$ 
($q$ $\lesssim$ 0.2) \citep{2017MNRAS.469.4649M,2018MNRAS.473L.101U}, which may reflect the nature of merging processes, the penetration of the 
secondary to the envelope of the primary and the dredge up of the primary's core material into the envelope. 

12. From the adopted parameter set of the best explosion model b18.3-high, the explosion of SN 1987A is likely to be a asymmetric bipolar (jetlike) explosion, 
which may be induced by magnetorotational effects \citep[e.g.,][]{2009ApJ...691.1360T,2013ApJ...764...10S,2014ApJ...785L..29M} 
or the combination of the neutrino heating and the magnetorotational effects \citep{2016ApJ...817..153S}. In order to deduce the explosion mechanism in a more 
robust way, observations of spatially resolved line emissions from iron or direct $\gamma$-ray lines from the decay of $^{44}$Ti are desirable. Recent observations 
of the 3D distribution of CO and SiO molecules \citep{2017ApJ...842L..24A} will shed light on the explosion mechanism. 

13. From the asymmetric bipolar explosions presented in this paper, NS kicks are expected as in \citet{2013A&A...552A.126W}. The absolute value of the estimated 
NS kick velocity, $v_{\rm NS}$, for the best model b18.3 is $\sim$ 300 km s$^{-1}$. The values for the other models are roughly in the range of 250--580 km s$^{-1}$ 
(except for the models b18.3-alpha1 and b18.3-ein1.5), which are consistent with the NS kick velocities deduced from the proper motions of young pulsars 
\citep[e.g.,][]{2005MNRAS.360..974H,2006ApJ...643..332F}. It is found that the direction of the NS kick is almost opposite to the bipolar (strongest) explosion axis. 
From the angles suggested by the best model b18.3-high, ($\theta$, $\chi$) = (130$^{\circ}$, 10$^{\circ}$), we predict the compact object of SN 1987A will be detected 
is the north part of the inner ejecta as opposed to the direction of the redshifted side of the explosion, which corresponds to the stronger explosion direction and it is 
directed to the south side (see Figure~\ref{fig:geo}). 

14.  As investigated in Paper~II, possible large density fluctuations with amplitude up to 50\% in the pre-supernova model could aid for the inner ejecta to penetrate 
through the helium layer due to strong RT instabilities. Hitherto, whether such large amplitude of fluctuations can be introduced in the density of pre-supernova models 
or not has not been unveiled because of the lack of appropriate theoretical modeling of multi-dimensional effects such as turbulent convection, in particular for the 
envelope at the last stage before the core-collapse. It is worth investigating such effects (partly) motivated by the recent explorations of the impact of pre-collapse 
asphericities on the core-collapse supernova explosions \citep[e.g.,][]{2015ApJ...799....5C,2017MNRAS.472..491M}. 

We plan to make use of the results of the model b18.3-high and n16.3-high as initial conditions of MHD simulations of 
the further evolution (up to $\sim$ 50 yr for SN 1987A)~(Orlando et al. 2019, in preparation), which is a natural extension of our previous investigations with spherically 
symmetric explosions \citep{2015ApJ...810..168O,2019A&A...622A..73O,2019NatAs...3..236M}. In the coming paper, we will discuss not only the X-ray emission (the 
light curve and the images) but also the distributions of CO and SiO molecules motivated by the recently observed 3D distributions of the molecules 
\citep{2017ApJ...842L..24A}. Additionally, we plan to investigate the molecule and/or dust formations in detail based on the 3D models in this paper 
in the near future (Ono et al., in preparation). 

\acknowledgments
We would like to thank  M. Barkov, D. C. Warren, and T. Nozawa for fruitful discussion on this work. 
The software used in this work was in part developed by the DOE NNSA-ASC OASCR Flash Center
at the University of Chicago. The numerical computations were carried out complementarily on XC40 (YITP, Kyoto University), 
Cray XC50 (Center for Computational Astrophysics, National Astronomical Observatory of Japan), HOKUSAI (RIKEN). 
This research also used computational resources of the K computer provided by the RIKEN Center for Computational Science 
through the HPCI System Research project (Project ID:hp180281) and the PRACE Research Infrastructure resource Marconi based in Italy at CINECA (PRACE Award N.2016153460). 
This work is supported by JSPS KAKENHI Grant Numbers JP26800141 and JP19H00693. MO and SN thank the supports from 
RIKEN Interdisciplinary Theoretical and Mathematical Sciences Program. SN also acknowledges the supports from Pioneering Program of RIKEN for Evolution of Matter in the Universe (r-EMU). 


\appendix

\section{Explosion asymmetries: initial radial velocity distributions} \label{sec:app1}

In the simulation, thermal and kinetic energies are artificially injected around the interface between the iron core and the silicon layer of 
the pre-supernova models. In order to initiate an asymmetric (non-spherical) explosion, initial radial velocities are distributed with arbitrary 
functions of $\theta$ in the spherical coordinates, $(r,\theta,\phi)$. 
In this paper, we assume the form of the initial radial velocities as $v_r \propto r \,f(\theta)$. 
Since observations of SN 1987A have shown that the inner ejecta is globally elliptical \citep{2010A&A...517A..51K,2013ApJ...768...89L,2016ApJ...833..147L}, 
bipolar-like explosions may be justified. Then, the following four cases for the shape of $f(\theta)$ ($0 \leq \theta \leq \pi$) are considered. 
In the four cases, the concentration of higher initial radial velocities around the polar axis is controlled by a parameter, 
$\beta \equiv v_{\rm pol}/ v_{\rm eq}$, where $v_{\rm pol}$ is the initial radial velocity on the polar axis ($\theta$ = 0) and $v_{\rm eq}$ 
is one on the equatorial plane ($\theta$ = $\pi$/2) at a same radius, $r$.

Case 1: a function with cosine (``cos" in Table~\ref{table:models}.) as,
\begin{equation}
f(\theta) = \frac{1 + \cos (2\theta)}{1 + \xi} 
\label{eq:vcos}
\end{equation}
where $\xi$ is a parameter related to $\beta$ with the relation, 
$\beta = (1+\xi)/ (1-\xi)$. This form of asymmetry was adopted in \citet{1998ApJ...495..413N} and Paper~I. 

Case 2: an exponential form (``exponential" in Table~\ref{table:models}.) as,
\begin{equation}
f(\theta) =  \exp\left( - \vartheta /d\right), \ \ \ d = \frac{\pi}{2 \ln \beta}, \ \ \ 
\vartheta =
\begin{cases}
\theta & \ \ \ (\theta \leq \pi/2) \\
\pi - \theta & \ \ \ (\theta > \pi/2)
\end{cases} 
.
\label{eq:vexp}
\end{equation}

Case 3: a power-law like form (``power" in Table~\ref{table:models}.) as,
\begin{equation}
f(\theta) = \left( 2 - \frac{2}{\pi} \, \vartheta \right)^{\gamma}, \ \ \ \gamma = \log_2 \beta \ \ \ 
\vartheta =
\begin{cases}
\theta & \ \ \ (\theta \leq \pi/2) \\
\pi - \theta & \ \ \ (\theta > \pi/2)
\end{cases} 
.
\label{eq:vpwr}
\end{equation}

Case 4: an elliptical form (``elliptical" in Table~\ref{table:models}.) as,
\begin{equation}
f(\theta) = \left(\beta^{-1} \cos^2 \theta + \beta \sin^2 \theta \right)^{-1/2}.
\label{eq:vell}
\end{equation}

We also introduce an asymmetry in the initial radial velocities across the equatorial plane ($x$--$y$). Such asymmetry could be 
introduced if an explosion is driven by a neutrino heating aided by the SASI of low-order unstable mode ($l = 1$) 
\citep[e.g.,][]{2006A&A...457..963S,2010PASJ...62L..49S,2013ApJ...770...66H,2016ApJ...817...72P} and could 
trigger a NS kick inferred from the observations of young supernova remnants \citep{2018ApJ...856...18K}. 
The initial radial velocities in the upper hemisphere are manually enhanced by multiplying the factor of $\alpha$ $\equiv$ $v_{\rm up}/v_{\rm down}$
and the velocities around the equatorial plane are smoothed so as not to introduce a jump across the plane. 
We denote the angle dependences after a normalization as $g(\theta)$, where the function is normalized in order for the maximum value to be unity. 
The angle dependences, $g(\theta)$, are shown in Figure~\ref{fig:shape}. The cases of $\beta = 2, 4, 8, 16$ with $\alpha = 1.5$ are displayed in the $x$--$z$ 
plane of the Cartesian coordinate system. As one can see, for example, the distribution of the elliptical case with $\beta$ = 16 (bottom right panel) invokes rather a jetlike explosion. 
Then, at this moment, we do not assume a specific explosion mechanism for this initial radial velocity distribution. 

\begin{figure*}[htb]
\begin{tabular}{cc}
\begin{minipage}{0.5\hsize}
\begin{center}
\includegraphics[width=7cm,keepaspectratio,clip]{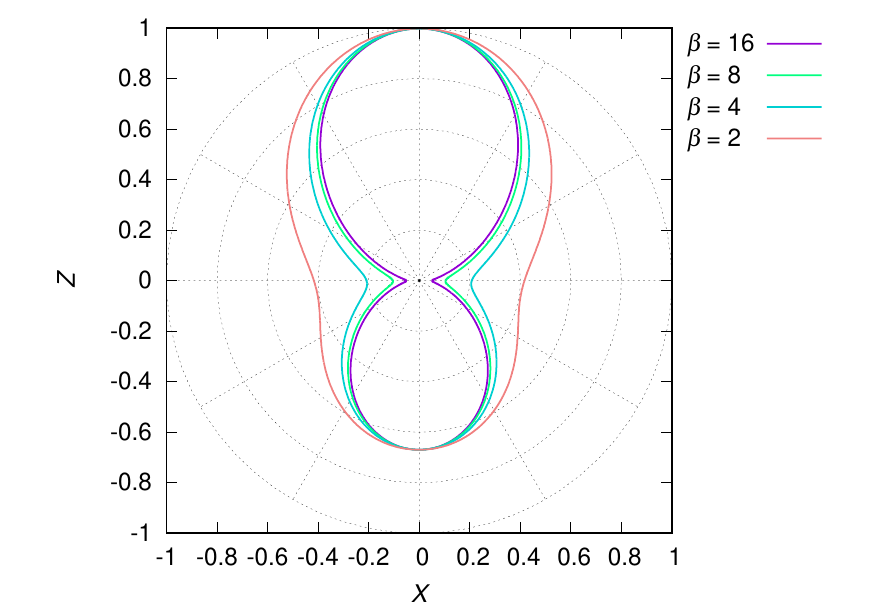}
\end{center}
\end{minipage}
\begin{minipage}{0.5\hsize}
\begin{center}
\includegraphics[width=7cm,keepaspectratio,clip]{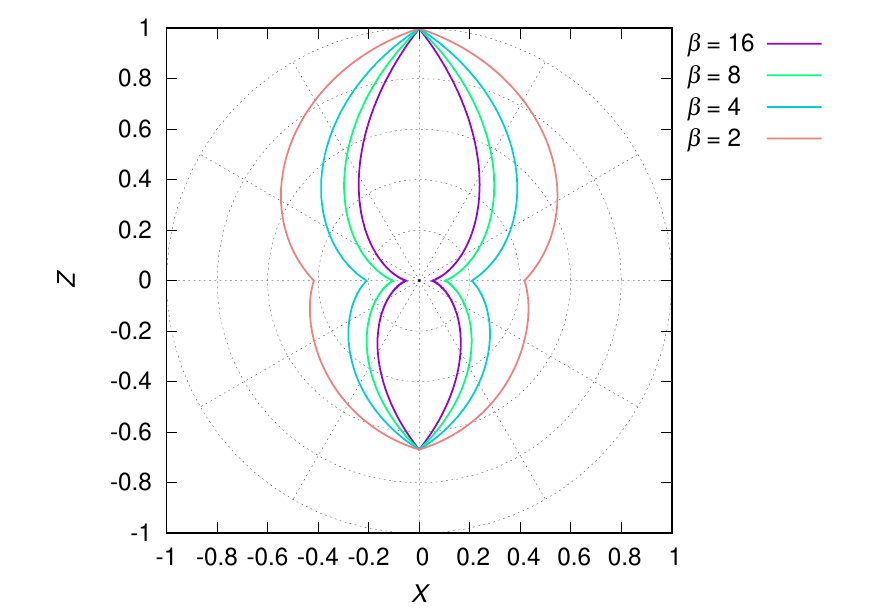}
\end{center}
\end{minipage}
\\
\begin{minipage}{0.5\hsize}
\begin{center}
\includegraphics[width=7cm,keepaspectratio,clip]{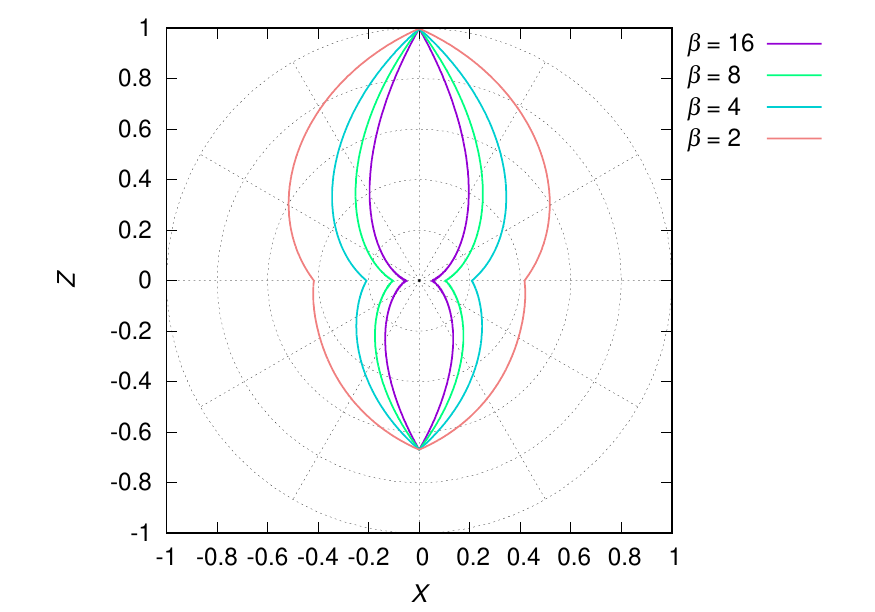}
\end{center}
\end{minipage}
\begin{minipage}{0.5\hsize}
\begin{center}
\includegraphics[width=7cm,keepaspectratio,clip]{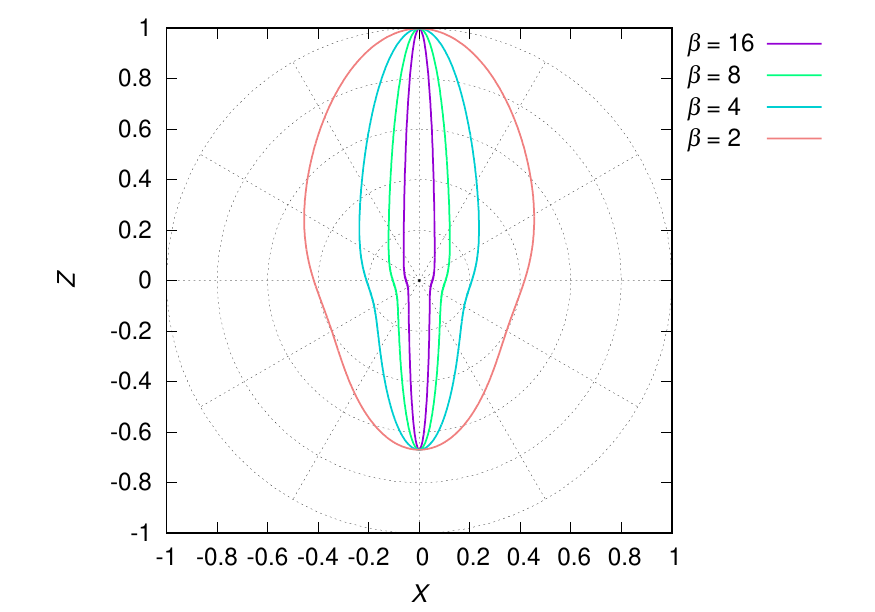}
\end{center}
\end{minipage}
\end{tabular}
\caption{Angle dependences of the initial radial velocities, $g(\theta)$. The curved surfaces of $r = g(\theta)$ are plotted in the $x$--$z$ plane of the Cartesian coordinate. 
Top left, top right, bottom left, and bottom right panels show the four cases of Eqs.~(\ref{eq:vcos}), (\ref{eq:vexp}), (\ref{eq:vpwr}), and (\ref{eq:vell}), respectively. Different 
solid lines represent different cases of $\beta \equiv v_{\rm pol}/v_{\rm eq}$. The case of $\alpha \equiv v_{\rm up}/v_{\rm down} = 1.5$ is shown.}
\label{fig:shape}
\end{figure*}

\section{Method for introducing the initial clumpy structures} \label{sec:app2}

In our previous study on the matter mixing based on two-dimensional hydrodynamic simulations (Paper~I), 
motivated by neutrino-driven explosions aided by SASI and/or convection, which have clumpy or bubble-like structures inside the supernova shock wave, fluctuations were 
introduced in the initial radial velocities by multiplying the following factor function of $\theta$: 
\begin{equation}
1 + \sum^{4}_{n=1} \frac{\epsilon}{2^{n-1}} \sin(m \, n \, \theta ),
\label{eq:clump1}
\end{equation}
where $\epsilon$ is the amplitude of the fluctuations and $m$ is an integer parameter. In the best model for SN 1987A in Paper~I (AM2), 
$m = 15$ was adopted. In this paper, the simulations are three-dimensional. Thus, we introduce such fluctuations with another function of ($\theta$, $\phi$). We utilize real 
spherical harmonics for the function:
\begin{equation}
Y^m_l(\theta,\phi) = \begin{cases} (-1)^m \sqrt{2} \sqrt{\frac{2l+1}{4\pi}\frac{(l-|m|)!}{(l+|m|)!}} P^{\,|m|}_l(\cos \theta) \sin(|m|\phi) & (m < 0) \vspace*{0.2cm} \\
\sqrt{\frac{2l+1}{4\pi}} P^{\,m}_l(\cos \theta) & (m = 0) \vspace*{0.2cm} \\
(-1)^m \sqrt{2} \sqrt{\frac{2l+1}{4\pi}\frac{(l-m)!}{(l+m)!}} P^{\,m}_l(\cos \theta) \cos(m \phi) & (m > 0).
\end{cases}
\label{eq:sphel_harm}
\end{equation}
With the function above, the following factor is multiplied to the initial radial velocities introduced in Appendix A. 
\begin{equation}
1 + \epsilon \left[ N \sum_{n = 1}^{4} \sum_{m = -l}^{l} \frac{A^m_l(\theta,\phi)}{n} \right] \ \ \ \ (l = n \cdot l_{\rm base}),
\label{eq:fluc}
\end{equation}
where $\epsilon$ is the amplitude of the fluctuations. In order to introduce fluctuations of similar sizes and amplitudes as in the best model (AM2) in Paper~I, $l_{\rm base} = 15$ 
and $\epsilon = 30 \%$ are adopted. 
It is noted that non-radial fluctuations with the amplitude of $\sim 30 \%$ could be introduced 
as seen in a 2D hydrodynamical simulation of a neutrino-driven core-collapse supernova explosion \citep[see Fig.~11 in][]{2010A&A...521A..38G}. 
$N$ is a normalization factor for the maximum value inside the square bracket in Eq.~(\ref{eq:fluc}) to be unity. 
The function, $A^m_l(\theta,\phi)$, is basically the function in Eq.~(\ref{eq:sphel_harm}), but 
depending on the numbers of $l$ and $m$, some values are arbitrarily set to be zero (some $m$ modes are arbitrarily selected) as follows:
%
\begin{equation}
A^m_l(\theta,\phi) = 
\begin{cases} 
\begin{cases} Y^m_l(\theta,\phi) & (m = 1, -3, 5, -7,..) \\ 0 & (\mbox{else}) \end{cases} & (l: \mbox{odd})  \vspace{0.5em} \\
\begin{cases} Y^m_l(\theta,\phi) & (m = 0, 2, -4, 6,..)  \\ 0 & (\mbox{else}) \end{cases} & (l: \mbox{even}).
\end{cases}
\end{equation}
There is no physical base for the selection of $m$ modes but it is noted that if we set $A^m_l(\theta,\phi)$ to be $Y^m_l(\theta,\phi)$, the distribution of the fluctuations seems 
to be unrealistic (nearly axisymmetric stripes are recognized). In this paper, we do not intend to discuss the effects of the specific form of the fluctuations 
but just the effects of the existence of such initial fluctuations is briefly argued by comparing one of the models with the fluctuations (b18.3-fid) and one without them (b18.3-clp0). 

%

\bibliography{ref}

\begin{thebibliography}{}
\expandafter\ifx\csname natexlab\endcsname\relax\def\natexlab#1{#1}\fi
\providecommand{\url}[1]{\href{#1}{#1}}
\providecommand{\dodoi}[1]{doi:~\href{http://doi.org/#1}{\nolinkurl{#1}}}
\providecommand{\doeprint}[1]{\href{http://ascl.net/#1}{\nolinkurl{http://ascl.net/#1}}}
\providecommand{\doarXiv}[1]{\href{https://arxiv.org/abs/#1}{\nolinkurl{https://arxiv.org/abs/#1}}}

\bibitem[{{Abell{\'a}n} {et~al.}(2017){Abell{\'a}n}, {Indebetouw}, {Marcaide},
  {Gabler}, {Fransson}, {Spyromilio}, {Burrows}, {Chevalier}, {Cigan},
  {Gaensler}, {Gomez}, {Janka}, {Kirshner}, {Larsson}, {Lundqvist}, {Matsuura},
  {McCray}, {Ng}, {Park}, {Roche}, {Staveley-Smith}, {van Loon}, {Wheeler}, \&
  {Woosley}}]{2017ApJ...842L..24A}
{Abell{\'a}n}, F.~J., {Indebetouw}, R., {Marcaide}, J.~M., {et~al.} 2017,
  \apjl, 842, L24, \dodoi{10.3847/2041-8213/aa784c}

\bibitem[{{Ahmad} {et~al.}(2006){Ahmad}, {Greene}, {Moore}, {Ghelberg}, {Ofan},
  {Paul}, \& {Kutschera}}]{2006PhRvC..74f5803A}
{Ahmad}, I., {Greene}, J.~P., {Moore}, E.~F., {et~al.} 2006, \prc, 74, 065803,
  \dodoi{10.1103/PhysRevC.74.065803}

\bibitem[{{Alp} {et~al.}(2018){Alp}, {Larsson}, {Fransson}, {Indebetouw},
  {Jerkstrand}, {Ahola}, {Burrows}, {Challis}, {Cigan}, {Cikota}, {Kirshner},
  {van Loon}, {Mattila}, {Ng}, {Park}, {Spyromilio}, {Woosley}, {Baes},
  {Bouchet}, {Chevalier}, {Frank}, {Gaensler}, {Gomez}, {Janka}, {Leibundgut},
  {Lundqvist}, {Marcaide}, {Matsuura}, {Sollerman}, {Sonneborn},
  {Staveley-Smith}, {Zanardo}, {Gabler}, {Taddia}, \&
  {Wheeler}}]{2018ApJ...864..174A}
{Alp}, D., {Larsson}, J., {Fransson}, C., {et~al.} 2018, \apj, 864, 174,
  \dodoi{10.3847/1538-4357/aad739}

\bibitem[{{Alp} {et~al.}(2019){Alp}, {Larsson}, {Maeda}, {Fransson},
  {Wongwathanarat}, {Gabler}, {Janka}, {Jerkstrand}, {Heger}, \&
  {Menon}}]{2019ApJ...882...22A}
{Alp}, D., {Larsson}, J., {Maeda}, K., {et~al.} 2019, \apj, 882, 22,
  \dodoi{10.3847/1538-4357/ab3395}

\bibitem[{{Arnett} {et~al.}(1989{\natexlab{a}}){Arnett}, {Fryxell}, \&
  {Mueller}}]{1989ApJ...341L..63A}
{Arnett}, D., {Fryxell}, B., \& {Mueller}, E. 1989{\natexlab{a}}, \apjl, 341,
  L63, \dodoi{10.1086/185458}

\bibitem[{{Arnett} {et~al.}(2009){Arnett}, {Meakin}, \&
  {Young}}]{2009ApJ...690.1715A}
{Arnett}, D., {Meakin}, C., \& {Young}, P.~A. 2009, \apj, 690, 1715,
  \dodoi{10.1088/0004-637X/690/2/1715}

\bibitem[{{Arnett} {et~al.}(1989{\natexlab{b}}){Arnett}, {Bahcall}, {Kirshner},
  \& {Woosley}}]{1989ARA&A..27..629A}
{Arnett}, W.~D., {Bahcall}, J.~N., {Kirshner}, R.~P., \& {Woosley}, S.~E.
  1989{\natexlab{b}}, \araa, 27, 629,
  \dodoi{10.1146/annurev.aa.27.090189.003213}

\bibitem[{{Arnett} \& {Meakin}(2011{\natexlab{a}})}]{2011ApJ...733...78A}
{Arnett}, W.~D., \& {Meakin}, C. 2011{\natexlab{a}}, \apj, 733, 78,
  \dodoi{10.1088/0004-637X/733/2/78}

\bibitem[{{Arnett} \& {Meakin}(2011{\natexlab{b}})}]{2011ApJ...741...33A}
---. 2011{\natexlab{b}}, \apj, 741, 33, \dodoi{10.1088/0004-637X/741/1/33}

\bibitem[{{Arnett} {et~al.}(2015){Arnett}, {Meakin}, {Viallet}, {Campbell},
  {Lattanzio}, \& {Moc{\'a}k}}]{2015ApJ...809...30A}
{Arnett}, W.~D., {Meakin}, C., {Viallet}, M., {et~al.} 2015, \apj, 809, 30,
  \dodoi{10.1088/0004-637X/809/1/30}

\bibitem[{{Arzoumanian} {et~al.}(2002){Arzoumanian}, {Chernoff}, \&
  {Cordes}}]{2002ApJ...568..289A}
{Arzoumanian}, Z., {Chernoff}, D.~F., \& {Cordes}, J.~M. 2002, \apj, 568, 289,
  \dodoi{10.1086/338805}

\bibitem[{{Balbus} \& {Hawley}(1998)}]{1998RvMP...70....1B}
{Balbus}, S.~A., \& {Hawley}, J.~F. 1998, Reviews of Modern Physics, 70, 1,
  \dodoi{10.1103/RevModPhys.70.1}

\bibitem[{{Bandiera}(1984)}]{1984A&A...139..368B}
{Bandiera}, R. 1984, \aap, 139, 368

\bibitem[{{Bazan} \& {Arnett}(1994)}]{1994ApJ...433L..41B}
{Bazan}, G., \& {Arnett}, D. 1994, \apjl, 433, L41, \dodoi{10.1086/187543}

\bibitem[{{Baz{\'a}n} \& {Arnett}(1998)}]{1998ApJ...496..316B}
{Baz{\'a}n}, G., \& {Arnett}, D. 1998, \apj, 496, 316, \dodoi{10.1086/305346}

\bibitem[{{Benz} \& {Thielemann}(1990)}]{1990ApJ...348L..17B}
{Benz}, W., \& {Thielemann}, F.-K. 1990, \apjl, 348, L17,
  \dodoi{10.1086/185620}

\bibitem[{{Bisnovatyi-Kogan}(1993)}]{1993A&AT....3..287B}
{Bisnovatyi-Kogan}, G.~S. 1993, Astronomical and Astrophysical Transactions, 3,
  287, \dodoi{10.1080/10556799308230566}

\bibitem[{{Blinnikov} {et~al.}(2000){Blinnikov}, {Lundqvist}, {Bartunov},
  {Nomoto}, \& {Iwamoto}}]{2000ApJ...532.1132B}
{Blinnikov}, S., {Lundqvist}, P., {Bartunov}, O., {Nomoto}, K., \& {Iwamoto},
  K. 2000, \apj, 532, 1132, \dodoi{10.1086/308588}

\bibitem[{{Blondin} {et~al.}(2003){Blondin}, {Mezzacappa}, \&
  {DeMarino}}]{2003ApJ...584..971B}
{Blondin}, J.~M., {Mezzacappa}, A., \& {DeMarino}, C. 2003, \apj, 584, 971,
  \dodoi{10.1086/345812}

\bibitem[{{Boggs} {et~al.}(2015){Boggs}, {Harrison}, {Miyasaka},
  {Grefenstette}, {Zoglauer}, {Fryer}, {Reynolds}, {Alexander}, {An}, {Barret},
  {Christensen}, {Craig}, {Forster}, {Giommi}, {Hailey}, {Hornstrup},
  {Kitaguchi}, {Koglin}, {Madsen}, {Mao}, {Mori}, {Perri}, {Pivovaroff},
  {Puccetti}, {Rana}, {Stern}, {Westergaard}, \& {Zhang}}]{2015Sci...348..670B}
{Boggs}, S.~E., {Harrison}, F.~A., {Miyasaka}, H., {et~al.} 2015, Science, 348,
  670, \dodoi{10.1126/science.aaa2259}

\bibitem[{{B{\"o}hm-Vitense}(1958)}]{1958ZA.....46..108B}
{B{\"o}hm-Vitense}, E. 1958, \zap, 46, 108

\bibitem[{{Brisken} {et~al.}(2003){Brisken}, {Fruchter}, {Goss}, {Herrnstein},
  \& {Thorsett}}]{2003AJ....126.3090B}
{Brisken}, W.~F., {Fruchter}, A.~S., {Goss}, W.~M., {Herrnstein}, R.~M., \&
  {Thorsett}, S.~E. 2003, \aj, 126, 3090, \dodoi{10.1086/379559}

\bibitem[{{Bruenn} {et~al.}(2013){Bruenn}, {Mezzacappa}, {Hix}, {Lentz},
  {Messer}, {Lingerfelt}, {Blondin}, {Endeve}, {Marronetti}, \&
  {Yakunin}}]{2013ApJ...767L...6B}
{Bruenn}, S.~W., {Mezzacappa}, A., {Hix}, W.~R., {et~al.} 2013, \apjl, 767, L6,
  \dodoi{10.1088/2041-8205/767/1/L6}

\bibitem[{{Bruenn} {et~al.}(2016){Bruenn}, {Lentz}, {Hix}, {Mezzacappa},
  {Harris}, {Messer}, {Endeve}, {Blondin}, {Chertkow}, {Lingerfelt},
  {Marronetti}, \& {Yakunin}}]{2016ApJ...818..123B}
{Bruenn}, S.~W., {Lentz}, E.~J., {Hix}, W.~R., {et~al.} 2016, \apj, 818, 123,
  \dodoi{10.3847/0004-637X/818/2/123}

\bibitem[{{Burrows}(2013)}]{2013RvMP...85..245B}
{Burrows}, A. 2013, Reviews of Modern Physics, 85, 245,
  \dodoi{10.1103/RevModPhys.85.245}

\bibitem[{{Burrows} {et~al.}(2007){Burrows}, {Dessart}, {Livne}, {Ott}, \&
  {Murphy}}]{2007ApJ...664..416B}
{Burrows}, A., {Dessart}, L., {Livne}, E., {Ott}, C.~D., \& {Murphy}, J. 2007,
  \apj, 664, 416, \dodoi{10.1086/519161}

\bibitem[{{Burrows} {et~al.}(1995{\natexlab{a}}){Burrows}, {Hayes}, \&
  {Fryxell}}]{1995ApJ...450..830B}
{Burrows}, A., {Hayes}, J., \& {Fryxell}, B.~A. 1995{\natexlab{a}}, \apj, 450,
  830, \dodoi{10.1086/176188}

\bibitem[{{Burrows} {et~al.}(2019){Burrows}, {Radice}, \&
  {Vartanyan}}]{2019MNRAS.485.3153B}
{Burrows}, A., {Radice}, D., \& {Vartanyan}, D. 2019, \mnras, 485, 3153,
  \dodoi{10.1093/mnras/stz543}

\bibitem[{{Burrows} {et~al.}(1995{\natexlab{b}}){Burrows}, {Krist}, {Hester},
  {Sahai}, {Trauger}, {Stapelfeldt}, {Gallagher}, {Ballester}, {Casertano},
  {Clarke}, {Crisp}, {Evans}, {Griffiths}, {Hoessel}, {Holtzman}, {Mould},
  {Scowen}, {Watson}, \& {Westphal}}]{1995ApJ...452..680B}
{Burrows}, C.~J., {Krist}, J., {Hester}, J.~J., {et~al.} 1995{\natexlab{b}},
  \apj, 452, 680, \dodoi{10.1086/176339}

\bibitem[{{Canuto} \& {Dubovikov}(1998)}]{1998ApJ...493..834C}
{Canuto}, V.~M., \& {Dubovikov}, M. 1998, \apj, 493, 834,
  \dodoi{10.1086/305141}

\bibitem[{{Chatterjee} {et~al.}(2005){Chatterjee}, {Vlemmings}, {Brisken},
  {Lazio}, {Cordes}, {Goss}, {Thorsett}, {Fomalont}, {Lyne}, \&
  {Kramer}}]{2005ApJ...630L..61C}
{Chatterjee}, S., {Vlemmings}, W.~H.~T., {Brisken}, W.~F., {et~al.} 2005,
  \apjl, 630, L61, \dodoi{10.1086/491701}

\bibitem[{{Chatzopoulos} {et~al.}(2016){Chatzopoulos}, {Couch}, {Arnett}, \&
  {Timmes}}]{2016ApJ...822...61C}
{Chatzopoulos}, E., {Couch}, S.~M., {Arnett}, W.~D., \& {Timmes}, F.~X. 2016,
  \apj, 822, 61, \dodoi{10.3847/0004-637X/822/2/61}

\bibitem[{{Chevalier}(1976)}]{1976ApJ...207..872C}
{Chevalier}, R.~A. 1976, \apj, 207, 872, \dodoi{10.1086/154557}

\bibitem[{{Chiavassa} {et~al.}(2010){Chiavassa}, {Haubois}, {Young}, {Plez},
  {Josselin}, {Perrin}, \& {Freytag}}]{2010A&A...515A..12C}
{Chiavassa}, A., {Haubois}, X., {Young}, J.~S., {et~al.} 2010, \aap, 515, A12,
  \dodoi{10.1051/0004-6361/200913907}

\bibitem[{{Chita} {et~al.}(2008){Chita}, {Langer}, {van Marle},
  {Garc{\'{\i}}a-Segura}, \& {Heger}}]{2008A&A...488L..37C}
{Chita}, S.~M., {Langer}, N., {van Marle}, A.~J., {Garc{\'{\i}}a-Segura}, G.,
  \& {Heger}, A. 2008, \aap, 488, L37, \dodoi{10.1051/0004-6361:200810087}

\bibitem[{{Cigan} {et~al.}(2019){Cigan}, {Matsuura}, {Gomez}, {Indebetouw},
  {Abell{\'a}n}, {Gabler}, {Richards}, {Alp}, {Davis}, {Janka}, {Spyromilio},
  {Barlow}, {Burrows}, {Dwek}, {Fransson}, {Gaensler}, {Larsson}, {Bouchet},
  {Lundqvist}, {Marcaide}, {Ng}, {Park}, {Roche}, {van Loon}, {Wheeler}, \&
  {Zanardo}}]{2019arXiv191002960C}
{Cigan}, P., {Matsuura}, M., {Gomez}, H.~L., {et~al.} 2019, arXiv e-prints,
  arXiv:1910.02960.
\newblock \doarXiv{1910.02960}

\bibitem[{{Claret}(2007)}]{2007A&A...475.1019C}
{Claret}, A. 2007, \aap, 475, 1019, \dodoi{10.1051/0004-6361:20078024}

\bibitem[{{Colgan} {et~al.}(1994){Colgan}, {Haas}, {Erickson}, {Lord}, \&
  {Hollenbach}}]{1994ApJ...427..874C}
{Colgan}, S.~W.~J., {Haas}, M.~R., {Erickson}, E.~F., {Lord}, S.~D., \&
  {Hollenbach}, D.~J. 1994, \apj, 427, 874, \dodoi{10.1086/174193}

\bibitem[{{Couch} {et~al.}(2015){Couch}, {Chatzopoulos}, {Arnett}, \&
  {Timmes}}]{2015ApJ...808L..21C}
{Couch}, S.~M., {Chatzopoulos}, E., {Arnett}, W.~D., \& {Timmes}, F.~X. 2015,
  \apjl, 808, L21, \dodoi{10.1088/2041-8205/808/1/L21}

\bibitem[{{Couch} \& {O'Connor}(2014)}]{2014ApJ...785..123C}
{Couch}, S.~M., \& {O'Connor}, E.~P. 2014, \apj, 785, 123,
  \dodoi{10.1088/0004-637X/785/2/123}

\bibitem[{{Couch} \& {Ott}(2013)}]{2013ApJ...778L...7C}
{Couch}, S.~M., \& {Ott}, C.~D. 2013, \apjl, 778, L7,
  \dodoi{10.1088/2041-8205/778/1/L7}

\bibitem[{{Couch} \& {Ott}(2015)}]{2015ApJ...799....5C}
---. 2015, \apj, 799, 5, \dodoi{10.1088/0004-637X/799/1/5}

\bibitem[{{Cristini} {et~al.}(2017){Cristini}, {Meakin}, {Hirschi}, {Arnett},
  {Georgy}, {Viallet}, \& {Walkington}}]{2017MNRAS.471..279C}
{Cristini}, A., {Meakin}, C., {Hirschi}, R., {et~al.} 2017, \mnras, 471, 279,
  \dodoi{10.1093/mnras/stx1535}

\bibitem[{{Crotts} \& {Heathcote}(1991)}]{1991Natur.350..683C}
{Crotts}, A.~P., \& {Heathcote}, S.~R. 1991, \nat, 350, 683,
  \dodoi{10.1038/350683a0}

\bibitem[{{Crotts} \& {Heathcote}(2000)}]{2000ApJ...528..426C}
{Crotts}, A.~P.~S., \& {Heathcote}, S.~R. 2000, \apj, 528, 426,
  \dodoi{10.1086/308141}

\bibitem[{{Decin} {et~al.}(2012){Decin}, {Cox}, {Royer}, {Van Marle},
  {Vandenbussche}, {Ladjal}, {Kerschbaum}, {Ottensamer}, {Barlow}, {Blommaert},
  {Gomez}, {Groenewegen}, {Lim}, {Swinyard}, {Waelkens}, \&
  {Tielens}}]{2012A&A...548A.113D}
{Decin}, L., {Cox}, N.~L.~J., {Royer}, P., {et~al.} 2012, \aap, 548, A113,
  \dodoi{10.1051/0004-6361/201219792}

\bibitem[{{Deng} \& {Xiong}(2008)}]{2008MNRAS.386.1979D}
{Deng}, L., \& {Xiong}, D.~R. 2008, \mnras, 386, 1979,
  \dodoi{10.1111/j.1365-2966.2008.12969.x}

\bibitem[{{Deng} {et~al.}(2006){Deng}, {Xiong}, \&
  {Chan}}]{2006ApJ...643..426D}
{Deng}, L., {Xiong}, D.~R., \& {Chan}, K.~L. 2006, \apj, 643, 426,
  \dodoi{10.1086/502707}

\bibitem[{{Dolence} {et~al.}(2013){Dolence}, {Burrows}, {Murphy}, \&
  {Nordhaus}}]{2013ApJ...765..110D}
{Dolence}, J.~C., {Burrows}, A., {Murphy}, J.~W., \& {Nordhaus}, J. 2013, \apj,
  765, 110, \dodoi{10.1088/0004-637X/765/2/110}

\bibitem[{{Dotani} {et~al.}(1987){Dotani}, {Hayashida}, {Inoue}, {Itoh},
  {Koyama}, {Makino}, {Mitsuda}, {Murakami}, {Oda}, {Ogawara}, {Takano},
  {Tanaka}, {Yoshida}, {Makishima}, {Ohashi}, {Kawai}, {Matsuoka}, {Hoshi},
  {Hayakawa}, {Kii}, {Kunieda}, {Nagase}, {Tawara}, {Hatsukade}, {Kitamoto},
  {Miyamoto}, {Tsunemi}, {Yamashita}, {Nakagawa}, {Yamauchi}, {Turner},
  {Pounds}, {Thomas}, {Stewart}, {Cruise}, {Patchett}, \&
  {Reading}}]{1987Natur.330..230D}
{Dotani}, T., {Hayashida}, K., {Inoue}, H., {et~al.} 1987, \nat, 330, 230,
  \dodoi{10.1038/330230a0}

\bibitem[{{Ebisuzaki} {et~al.}(1989){Ebisuzaki}, {Shigeyama}, \&
  {Nomoto}}]{1989ApJ...344L..65E}
{Ebisuzaki}, T., {Shigeyama}, T., \& {Nomoto}, K. 1989, \apjl, 344, L65,
  \dodoi{10.1086/185532}

\bibitem[{{Ellinger} {et~al.}(2012){Ellinger}, {Young}, {Fryer}, \&
  {Rockefeller}}]{2012ApJ...755..160E}
{Ellinger}, C.~I., {Young}, P.~A., {Fryer}, C.~L., \& {Rockefeller}, G. 2012,
  \apj, 755, 160, \dodoi{10.1088/0004-637X/755/2/160}

\bibitem[{{Faucher-Gigu{\`e}re} \& {Kaspi}(2006)}]{2006ApJ...643..332F}
{Faucher-Gigu{\`e}re}, C.-A., \& {Kaspi}, V.~M. 2006, \apj, 643, 332,
  \dodoi{10.1086/501516}

\bibitem[{{Ferrand} {et~al.}(2019){Ferrand}, {Warren}, {Ono}, {Nagataki},
  {R{\"o}pke}, \& {Seitenzahl}}]{2019ApJ...877..136F}
{Ferrand}, G., {Warren}, D.~C., {Ono}, M., {et~al.} 2019, \apj, 877, 136,
  \dodoi{10.3847/1538-4357/ab1a3d}

\bibitem[{{France} {et~al.}(2011){France}, {McCray}, {Penton}, {Kirshner},
  {Challis}, {Laming}, {Bouchet}, {Chevalier}, {Garnavich}, {Fransson}, {Heng},
  {Larsson}, {Lawrence}, {Lundqvist}, {Panagia}, {Pun}, {Smith}, {Sollerman},
  {Sonneborn}, {Sugerman}, \& {Wheeler}}]{2011ApJ...743..186F}
{France}, K., {McCray}, R., {Penton}, S.~V., {et~al.} 2011, \apj, 743, 186,
  \dodoi{10.1088/0004-637X/743/2/186}

\bibitem[{{Fransson} {et~al.}(1989){Fransson}, {Cassatella}, {Gilmozzi},
  {Kirshner}, {Panagia}, {Sonneborn}, \& {Wamsteker}}]{1989ApJ...336..429F}
{Fransson}, C., {Cassatella}, A., {Gilmozzi}, R., {et~al.} 1989, \apj, 336,
  429, \dodoi{10.1086/167022}

\bibitem[{{Fryxell} {et~al.}(1991){Fryxell}, {Mueller}, \&
  {Arnett}}]{1991ApJ...367..619F}
{Fryxell}, B., {Mueller}, E., \& {Arnett}, D. 1991, \apj, 367, 619,
  \dodoi{10.1086/169657}

\bibitem[{{Fryxell} {et~al.}(2000){Fryxell}, {Olson}, {Ricker}, {Timmes},
  {Zingale}, {Lamb}, {MacNeice}, {Rosner}, {Truran}, \&
  {Tufo}}]{2000ApJS..131..273F}
{Fryxell}, B., {Olson}, K., {Ricker}, P., {et~al.} 2000, \apjs, 131, 273,
  \dodoi{10.1086/317361}

\bibitem[{{Fujimoto} {et~al.}(2011){Fujimoto}, {Kotake}, {Hashimoto}, {Ono}, \&
  {Ohnishi}}]{2011ApJ...738...61F}
{Fujimoto}, S.-i., {Kotake}, K., {Hashimoto}, M.-a., {Ono}, M., \& {Ohnishi},
  N. 2011, \apj, 738, 61, \dodoi{10.1088/0004-637X/738/1/61}

\bibitem[{{Gawryszczak} {et~al.}(2010){Gawryszczak}, {Guzman}, {Plewa}, \&
  {Kifonidis}}]{2010A&A...521A..38G}
{Gawryszczak}, A., {Guzman}, J., {Plewa}, T., \& {Kifonidis}, K. 2010, \aap,
  521, A38, \dodoi{10.1051/0004-6361/200913431}

\bibitem[{{Grebenev} {et~al.}(2012){Grebenev}, {Lutovinov}, {Tsygankov}, \&
  {Winkler}}]{2012Natur.490..373G}
{Grebenev}, S.~A., {Lutovinov}, A.~A., {Tsygankov}, S.~S., \& {Winkler}, C.
  2012, \nat, 490, 373, \dodoi{10.1038/nature11473}

\bibitem[{{Grossman} {et~al.}(1993){Grossman}, {Narayan}, \&
  {Arnett}}]{1993ApJ...407..284G}
{Grossman}, S.~A., {Narayan}, R., \& {Arnett}, D. 1993, \apj, 407, 284,
  \dodoi{10.1086/172513}

\bibitem[{{Haas} {et~al.}(1990){Haas}, {Colgan}, {Erickson}, {Lord}, {Burton},
  \& {Hollenbach}}]{1990ApJ...360..257H}
{Haas}, M.~R., {Colgan}, S.~W.~J., {Erickson}, E.~F., {et~al.} 1990, \apj, 360,
  257, \dodoi{10.1086/169115}

\bibitem[{{Hachisu} {et~al.}(1990){Hachisu}, {Matsuda}, {Nomoto}, \&
  {Shigeyama}}]{1990ApJ...358L..57H}
{Hachisu}, I., {Matsuda}, T., {Nomoto}, K., \& {Shigeyama}, T. 1990, \apjl,
  358, L57, \dodoi{10.1086/185779}

\bibitem[{{Hachisu} {et~al.}(1992){Hachisu}, {Matsuda}, {Nomoto}, \&
  {Shigeyama}}]{1992ApJ...390..230H}
---. 1992, \apj, 390, 230, \dodoi{10.1086/171274}

\bibitem[{{Hammer} {et~al.}(2010){Hammer}, {Janka}, \&
  {M{\"u}ller}}]{2010ApJ...714.1371H}
{Hammer}, N.~J., {Janka}, H.-T., \& {M{\"u}ller}, E. 2010, \apj, 714, 1371,
  \dodoi{10.1088/0004-637X/714/2/1371}

\bibitem[{{Handy} {et~al.}(2014){Handy}, {Plewa}, \&
  {Odrzywo{\l}ek}}]{2014ApJ...783..125H}
{Handy}, T., {Plewa}, T., \& {Odrzywo{\l}ek}, A. 2014, \apj, 783, 125,
  \dodoi{10.1088/0004-637X/783/2/125}

\bibitem[{{Hanke} {et~al.}(2013){Hanke}, {M{\"u}ller}, {Wongwathanarat},
  {Marek}, \& {Janka}}]{2013ApJ...770...66H}
{Hanke}, F., {M{\"u}ller}, B., {Wongwathanarat}, A., {Marek}, A., \& {Janka},
  H.-T. 2013, \apj, 770, 66, \dodoi{10.1088/0004-637X/770/1/66}

\bibitem[{{Hanuschik} {et~al.}(1988){Hanuschik}, {Thimm}, \&
  {Dachs}}]{1988MNRAS.234P..41H}
{Hanuschik}, R.~W., {Thimm}, G., \& {Dachs}, J. 1988, \mnras, 234, 41P,
  \dodoi{10.1093/mnras/234.1.41P}

\bibitem[{{Heger} \& {Langer}(1998)}]{1998A&A...334..210H}
{Heger}, A., \& {Langer}, N. 1998, \aap, 334, 210

\bibitem[{{Heger} {et~al.}(2005){Heger}, {Woosley}, \&
  {Spruit}}]{2005ApJ...626..350H}
{Heger}, A., {Woosley}, S.~E., \& {Spruit}, H.~C. 2005, \apj, 626, 350,
  \dodoi{10.1086/429868}

\bibitem[{{Herant} \& {Benz}(1991)}]{1991ApJ...370L..81H}
{Herant}, M., \& {Benz}, W. 1991, \apjl, 370, L81, \dodoi{10.1086/185982}

\bibitem[{{Herant} \& {Benz}(1992)}]{1992ApJ...387..294H}
---. 1992, \apj, 387, 294, \dodoi{10.1086/171081}

\bibitem[{{Herant} {et~al.}(1994){Herant}, {Benz}, {Hix}, {Fryer}, \&
  {Colgate}}]{1994ApJ...435..339H}
{Herant}, M., {Benz}, W., {Hix}, W.~R., {Fryer}, C.~L., \& {Colgate}, S.~A.
  1994, \apj, 435, 339, \dodoi{10.1086/174817}

\bibitem[{{Herwig} {et~al.}(2014){Herwig}, {Woodward}, {Lin}, {Knox}, \&
  {Fryer}}]{2014ApJ...792L...3H}
{Herwig}, F., {Woodward}, P.~R., {Lin}, P.-H., {Knox}, M., \& {Fryer}, C. 2014,
  \apjl, 792, L3, \dodoi{10.1088/2041-8205/792/1/L3}

\bibitem[{{Hillebrandt} \& {Meyer}(1989)}]{1989A&A...219L...3H}
{Hillebrandt}, W., \& {Meyer}, F. 1989, \aap, 219, L3

\bibitem[{{Hobbs} {et~al.}(2005){Hobbs}, {Lorimer}, {Lyne}, \&
  {Kramer}}]{2005MNRAS.360..974H}
{Hobbs}, G., {Lorimer}, D.~R., {Lyne}, A.~G., \& {Kramer}, M. 2005, \mnras,
  360, 974, \dodoi{10.1111/j.1365-2966.2005.09087.x}

\bibitem[{{Holland-Ashford} {et~al.}(2019){Holland-Ashford}, {Lopez}, \&
  {Auchettl}}]{2019arXiv190406357H}
{Holland-Ashford}, T., {Lopez}, L.~A., \& {Auchettl}, K. 2019, arXiv e-prints.
\newblock \doarXiv{1904.06357}

\bibitem[{{Hotta} {et~al.}(2014){Hotta}, {Rempel}, \&
  {Yokoyama}}]{2014ApJ...786...24H}
{Hotta}, H., {Rempel}, M., \& {Yokoyama}, T. 2014, \apj, 786, 24,
  \dodoi{10.1088/0004-637X/786/1/24}

\bibitem[{{Hotta} {et~al.}(2015){Hotta}, {Rempel}, \&
  {Yokoyama}}]{2015ApJ...798...51H}
---. 2015, \apj, 798, 51, \dodoi{10.1088/0004-637X/798/1/51}

\bibitem[{{Humphreys} \& {McElroy}(1984)}]{1984ApJ...284..565H}
{Humphreys}, R.~M., \& {McElroy}, D.~B. 1984, \apj, 284, 565,
  \dodoi{10.1086/162439}

\bibitem[{{Hungerford} {et~al.}(2005){Hungerford}, {Fryer}, \&
  {Rockefeller}}]{2005ApJ...635..487H}
{Hungerford}, A.~L., {Fryer}, C.~L., \& {Rockefeller}, G. 2005, \apj, 635, 487,
  \dodoi{10.1086/497323}

\bibitem[{{Hungerford} {et~al.}(2003){Hungerford}, {Fryer}, \&
  {Warren}}]{2003ApJ...594..390H}
{Hungerford}, A.~L., {Fryer}, C.~L., \& {Warren}, M.~S. 2003, \apj, 594, 390,
  \dodoi{10.1086/376776}

\bibitem[{{Ivanova} {et~al.}(2002){Ivanova}, {Podsiadlowski}, \&
  {Spruit}}]{2002MNRAS.334..819I}
{Ivanova}, N., {Podsiadlowski}, P., \& {Spruit}, H. 2002, \mnras, 334, 819,
  \dodoi{10.1046/j.1365-8711.2002.05543.x}

\bibitem[{{Janka}(2012)}]{2012ARNPS..62..407J}
{Janka}, H.-T. 2012, Annual Review of Nuclear and Particle Science, 62, 407,
  \dodoi{10.1146/annurev-nucl-102711-094901}

\bibitem[{{Janka} {et~al.}(2012){Janka}, {Hanke}, {H{\"u}depohl}, {Marek},
  {M{\"u}ller}, \& {Obergaulinger}}]{2012PTEP.2012aA309J}
{Janka}, H.-T., {Hanke}, F., {H{\"u}depohl}, L., {et~al.} 2012, Progress of
  Theoretical and Experimental Physics, 2012, 01A309,
  \dodoi{10.1093/ptep/pts067}

\bibitem[{{Jerkstrand} {et~al.}(2015){Jerkstrand}, {Timmes}, {Magkotsios},
  {Sim}, {Fransson}, {Spyromilio}, {M{\"u}ller}, {Heger}, {Sollerman}, \&
  {Smartt}}]{2015ApJ...807..110J}
{Jerkstrand}, A., {Timmes}, F.~X., {Magkotsios}, G., {et~al.} 2015, \apj, 807,
  110, \dodoi{10.1088/0004-637X/807/1/110}

\bibitem[{{Joggerst} {et~al.}(2010{\natexlab{a}}){Joggerst}, {Almgren}, {Bell},
  {Heger}, {Whalen}, \& {Woosley}}]{2010ApJ...709...11J}
{Joggerst}, C.~C., {Almgren}, A., {Bell}, J., {et~al.} 2010{\natexlab{a}},
  \apj, 709, 11, \dodoi{10.1088/0004-637X/709/1/11}

\bibitem[{{Joggerst} {et~al.}(2010{\natexlab{b}}){Joggerst}, {Almgren}, \&
  {Woosley}}]{2010ApJ...723..353J}
{Joggerst}, C.~C., {Almgren}, A., \& {Woosley}, S.~E. 2010{\natexlab{b}}, \apj,
  723, 353, \dodoi{10.1088/0004-637X/723/1/353}

\bibitem[{{Joggerst} {et~al.}(2009){Joggerst}, {Woosley}, \&
  {Heger}}]{2009ApJ...693.1780J}
{Joggerst}, C.~C., {Woosley}, S.~E., \& {Heger}, A. 2009, \apj, 693, 1780,
  \dodoi{10.1088/0004-637X/693/2/1780}

\bibitem[{{Justham} {et~al.}(2014){Justham}, {Podsiadlowski}, \&
  {Vink}}]{2014ApJ...796..121J}
{Justham}, S., {Podsiadlowski}, P., \& {Vink}, J.~S. 2014, \apj, 796, 121,
  \dodoi{10.1088/0004-637X/796/2/121}

\bibitem[{{Katsuda} {et~al.}(2018){Katsuda}, {Morii}, {Janka},
  {Wongwathanarat}, {Nakamura}, {Kotake}, {Mori}, {M{\"u}ller}, {Takiwaki},
  {Tanaka}, {Tominaga}, \& {Tsunemi}}]{2018ApJ...856...18K}
{Katsuda}, S., {Morii}, M., {Janka}, H.-T., {et~al.} 2018, \apj, 856, 18,
  \dodoi{10.3847/1538-4357/aab092}

\bibitem[{{Kifonidis} {et~al.}(2000){Kifonidis}, {Plewa}, {Janka}, \&
  {M{\"u}ller}}]{2000ApJ...531L.123K}
{Kifonidis}, K., {Plewa}, T., {Janka}, H.-T., \& {M{\"u}ller}, E. 2000, \apjl,
  531, L123, \dodoi{10.1086/312541}

\bibitem[{{Kifonidis} {et~al.}(2003){Kifonidis}, {Plewa}, {Janka}, \&
  {M{\"u}ller}}]{2003A&A...408..621K}
---. 2003, \aap, 408, 621, \dodoi{10.1051/0004-6361:20030863}

\bibitem[{{Kifonidis} {et~al.}(2006){Kifonidis}, {Plewa}, {Scheck}, {Janka}, \&
  {M{\"u}ller}}]{2006A&A...453..661K}
{Kifonidis}, K., {Plewa}, T., {Scheck}, L., {Janka}, H.-T., \& {M{\"u}ller}, E.
  2006, \aap, 453, 661, \dodoi{10.1051/0004-6361:20054512}

\bibitem[{{Kippenhahn} \& {Weigert}(1990)}]{1990sse..book.....K}
{Kippenhahn}, R., \& {Weigert}, A. 1990, {Stellar Structure and Evolution}
  (Springer-Verlag Berlin Heidelberg), 192

\bibitem[{{Kj{\ae}r} {et~al.}(2010){Kj{\ae}r}, {Leibundgut}, {Fransson},
  {Jerkstrand}, \& {Spyromilio}}]{2010A&A...517A..51K}
{Kj{\ae}r}, K., {Leibundgut}, B., {Fransson}, C., {Jerkstrand}, A., \&
  {Spyromilio}, J. 2010, \aap, 517, A51, \dodoi{10.1051/0004-6361/201014538}

\bibitem[{{Kotake} {et~al.}(2004){Kotake}, {Sawai}, {Yamada}, \&
  {Sato}}]{2004ApJ...608..391K}
{Kotake}, K., {Sawai}, H., {Yamada}, S., \& {Sato}, K. 2004, \apj, 608, 391,
  \dodoi{10.1086/392530}

\bibitem[{{Kotake} {et~al.}(2012{\natexlab{a}}){Kotake}, {Sumiyoshi}, {Yamada},
  {Takiwaki}, {Kuroda}, {Suwa}, \& {Nagakura}}]{2012PTEP.2012aA301K}
{Kotake}, K., {Sumiyoshi}, K., {Yamada}, S., {et~al.} 2012{\natexlab{a}},
  Progress of Theoretical and Experimental Physics, 2012, 01A301,
  \dodoi{10.1093/ptep/pts009}

\bibitem[{{Kotake} {et~al.}(2012{\natexlab{b}}){Kotake}, {Takiwaki}, {Suwa},
  {Iwakami Nakano}, {Kawagoe}, {Masada}, \& {Fujimoto}}]{2012AdAst2012E..39K}
{Kotake}, K., {Takiwaki}, T., {Suwa}, Y., {et~al.} 2012{\natexlab{b}}, Advances
  in Astronomy, 2012, 428757, \dodoi{10.1155/2012/428757}

\bibitem[{{Kozma} \& {Fransson}(1998)}]{1998ApJ...497..431K}
{Kozma}, C., \& {Fransson}, C. 1998, \apj, 497, 431, \dodoi{10.1086/305452}

\bibitem[{{Kuroda} {et~al.}(2012){Kuroda}, {Kotake}, \&
  {Takiwaki}}]{2012ApJ...755...11K}
{Kuroda}, T., {Kotake}, K., \& {Takiwaki}, T. 2012, \apj, 755, 11,
  \dodoi{10.1088/0004-637X/755/1/11}

\bibitem[{{Langer} {et~al.}(1985){Langer}, {El Eid}, \&
  {Fricke}}]{1985A&A...145..179L}
{Langer}, N., {El Eid}, M.~F., \& {Fricke}, K.~J. 1985, \aap, 145, 179

\bibitem[{{Larsson} {et~al.}(2013){Larsson}, {Fransson}, {Kjaer}, {Jerkstrand},
  {Kirshner}, {Leibundgut}, {Lundqvist}, {Mattila}, {McCray}, {Sollerman},
  {Spyromilio}, \& {Wheeler}}]{2013ApJ...768...89L}
{Larsson}, J., {Fransson}, C., {Kjaer}, K., {et~al.} 2013, \apj, 768, 89,
  \dodoi{10.1088/0004-637X/768/1/89}

\bibitem[{{Larsson} {et~al.}(2016){Larsson}, {Fransson}, {Spyromilio},
  {Leibundgut}, {Challis}, {Chevalier}, {France}, {Jerkstrand}, {Kirshner},
  {Lundqvist}, {Matsuura}, {McCray}, {Smith}, {Sollerman}, {Garnavich}, {Heng},
  {Lawrence}, {Mattila}, {Migotto}, {Sonneborn}, {Taddia}, \&
  {Wheeler}}]{2016ApJ...833..147L}
{Larsson}, J., {Fransson}, C., {Spyromilio}, J., {et~al.} 2016, \apj, 833, 147,
  \dodoi{10.3847/1538-4357/833/2/147}

\bibitem[{{Larsson} {et~al.}(2019){Larsson}, {Fransson}, {Alp}, {Challis},
  {Chevalier}, {France}, {Kirshner}, {Lawrence}, {Leibundgut}, {Lundqvist},
  {Mattila}, {Migotto}, {Sollerman}, {Sonneborn}, {Spyromilio}, {Suntzeff}, \&
  {Wheeler}}]{2019arXiv191009582L}
{Larsson}, J., {Fransson}, C., {Alp}, D., {et~al.} 2019, arXiv e-prints,
  arXiv:1910.09582.
\newblock \doarXiv{1910.09582}

\bibitem[{{Li}(2012)}]{2012ApJ...756...37L}
{Li}, Y. 2012, \apj, 756, 37, \dodoi{10.1088/0004-637X/756/1/37}

\bibitem[{{Li} \& {Yang}(2007)}]{2007MNRAS.375..388L}
{Li}, Y., \& {Yang}, J.~Y. 2007, \mnras, 375, 388,
  \dodoi{10.1111/j.1365-2966.2006.11319.x}

\bibitem[{{Lundqvist} \& {Fransson}(1996)}]{1996ApJ...464..924L}
{Lundqvist}, P., \& {Fransson}, C. 1996, \apj, 464, 924, \dodoi{10.1086/177380}

\bibitem[{{MacNeice} {et~al.}(2000){MacNeice}, {Olson}, {Mobarry}, {de
  Fainchtein}, \& {Packer}}]{2000CoPhC.126..330M}
{MacNeice}, P., {Olson}, K.~M., {Mobarry}, C., {de Fainchtein}, R., \&
  {Packer}, C. 2000, Computer Physics Communications, 126, 330,
  \dodoi{10.1016/S0010-4655(99)00501-9}

\bibitem[{{Mao} {et~al.}(2015){Mao}, {Ono}, {Nagataki}, {Hashimoto}, {Ito},
  {Matsumoto}, {Dainotti}, \& {Lee}}]{2015ApJ...808..164M}
{Mao}, J., {Ono}, M., {Nagataki}, S., {et~al.} 2015, \apj, 808, 164,
  \dodoi{10.1088/0004-637X/808/2/164}

\bibitem[{{Marek} \& {Janka}(2009)}]{2009ApJ...694..664M}
{Marek}, A., \& {Janka}, H.-T. 2009, \apj, 694, 664,
  \dodoi{10.1088/0004-637X/694/1/664}

\bibitem[{{Mattila} {et~al.}(2010){Mattila}, {Lundqvist}, {Gr{\"o}ningsson},
  {Meikle}, {Stathakis}, {Fransson}, \& {Cannon}}]{2010ApJ...717.1140M}
{Mattila}, S., {Lundqvist}, P., {Gr{\"o}ningsson}, P., {et~al.} 2010, \apj,
  717, 1140, \dodoi{10.1088/0004-637X/717/2/1140}

\bibitem[{{Matz} {et~al.}(1988){Matz}, {Share}, {Leising}, {Chupp}, {Vestrand},
  {Purcell}, {Strickman}, \& {Reppin}}]{1988Natur.331..416M}
{Matz}, S.~M., {Share}, G.~H., {Leising}, M.~D., {et~al.} 1988, \nat, 331, 416,
  \dodoi{10.1038/331416a0}

\bibitem[{{McCray}(1993)}]{1993ARA&A..31..175M}
{McCray}, R. 1993, \araa, 31, 175, \dodoi{10.1146/annurev.aa.31.090193.001135}

\bibitem[{{McCray} \& {Fransson}(2016)}]{2016ARA&A..54...19M}
{McCray}, R., \& {Fransson}, C. 2016, \araa, 54, 19,
  \dodoi{10.1146/annurev-astro-082615-105405}

\bibitem[{{Meakin} \& {Arnett}(2006)}]{2006ApJ...637L..53M}
{Meakin}, C.~A., \& {Arnett}, D. 2006, \apjl, 637, L53, \dodoi{10.1086/500544}

\bibitem[{{Meakin} \& {Arnett}(2007{\natexlab{a}})}]{2007ApJ...665..690M}
---. 2007{\natexlab{a}}, \apj, 665, 690, \dodoi{10.1086/519372}

\bibitem[{{Meakin} \& {Arnett}(2007{\natexlab{b}})}]{2007ApJ...667..448M}
---. 2007{\natexlab{b}}, \apj, 667, 448, \dodoi{10.1086/520318}

\bibitem[{{Menon} \& {Heger}(2017)}]{2017MNRAS.469.4649M}
{Menon}, A., \& {Heger}, A. 2017, \mnras, 469, 4649.
\newblock \doarXiv{1703.04918}

\bibitem[{{Menon} {et~al.}(2019){Menon}, {Utrobin}, \&
  {Heger}}]{2019MNRAS.482..438M}
{Menon}, A., {Utrobin}, V., \& {Heger}, A. 2019, \mnras, 482, 438,
  \dodoi{10.1093/mnras/sty2647}

\bibitem[{{Miceli} {et~al.}(2019){Miceli}, {Orlando}, {Burrows}, {Frank},
  {Argiroffi}, {Reale}, {Peres}, {Petruk}, \& {Bocchino}}]{2019NatAs...3..236M}
{Miceli}, M., {Orlando}, S., {Burrows}, D.~N., {et~al.} 2019, Nature Astronomy,
  3, 236, \dodoi{10.1038/s41550-018-0677-8}

\bibitem[{{Moc{\'a}k} {et~al.}(2018){Moc{\'a}k}, {Meakin}, {Campbell}, \&
  {Arnett}}]{2018MNRAS.481.2918M}
{Moc{\'a}k}, M., {Meakin}, C., {Campbell}, S.~W., \& {Arnett}, W.~D. 2018,
  \mnras, 481, 2918, \dodoi{10.1093/mnras/sty2392}

\bibitem[{{Montalb{\'a}n} {et~al.}(2013){Montalb{\'a}n}, {Miglio}, {Noels},
  {Dupret}, {Scuflaire}, \& {Ventura}}]{2013ApJ...766..118M}
{Montalb{\'a}n}, J., {Miglio}, A., {Noels}, A., {et~al.} 2013, \apj, 766, 118,
  \dodoi{10.1088/0004-637X/766/2/118}

\bibitem[{{Montarg{\`e}s} {et~al.}(2014){Montarg{\`e}s}, {Kervella}, {Perrin},
  {Ohnaka}, {Chiavassa}, {Ridgway}, \& {Lacour}}]{2014A&A...572A..17M}
{Montarg{\`e}s}, M., {Kervella}, P., {Perrin}, G., {et~al.} 2014, \aap, 572,
  A17, \dodoi{10.1051/0004-6361/201423538}

\bibitem[{{Morris} \& {Podsiadlowski}(2007)}]{2007Sci...315.1103M}
{Morris}, T., \& {Podsiadlowski}, P. 2007, Science, 315, 1103,
  \dodoi{10.1126/science.1136351}

\bibitem[{{Morris} \& {Podsiadlowski}(2009)}]{2009MNRAS.399..515M}
---. 2009, \mnras, 399, 515, \dodoi{10.1111/j.1365-2966.2009.15114.x}

\bibitem[{{M{\"o}sta} {et~al.}(2014){M{\"o}sta}, {Richers}, {Ott}, {Haas},
  {Piro}, {Boydstun}, {Abdikamalov}, {Reisswig}, \&
  {Schnetter}}]{2014ApJ...785L..29M}
{M{\"o}sta}, P., {Richers}, S., {Ott}, C.~D., {et~al.} 2014, \apjl, 785, L29,
  \dodoi{10.1088/2041-8205/785/2/L29}

\bibitem[{{Mueller} {et~al.}(1991){Mueller}, {Fryxell}, \&
  {Arnett}}]{1991A&A...251..505M}
{Mueller}, E., {Fryxell}, B., \& {Arnett}, D. 1991, \aap, 251, 505

\bibitem[{{M{\"u}ller}(2016)}]{2016PASA...33...48M}
{M{\"u}ller}, B. 2016, \pasa, 33, e048, \dodoi{10.1017/pasa.2016.40}

\bibitem[{{M{\"u}ller} {et~al.}(2012{\natexlab{a}}){M{\"u}ller}, {Janka}, \&
  {Heger}}]{2012ApJ...761...72M}
{M{\"u}ller}, B., {Janka}, H.-T., \& {Heger}, A. 2012{\natexlab{a}}, \apj, 761,
  72, \dodoi{10.1088/0004-637X/761/1/72}

\bibitem[{{M{\"u}ller} {et~al.}(2012{\natexlab{b}}){M{\"u}ller}, {Janka}, \&
  {Marek}}]{2012ApJ...756...84M}
{M{\"u}ller}, B., {Janka}, H.-T., \& {Marek}, A. 2012{\natexlab{b}}, \apj, 756,
  84, \dodoi{10.1088/0004-637X/756/1/84}

\bibitem[{{M{\"u}ller} {et~al.}(2017){M{\"u}ller}, {Melson}, {Heger}, \&
  {Janka}}]{2017MNRAS.472..491M}
{M{\"u}ller}, B., {Melson}, T., {Heger}, A., \& {Janka}, H.-T. 2017, \mnras,
  472, 491, \dodoi{10.1093/mnras/stx1962}

\bibitem[{{M{\"u}ller} {et~al.}(2016){M{\"u}ller}, {Viallet}, {Heger}, \&
  {Janka}}]{2016ApJ...833..124M}
{M{\"u}ller}, B., {Viallet}, M., {Heger}, A., \& {Janka}, H.-T. 2016, \apj,
  833, 124, \dodoi{10.3847/1538-4357/833/1/124}

\bibitem[{{M{\"u}ller} {et~al.}(2012{\natexlab{c}}){M{\"u}ller}, {Janka}, \&
  {Wongwathanarat}}]{2012A&A...537A..63M}
{M{\"u}ller}, E., {Janka}, H.-T., \& {Wongwathanarat}, A. 2012{\natexlab{c}},
  \aap, 537, A63, \dodoi{10.1051/0004-6361/201117611}

\bibitem[{{Nadyozhin}(1994)}]{1994ApJS...92..527N}
{Nadyozhin}, D.~K. 1994, \apjs, 92, 527, \dodoi{10.1086/192008}

\bibitem[{{Nagakura} {et~al.}(2017){Nagakura}, {Iwakami}, {Furusawa},
  {Sumiyoshi}, {Yamada}, {Matsufuru}, \& {Imakura}}]{2017ApJS..229...42N}
{Nagakura}, H., {Iwakami}, W., {Furusawa}, S., {et~al.} 2017, \apjs, 229, 42,
  \dodoi{10.3847/1538-4365/aa69ea}

\bibitem[{{Nagataki}(2000)}]{2000ApJS..127..141N}
{Nagataki}, S. 2000, \apjs, 127, 141, \dodoi{10.1086/313317}

\bibitem[{{Nagataki} {et~al.}(1997){Nagataki}, {Hashimoto}, {Sato}, \&
  {Yamada}}]{1997ApJ...486.1026N}
{Nagataki}, S., {Hashimoto}, M.-a., {Sato}, K., \& {Yamada}, S. 1997, \apj,
  486, 1026, \dodoi{10.1086/304565}

\bibitem[{{Nagataki} {et~al.}(1998{\natexlab{a}}){Nagataki}, {Hashimoto},
  {Sato}, {Yamada}, \& {Mochizuki}}]{1998ApJ...492L..45N}
{Nagataki}, S., {Hashimoto}, M.-a., {Sato}, K., {Yamada}, S., \& {Mochizuki},
  Y.~S. 1998{\natexlab{a}}, \apjl, 492, L45, \dodoi{10.1086/311089}

\bibitem[{{Nagataki} {et~al.}(1998{\natexlab{b}}){Nagataki}, {Shimizu}, \&
  {Sato}}]{1998ApJ...495..413N}
{Nagataki}, S., {Shimizu}, T.~M., \& {Sato}, K. 1998{\natexlab{b}}, \\apj, 495,
  413, \dodoi{10.1086/305258}

\bibitem[{{Nakamura} {et~al.}(2014){Nakamura}, {Takiwaki}, {Kotake}, \&
  {Nishimura}}]{2014ApJ...782...91N}
{Nakamura}, K., {Takiwaki}, T., {Kotake}, K., \& {Nishimura}, N. 2014, \apj,
  782, 91, \dodoi{10.1088/0004-637X/782/2/91}

\bibitem[{{Nomoto} \& {Hashimoto}(1988)}]{1988PhR...163...13N}
{Nomoto}, K., \& {Hashimoto}, M. 1988, \physrep, 163, 13,
  \dodoi{10.1016/0370-1573(88)90032-4}

\bibitem[{{Nordhaus} {et~al.}(2012){Nordhaus}, {Brandt}, {Burrows}, \&
  {Almgren}}]{2012MNRAS.423.1805N}
{Nordhaus}, J., {Brandt}, T.~D., {Burrows}, A., \& {Almgren}, A. 2012, \mnras,
  423, 1805, \dodoi{10.1111/j.1365-2966.2012.21002.x}

\bibitem[{{Nordhaus} {et~al.}(2010{\natexlab{a}}){Nordhaus}, {Brandt},
  {Burrows}, {Livne}, \& {Ott}}]{2010PhRvD..82j3016N}
{Nordhaus}, J., {Brandt}, T.~D., {Burrows}, A., {Livne}, E., \& {Ott}, C.~D.
  2010{\natexlab{a}}, \prd, 82, 103016, \dodoi{10.1103/PhysRevD.82.103016}

\bibitem[{{Nordhaus} {et~al.}(2010{\natexlab{b}}){Nordhaus}, {Burrows},
  {Almgren}, \& {Bell}}]{2010ApJ...720..694N}
{Nordhaus}, J., {Burrows}, A., {Almgren}, A., \& {Bell}, J. 2010{\natexlab{b}},
  \apj, 720, 694, \dodoi{10.1088/0004-637X/720/1/694}

\bibitem[{{Ono} {et~al.}(2013){Ono}, {Nagataki}, {Ito}, {Lee}, {Mao},
  {Hashimoto}, \& {Tolstov}}]{2013ApJ...773..161O}
{Ono}, M., {Nagataki}, S., {Ito}, H., {et~al.} 2013, \apj, 773, 161,
  \dodoi{10.1088/0004-637X/773/2/161}

\bibitem[{{Orlando} {et~al.}(2015){Orlando}, {Miceli}, {Pumo}, \&
  {Bocchino}}]{2015ApJ...810..168O}
{Orlando}, S., {Miceli}, M., {Pumo}, M.~L., \& {Bocchino}, F. 2015, \apj, 810,
  168, \dodoi{10.1088/0004-637X/810/2/168}

\bibitem[{{Orlando} {et~al.}(2019){Orlando}, {Miceli}, {Petruk}, {Ono},
  {Nagataki}, {Aloy}, {Mimica}, {Lee}, {Bocchino}, {Peres}, \&
  {Guarrasi}}]{2019A&A...622A..73O}
{Orlando}, S., {Miceli}, M., {Petruk}, O., {et~al.} 2019, \aap, 622, A73,
  \dodoi{10.1051/0004-6361/201834487}

\bibitem[{{Ott} {et~al.}(2013){Ott}, {Abdikamalov}, {M{\"o}sta}, {Haas},
  {Drasco}, {O'Connor}, {Reisswig}, {Meakin}, \&
  {Schnetter}}]{2013ApJ...768..115O}
{Ott}, C.~D., {Abdikamalov}, E., {M{\"o}sta}, P., {et~al.} 2013, \apj, 768,
  115, \dodoi{10.1088/0004-637X/768/2/115}

\bibitem[{{Pan} {et~al.}(2016){Pan}, {Liebend{\"o}rfer}, {Hempel}, \&
  {Thielemann}}]{2016ApJ...817...72P}
{Pan}, K.-C., {Liebend{\"o}rfer}, M., {Hempel}, M., \& {Thielemann}, F.-K.
  2016, \apj, 817, 72, \dodoi{10.3847/0004-637X/817/1/72}

\bibitem[{{Pasquali} {et~al.}(2000){Pasquali}, {Nota}, {Langer},
  {Schulte-Ladbeck}, \& {Clampin}}]{2000AJ....119.1352P}
{Pasquali}, A., {Nota}, A., {Langer}, N., {Schulte-Ladbeck}, R.~E., \&
  {Clampin}, M. 2000, \aj, 119, 1352, \dodoi{10.1086/301257}

\bibitem[{{Pinto} \& {Woosley}(1988)}]{1988ApJ...329..820P}
{Pinto}, P.~A., \& {Woosley}, S.~E. 1988, \apj, 329, 820,
  \dodoi{10.1086/166426}

\bibitem[{{Podsiadlowski}(1992)}]{1992PASP..104..717P}
{Podsiadlowski}, P. 1992, \pasp, 104, 717, \dodoi{10.1086/133043}

\bibitem[{{Podsiadlowski} {et~al.}(1992){Podsiadlowski}, {Joss}, \&
  {Hsu}}]{1992ApJ...391..246P}
{Podsiadlowski}, P., {Joss}, P.~C., \& {Hsu}, J.~J.~L. 1992, \apj, 391, 246,
  \dodoi{10.1086/171341}

\bibitem[{{Podsiadlowski} {et~al.}(1990){Podsiadlowski}, {Joss}, \&
  {Rappaport}}]{1990A&A...227L...9P}
{Podsiadlowski}, P., {Joss}, P.~C., \& {Rappaport}, S. 1990, \aap, 227, L9

\bibitem[{{Portegies Zwart} \& {van den Heuvel}(2016)}]{2016MNRAS.456.3401P}
{Portegies Zwart}, S.~F., \& {van den Heuvel}, E.~P.~J. 2016, \mnras, 456,
  3401, \dodoi{10.1093/mnras/stv2787}

\bibitem[{{Potter} {et~al.}(2014){Potter}, {Staveley-Smith}, {Reville}, {Ng},
  {Bicknell}, {Sutherland}, \& {Wagner}}]{2014ApJ...794..174P}
{Potter}, T.~M., {Staveley-Smith}, L., {Reville}, B., {et~al.} 2014, \apj, 794,
  174, \dodoi{10.1088/0004-637X/794/2/174}

\bibitem[{{Radice} {et~al.}(2017){Radice}, {Burrows}, {Vartanyan}, {Skinner},
  \& {Dolence}}]{2017ApJ...850...43R}
{Radice}, D., {Burrows}, A., {Vartanyan}, D., {Skinner}, M.~A., \& {Dolence},
  J.~C. 2017, \apj, 850, 43, \dodoi{10.3847/1538-4357/aa92c5}

\bibitem[{{Radice} {et~al.}(2016){Radice}, {Ott}, {Abdikamalov}, {Couch},
  {Haas}, \& {Schnetter}}]{2016ApJ...820...76R}
{Radice}, D., {Ott}, C.~D., {Abdikamalov}, E., {et~al.} 2016, \apj, 820, 76,
  \dodoi{10.3847/0004-637X/820/1/76}

\bibitem[{{Rogers} {et~al.}(2006){Rogers}, {Glatzmaier}, \&
  {Jones}}]{2006ApJ...653..765R}
{Rogers}, T.~M., {Glatzmaier}, G.~A., \& {Jones}, C.~A. 2006, \apj, 653, 765,
  \dodoi{10.1086/508482}

\bibitem[{{Saio} {et~al.}(1988{\natexlab{a}}){Saio}, {Kato}, \&
  {Nomoto}}]{1988ApJ...331..388S}
{Saio}, H., {Kato}, M., \& {Nomoto}, K. 1988{\natexlab{a}}, \apj, 331, 388,
  \dodoi{10.1086/166565}

\bibitem[{{Saio} {et~al.}(1988{\natexlab{b}}){Saio}, {Nomoto}, \&
  {Kato}}]{1988Natur.334..508S}
{Saio}, H., {Nomoto}, K., \& {Kato}, M. 1988{\natexlab{b}}, \nat, 334, 508,
  \dodoi{10.1038/334508a0}

\bibitem[{{Sawai} {et~al.}(2005){Sawai}, {Kotake}, \&
  {Yamada}}]{2005ApJ...631..446S}
{Sawai}, H., {Kotake}, K., \& {Yamada}, S. 2005, \apj, 631, 446,
  \dodoi{10.1086/432529}

\bibitem[{{Sawai} \& {Yamada}(2016)}]{2016ApJ...817..153S}
{Sawai}, H., \& {Yamada}, S. 2016, \apj, 817, 153,
  \dodoi{10.3847/0004-637X/817/2/153}

\bibitem[{{Sawai} {et~al.}(2013){Sawai}, {Yamada}, {Kotake}, \&
  {Suzuki}}]{2013ApJ...764...10S}
{Sawai}, H., {Yamada}, S., {Kotake}, K., \& {Suzuki}, H. 2013, \apj, 764, 10,
  \dodoi{10.1088/0004-637X/764/1/10}

\bibitem[{{Scheck} {et~al.}(2008){Scheck}, {Janka}, {Foglizzo}, \&
  {Kifonidis}}]{2008A&A...477..931S}
{Scheck}, L., {Janka}, H.-T., {Foglizzo}, T., \& {Kifonidis}, K. 2008, \aap,
  477, 931, \dodoi{10.1051/0004-6361:20077701}

\bibitem[{{Scheck} {et~al.}(2006){Scheck}, {Kifonidis}, {Janka}, \&
  {M{\"u}ller}}]{2006A&A...457..963S}
{Scheck}, L., {Kifonidis}, K., {Janka}, H.-T., \& {M{\"u}ller}, E. 2006, \aap,
  457, 963, \dodoi{10.1051/0004-6361:20064855}

\bibitem[{{Scheck} {et~al.}(2004){Scheck}, {Plewa}, {Janka}, {Kifonidis}, \&
  {M{\"u}ller}}]{2004PhRvL..92a1103S}
{Scheck}, L., {Plewa}, T., {Janka}, H.-T., {Kifonidis}, K., \& {M{\"u}ller}, E.
  2004, Physical Review Letters, 92, 011103,
  \dodoi{10.1103/PhysRevLett.92.011103}

\bibitem[{{Sedov}(1959)}]{1959sdmm.book.....S}
{Sedov}, L.~I. 1959, {Similarity and Dimensional Methods in Mechanics}

\bibitem[{{Shigeyama} \& {Nomoto}(1990)}]{1990ApJ...360..242S}
{Shigeyama}, T., \& {Nomoto}, K. 1990, \apj, 360, 242, \dodoi{10.1086/169114}

\bibitem[{{Silva Aguirre} {et~al.}(2011){Silva Aguirre}, {Ballot}, {Serenelli},
  \& {Weiss}}]{2011A&A...529A..63S}
{Silva Aguirre}, V., {Ballot}, J., {Serenelli}, A.~M., \& {Weiss}, A. 2011,
  \aap, 529, A63, \dodoi{10.1051/0004-6361/201015847}

\bibitem[{{Sinnott} {et~al.}(2013){Sinnott}, {Welch}, {Rest}, {Sutherland}, \&
  {Bergmann}}]{2013ApJ...767...45S}
{Sinnott}, B., {Welch}, D.~L., {Rest}, A., {Sutherland}, P.~G., \& {Bergmann},
  M. 2013, \apj, 767, 45, \dodoi{10.1088/0004-637X/767/1/45}

\bibitem[{{Smartt}(2009)}]{2009ARA&A..47...63S}
{Smartt}, S.~J. 2009, \araa, 47, 63,
  \dodoi{10.1146/annurev-astro-082708-101737}

\bibitem[{{Smith}(2007)}]{2007AJ....133.1034S}
{Smith}, N. 2007, \aj, 133, 1034, \dodoi{10.1086/510838}

\bibitem[{{Smith} \& {Arnett}(2014)}]{2014ApJ...785...82S}
{Smith}, N., \& {Arnett}, W.~D. 2014, \apj, 785, 82,
  \dodoi{10.1088/0004-637X/785/2/82}

\bibitem[{{Socrates} {et~al.}(2005){Socrates}, {Blaes}, {Hungerford}, \&
  {Fryer}}]{2005ApJ...632..531S}
{Socrates}, A., {Blaes}, O., {Hungerford}, A., \& {Fryer}, C.~L. 2005, \apj,
  632, 531, \dodoi{10.1086/431786}

\bibitem[{{Spruit}(1992)}]{1992A&A...253..131S}
{Spruit}, H.~C. 1992, \aap, 253, 131

\bibitem[{{Spruit}(2013)}]{2013A&A...552A..76S}
---. 2013, \aap, 552, A76, \dodoi{10.1051/0004-6361/201220575}

\bibitem[{{Stathakis} {et~al.}(1991){Stathakis}, {Dopita}, {Cannon}, \&
  {Sadler}}]{1991supe.conf...95S}
{Stathakis}, R.~A., {Dopita}, M.~A., {Cannon}, R.~D., \& {Sadler}, E.~M. 1991,
  in Supernovae, ed. S.~E. {Woosley}, 95

\bibitem[{{Sugerman} {et~al.}(2005{\natexlab{a}}){Sugerman}, {Crotts},
  {Kunkel}, {Heathcote}, \& {Lawrence}}]{2005ApJ...627..888S}
{Sugerman}, B.~E.~K., {Crotts}, A.~P.~S., {Kunkel}, W.~E., {Heathcote}, S.~R.,
  \& {Lawrence}, S.~S. 2005{\natexlab{a}}, \apj, 627, 888,
  \dodoi{10.1086/430396}

\bibitem[{{Sugerman} {et~al.}(2005{\natexlab{b}}){Sugerman}, {Crotts},
  {Kunkel}, {Heathcote}, \& {Lawrence}}]{2005ApJS..159...60S}
---. 2005{\natexlab{b}}, \apjs, 159, 60, \dodoi{10.1086/430408}

\bibitem[{{Sukhbold} {et~al.}(2016){Sukhbold}, {Ertl}, {Woosley}, {Brown}, \&
  {Janka}}]{2016ApJ...821...38S}
{Sukhbold}, T., {Ertl}, T., {Woosley}, S.~E., {Brown}, J.~M., \& {Janka}, H.-T.
  2016, \apj, 821, 38, \dodoi{10.3847/0004-637X/821/1/38}

\bibitem[{{Sunyaev} {et~al.}(1987){Sunyaev}, {Kaniovsky}, {Efremov},
  {Gilfanov}, {Churazov}, {Grebenev}, {Kuznetsov}, {Melioranskiy},
  {Yamburenko}, {Yunin}, {Stepanov}, {Chulkov}, {Pappe}, {Boyarskiy},
  {Gavrilova}, {Loznikov}, {Prudkoglyad}, {Rodin}, {Reppin}, {Pietsch},
  {Engelhauser}, {Truemper}, {Voges}, {Kendziorra}, {Bezler}, {Staubert},
  {Brinkman}, {Heise}, {Mels}, {Jager}, {Skinner}, {Al-Emam}, {Patterson},
  {Willmore}, {Gilfanov}, \& {Churazov}}]{1987Natur.330..227S}
{Sunyaev}, R., {Kaniovsky}, A., {Efremov}, V., {et~al.} 1987, \nat, 330, 227,
  \dodoi{10.1038/330227a0}

\bibitem[{{Suwa} {et~al.}(2010){Suwa}, {Kotake}, {Takiwaki}, {Whitehouse},
  {Liebend{\~A}-rfer}, \& {Sato}}]{2010PASJ...62L..49S}
{Suwa}, Y., {Kotake}, K., {Takiwaki}, T., {et~al.} 2010, \pasj, 62, L49,
  \dodoi{10.1093/pasj/62.6.L49}

\bibitem[{{Takiwaki} {et~al.}(2009){Takiwaki}, {Kotake}, \&
  {Sato}}]{2009ApJ...691.1360T}
{Takiwaki}, T., {Kotake}, K., \& {Sato}, K. 2009, \apj, 691, 1360,
  \dodoi{10.1088/0004-637X/691/2/1360}

\bibitem[{{Takiwaki} {et~al.}(2012){Takiwaki}, {Kotake}, \&
  {Suwa}}]{2012ApJ...749...98T}
{Takiwaki}, T., {Kotake}, K., \& {Suwa}, Y. 2012, \apj, 749, 98,
  \dodoi{10.1088/0004-637X/749/2/98}

\bibitem[{{Takiwaki} {et~al.}(2014){Takiwaki}, {Kotake}, \&
  {Suwa}}]{2014ApJ...786...83T}
---. 2014, \apj, 786, 83, \dodoi{10.1088/0004-637X/786/2/83}

\bibitem[{{Timmes} \& {Swesty}(2000)}]{2000ApJS..126..501T}
{Timmes}, F.~X., \& {Swesty}, F.~D. 2000, \apjs, 126, 501,
  \dodoi{10.1086/313304}

\bibitem[{{Tziamtzis} {et~al.}(2011){Tziamtzis}, {Lundqvist},
  {Gr{\"o}ningsson}, \& {Nasoudi-Shoar}}]{2011A&A...527A..35T}
{Tziamtzis}, A., {Lundqvist}, P., {Gr{\"o}ningsson}, P., \& {Nasoudi-Shoar}, S.
  2011, \aap, 527, A35, \dodoi{10.1051/0004-6361/201015576}

\bibitem[{{Urushibata} {et~al.}(2018){Urushibata}, {Takahashi}, {Umeda}, \&
  {Yoshida}}]{2018MNRAS.473L.101U}
{Urushibata}, T., {Takahashi}, K., {Umeda}, H., \& {Yoshida}, T. 2018, \mnras,
  473, L101, \dodoi{10.1093/mnrasl/slx166}

\bibitem[{{Utrobin} {et~al.}(1995){Utrobin}, {Chugai}, \&
  {Andronova}}]{1995A&A...295..129U}
{Utrobin}, V.~P., {Chugai}, N.~N., \& {Andronova}, A.~A. 1995, \aap, 295, 129

\bibitem[{{Utrobin} {et~al.}(2015){Utrobin}, {Wongwathanarat}, {Janka}, \&
  {M{\"u}ller}}]{2015A&A...581A..40U}
{Utrobin}, V.~P., {Wongwathanarat}, A., {Janka}, H.-T., \& {M{\"u}ller}, E.
  2015, \aap, 581, A40, \dodoi{10.1051/0004-6361/201425513}

\bibitem[{{Utrobin} {et~al.}(2019){Utrobin}, {Wongwathanarat}, {Janka},
  {M{\"u}ller}, {Ertl}, \& {Woosley}}]{2019A&A...624A.116U}
{Utrobin}, V.~P., {Wongwathanarat}, A., {Janka}, H.-T., {et~al.} 2019, \aap,
  624, A116, \dodoi{10.1051/0004-6361/201834976}

\bibitem[{{Varani} {et~al.}(1990){Varani}, {Meikle}, {Spyromilio}, \&
  {Allen}}]{1990MNRAS.245..570V}
{Varani}, G.~F., {Meikle}, W.~P.~S., {Spyromilio}, J., \& {Allen}, D.~A. 1990,
  \mnras, 245, 570

\bibitem[{{Vartanyan} {et~al.}(2019){Vartanyan}, {Burrows}, {Radice},
  {Skinner}, \& {Dolence}}]{2019MNRAS.482..351V}
{Vartanyan}, D., {Burrows}, A., {Radice}, D., {Skinner}, M.~A., \& {Dolence},
  J. 2019, \mnras, 482, 351, \dodoi{10.1093/mnras/sty2585}

\bibitem[{{Viallet} {et~al.}(2013){Viallet}, {Meakin}, {Arnett}, \&
  {Moc{\'a}k}}]{2013ApJ...769....1V}
{Viallet}, M., {Meakin}, C., {Arnett}, D., \& {Moc{\'a}k}, M. 2013, \apj, 769,
  1, \dodoi{10.1088/0004-637X/769/1/1}

\bibitem[{{Viallet} {et~al.}(2015){Viallet}, {Meakin}, {Prat}, \&
  {Arnett}}]{2015A&A...580A..61V}
{Viallet}, M., {Meakin}, C., {Prat}, V., \& {Arnett}, D. 2015, \aap, 580, A61,
  \dodoi{10.1051/0004-6361/201526294}

\bibitem[{{Walborn} {et~al.}(1987){Walborn}, {Lasker}, {Laidler}, \&
  {Chu}}]{1987ApJ...321L..41W}
{Walborn}, N.~R., {Lasker}, B.~M., {Laidler}, V.~G., \& {Chu}, Y.-H. 1987,
  \apjl, 321, L41, \dodoi{10.1086/185002}

\bibitem[{{Wampler} {et~al.}(1990){Wampler}, {Wang}, {Baade}, {Banse},
  {D'Odorico}, {Gouiffes}, \& {Tarenghi}}]{1990ApJ...362L..13W}
{Wampler}, E.~J., {Wang}, L., {Baade}, D., {et~al.} 1990, \apjl, 362, L13,
  \dodoi{10.1086/185836}

\bibitem[{{Weaver} {et~al.}(1978){Weaver}, {Zimmerman}, \&
  {Woosley}}]{1978ApJ...225.1021W}
{Weaver}, T.~A., {Zimmerman}, G.~B., \& {Woosley}, S.~E. 1978, \apj, 225, 1021,
  \dodoi{10.1086/156569}

\bibitem[{{Weiss} {et~al.}(1988){Weiss}, {Hillebrandt}, \&
  {Truran}}]{1988A&A...197L..11W}
{Weiss}, A., {Hillebrandt}, W., \& {Truran}, J.~W. 1988, \aap, 197, L11

\bibitem[{{West} {et~al.}(1987){West}, {Lauberts}, {Jorgensen}, \&
  {Schuster}}]{1987A&A...177L...1W}
{West}, R.~M., {Lauberts}, A., {Jorgensen}, H.~E., \& {Schuster}, H.~E. 1987,
  \aap, 177, L1

\bibitem[{{Wongwathanarat} {et~al.}(2010){Wongwathanarat}, {Janka}, \&
  {M{\"u}ller}}]{2010ApJ...725L.106W}
{Wongwathanarat}, A., {Janka}, H.-T., \& {M{\"u}ller}, E. 2010, \apjl, 725,
  L106, \dodoi{10.1088/2041-8205/725/1/L106}

\bibitem[{{Wongwathanarat} {et~al.}(2013){Wongwathanarat}, {Janka}, \&
  {M{\"u}ller}}]{2013A&A...552A.126W}
---. 2013, \aap, 552, A126, \dodoi{10.1051/0004-6361/201220636}

\bibitem[{{Wongwathanarat} {et~al.}(2015){Wongwathanarat}, {M{\"u}ller}, \&
  {Janka}}]{2015A&A...577A..48W}
{Wongwathanarat}, A., {M{\"u}ller}, E., \& {Janka}, H.-T. 2015, \aap, 577, A48,
  \dodoi{10.1051/0004-6361/201425025}

\bibitem[{{Woosley}(1987)}]{1987IAUS..125..255W}
{Woosley}, S.~E. 1987, in IAU Symposium, Vol. 125, The Origin and Evolution of
  Neutron Stars, ed. D.~J. {Helfand} \& J.-H. {Huang}, 255--270

\bibitem[{{Woosley}(1988)}]{1988ApJ...330..218W}
{Woosley}, S.~E. 1988, \apj, 330, 218, \dodoi{10.1086/166468}

\bibitem[{{Woosley} {et~al.}(2002){Woosley}, {Heger}, \&
  {Weaver}}]{2002RvMP...74.1015W}
{Woosley}, S.~E., {Heger}, A., \& {Weaver}, T.~A. 2002, Reviews of Modern
  Physics, 74, 1015, \dodoi{10.1103/RevModPhys.74.1015}

\bibitem[{{Woosley} {et~al.}(1988){Woosley}, {Pinto}, \&
  {Ensman}}]{1988ApJ...324..466W}
{Woosley}, S.~E., {Pinto}, P.~A., \& {Ensman}, L. 1988, \apj, 324, 466,
  \dodoi{10.1086/165908}

\bibitem[{{Xiong}(1977)}]{1977AcASn..18...86X}
{Xiong}, D.-R. 1977, Acta Astronomica Sinica, 18, 86

\bibitem[{{Xiong} {et~al.}(1997){Xiong}, {Cheng}, \&
  {Deng}}]{1997ApJS..108..529X}
{Xiong}, D.~R., {Cheng}, Q.~L., \& {Deng}, L. 1997, \apjs, 108, 529,
  \dodoi{10.1086/312959}

\bibitem[{{Yamada} \& {Sato}(1991)}]{1991ApJ...382..594Y}
{Yamada}, S., \& {Sato}, K. 1991, \apj, 382, 594, \dodoi{10.1086/170746}

\bibitem[{{Yoshida} {et~al.}(2019){Yoshida}, {Takiwaki}, {Kotake}, {Takahashi},
  {Nakamura}, \& {Umeda}}]{2019ApJ...881...16Y}
{Yoshida}, T., {Takiwaki}, T., {Kotake}, K., {et~al.} 2019, \apj, 881, 16,
  \dodoi{10.3847/1538-4357/ab2b9d}

\bibitem[{{Zaussinger} \& {Spruit}(2013)}]{2013A&A...554A.119Z}
{Zaussinger}, F., \& {Spruit}, H.~C. 2013, \aap, 554, A119,
  \dodoi{10.1051/0004-6361/201220573}

\bibitem[{{Zhang}(2013)}]{2013ApJS..205...18Z}
{Zhang}, Q.~S. 2013, \apjs, 205, 18, \dodoi{10.1088/0067-0049/205/2/18}

\bibitem[{{Zhang}(2016)}]{2016ApJ...818..146Z}
---. 2016, \apj, 818, 146, \dodoi{10.3847/0004-637X/818/2/146}

\end{thebibliography}

\end{document}